\documentclass[12pt,a4paper]{article}

\usepackage[margin=2.5cm]{geometry}
\usepackage[T1]{fontenc}
\usepackage[utf8]{inputenc}
\usepackage{caption}

\newcommand{\TableDescription}[1]{%
    \par
    \begin{minipage}{\textwidth}
        \footnotesize
        \raggedright
        #1
    \end{minipage}
    \par\vspace{0.6em}
}

\newcommand{\FMcell}[1]{%
  \begin{tabular}[t]{@{}c@{}}
  #1
  \end{tabular}%
}
\usepackage{pgfplots}
\pgfplotsset{compat=1.18}
\usepackage{rotating}
\usepackage{ragged2e}

\usepackage{microtype}
\usepackage{setspace}
\usepackage{parskip}
\usepackage{titlesec}
\usepackage{titling}
\usepackage{abstract}
\usepackage{booktabs}
\usepackage{tabularx}
\usepackage{array}
\newcolumntype{L}[1]{>{\raggedright\arraybackslash}p{#1}}
\newcolumntype{C}[1]{>{\centering\arraybackslash}p{#1}}
\newcolumntype{R}[1]{>{\raggedleft\arraybackslash}p{#1}}
\usepackage{longtable}
\usepackage{multirow}
\usepackage{graphicx}
\usepackage{float}
\usepackage{caption}
\usepackage{subcaption}
\usepackage{amsmath}
\usepackage{amssymb}
\usepackage{mathtools}
\usepackage[authoryear]{natbib}
\usepackage[hidelinks,colorlinks=true,linkcolor=darkblue,citecolor=darkblue,urlcolor=darkblue]{hyperref}
\usepackage[table]{xcolor}
\usepackage{fancyhdr}
\usepackage{enumitem}
\usepackage{footnote}
\usepackage{url}
\usepackage[T1]{fontenc}
\hypersetup{
    colorlinks=true,
    citecolor=darkblue
}

\definecolor{darkblue}{RGB}{30,64,87}
\definecolor{lightblue}{RGB}{200,216,232}
\definecolor{rulecolor}{RGB}{30,64,87}
\definecolor{tableheader}{RGB}{197,218,237}
\definecolor{rowalt}{RGB}{239,245,251}
\definecolor{techgreen}{RGB}{235,245,235}
\definecolor{infrorange}{RGB}{255,243,224}
\definecolor{socpurple}{RGB}{243,229,245}
\definecolor{govblue}{RGB}{232,234,246}

\titleformat{\section}
  {\Large\bfseries\color{black}}
  {\thesection.}{0.6em}{}

\titleformat{\subsection}
  {\large\bfseries\color{black}}
  {\thesubsection}{0.6em}{}

\titlespacing*{\section}{0pt}{18pt}{8pt}
\titlespacing*{\subsection}{0pt}{12pt}{4pt}

\newcolumntype{L}[1]{>{\raggedright\arraybackslash}p{#1}}
\newcolumntype{C}[1]{>{\centering\arraybackslash}p{#1}}
\newcolumntype{R}[1]{>{\raggedleft\arraybackslash}p{#1}}

\title{%
  \vspace{-1cm}
  {\LARGE\bfseries\color{black} Are AI Risks Priced in the U.S. Stock Market?}\\[6pt]
  {\large\itshape Evidence from Financial News Factors}
}
\author{}
\date{}

\begin{document}

\maketitle
\vspace{-1.5cm}
\begin{center}
\end{center}
\begin{center}
\large Yanhui (Patti) Shen
\end{center}
\thispagestyle{fancy}
\vspace{0cm}
\vspace{0.8cm}

\begin{abstract}
\noindent
This paper asks whether firms' exposures to news about different types of AI risk are priced in U.S.\ stock returns. Using AI and risk keywords, I identify 7,787 Wall Street Journal articles from January 2016 to December 2025. I combine latent Dirichlet allocation (LDA) with the Domain Taxonomy in the MIT AI Risk Repository to construct four news-based systematic risk factors. I estimate betas to factor innovations and test pricing with univariate portfolio analysis and Fama--MacBeth regressions. Only the taxonomy-mapped Misinformation factor (D3) is robustly priced. Its high-minus-low beta portfolio earns monthly alphas of 0.49\%--0.57\%, and the estimated D3 price of risk is positive and statistically significant across beta-estimation windows, conventional factor and industry controls, alternative innovation models, and the pre-ChatGPT subsample. The other factors are not reliably priced, indicating that AI-risk pricing is domain-specific.
\end{abstract}

\vspace{0.3cm}

\section{Introduction}
\label{sec:introduction}

Artificial intelligence (AI) has advanced rapidly since early 2010s, with breakthroughs in deep learning around 2012 and the emergence of large language models after 2017 \citep{krizhevsky2012imagenet,he2016deep,vaswani2017attention}. These advances have expanded the role of AI in healthcare, finance, transportation, scientific research, education, and public services \citep{oecd2019artificial}. 
However, the diffusion of AI also raises concerns about privacy, misinformation, malicious use, job displacement, and environmental costs.
Governments and international organizations have responded with major policy initiatives, including the OECD AI Principles, UNESCO's Recommendation on the Ethics of AI, and the EU AI Act \citep{oecd2019recommendation,unesco2021recommendation,europeanunion2024artificial}. These developments suggest that AI is not only a driver of technological progress but also a growing source of economic risk and policy uncertainty.

In this paper, I examine whether firms' exposures to news about different types of AI risk are priced in the cross-section of U.S.\ stock returns. I classify these risk types using the Domain Taxonomy in the MIT AI Risk Repository \citep{slattery2026airisk} and assess which domains carry a nonzero price of risk. The taxonomy contains seven domains:  Discrimination and toxicity (D1), Privacy and security (D2), Misinformation (D3), Malicious actors and misuse (D4), Human–computer interaction (D5), Socioeconomic and environmental harm (D6), and AI system safety, failures, and limitations (D7).

My main hypothesis is that at least one empirically active, domain-specific AI-risk factor carries a nonzero price of risk. Not every type of AI risk should carry a risk premium. Firm-specific harms can be diversified away and therefore should not command a risk premium \citep{sharpe1964capital,ross1976arbitrage}. By contrast, a domain-specific AI-risk factor can be priced only if it captures a common, non-diversifiable state that affects firm payoffs or investor demand. Because the seven domains differ in their potential to capture such common states, the hypothesis does not require every domain to be priced. Section~\ref{sec:conceptual_framework} develops the domain-specific predictions underlying this hypothesis.

To test this hypothesis, I use 7,787 AI-risk articles published by The Wall Street Journal between January 2016 and December 2025. First, I use latent Dirichlet allocation (LDA) \citep{blei2003latent} to identify 18 recurring narratives in the news sample. I retain 14 topics with clear AI-risk content based on a manual review of the dominant articles for each topic, supplemented by a machine assessment using GPT-5.6 Sol. Next, I map the 14 retained topics to the 24 subdomains in the MIT taxonomy using semantic similarity and then aggregate the subdomain-level mapping weights into the seven risk domains. This mapping yields four active daily factors: Privacy and security (D2), Misinformation (D3), Malicious actors and misuse (D4), and Socioeconomic and environmental harm (D6). For each active domain, I construct a daily news factor by summing the LDA topic proportions of all articles published that day, weighted by the corresponding topic-to-domain mapping weights. Finally, I estimate firm-level betas to innovations in these daily factors using rolling daily returns and test whether these betas are priced using univariate portfolio analysis and Fama--MacBeth regressions.

The asset-pricing results show clear differences across the four AI-risk domains. Joint tests reject the hypothesis that the prices of risk for D2, D3, D4, and D6 are all zero. Among the four domains, D3 is the only one with a consistently positive and statistically significant price of risk. In the reported three-month beta specification, the long-short portfolio that buys high-D3-beta stocks and sells low-D3-beta stocks earns monthly risk-adjusted alphas of 0.49\% to 0.57\%. The estimated D3 price of risk is positive and statistically significant in all 30 Fama--MacBeth specifications. These specifications include three-, six-, and twelve-month beta-estimation windows and five conventional factor-control models: CAPM, the Fama--French three-factor model (FF3), the Carhart four-factor model (FF3 plus momentum), the Fama--French five-factor model (FF5), and FF5 plus momentum. Each specification is estimated both with and without Fama--French ten-industry controls.
By contrast, D2, D4, and D6 do not show reliable individual pricing evidence.

The empirical content of D3 helps interpret its positive price of risk. To interpret the time-series variation in D3, I conduct a narrative analysis of the news events associated with its prominent peaks, based on the articles making the largest contributions to each peak. The analysis shows that these peaks are primarily driven by news coverage of major developments in AI governance and policy, including the development and passage of the EU AI Act, U.S.\ executive orders and state legislation on AI, and voluntary AI safety commitments by technology firms.
These events suggest that increases in D3 tend to signal greater policy attention, greater regulatory clarity, and lower uncertainty about the AI-policy environment. This interpretation connects to \citet{pastor2013political}, who show that government-policy uncertainty affects stock valuations and risk premia. Because the asset-pricing tests use factor innovations, a positive D3 innovation can therefore be interpreted as an unexpected improvement in AI-policy clarity. High-D3-beta stocks perform relatively well when AI-policy clarity improves and uncertainty declines, but provide less protection when policy clarity is low and uncertainty is high. Investors require compensation for this exposure, which is consistent with the intertemporal asset pricing framework of \citet{merton1973intertemporal}.

The firm-characteristic analysis provides complementary evidence on which firms carry D3 exposure. High-D3-beta firms are associated with higher investment rates, and this relationship remains statistically significant after controlling for other firm characteristics and industry effects. This finding is consistent with \citet{gulen2016policy}, who document a negative relationship between policy uncertainty and corporate investment. If positive D3 innovations tend to signal greater AI-policy clarity and lower uncertainty, investment-intensive firms may be particularly sensitive to changes in the AI-policy environment. Taken together, the peak-event and firm-characteristic evidence suggests that the AI-policy environment is a plausible economic channel linking D3 innovations to firm payoffs or investor demand.

This paper contributes to three lines of literature. First, it extends the literature on intertemporal asset pricing and macroeconomic risk factors to domain-specific AI risk. \citet{merton1973intertemporal} shows that expected returns can compensate investors for exposure to unfavorable stochastic shifts in the investment opportunity set. \citet{chen1986economic} test whether exposures to innovations in five macroeconomic state variables are priced in the cross-section of stock returns, while \citet{bali2017economic} show that exposure to economic uncertainty is priced. I test whether firms' exposures to four domain-specific AI-risk news factors command nonzero cross-sectional risk premia. 
To the best of my knowledge, this is the first beta-pricing test of AI-risk news factors defined using an external taxonomy.

Second, this paper contributes to the literature on AI, firm outcomes, and asset prices. \citet{pastor2009technological} show that uncertainty about a new technology can become systematic as adoption spreads, providing a theoretical foundation for asking whether AI risk is priced. Existing empirical work studies AI primarily from the firm side. \citet{babina2023systematic} document that AI investment increases firms' market betas, \citet{babina2024artificial} show that AI-investing firms grow primarily through product innovation, and \citet{eisfeldt2026generative} construct workforce-based exposure to generative AI and document valuation effects following the release of ChatGPT. These studies measure AI investment or exposure at the firm level and examine how such characteristics relate to market risk, growth, or valuation. I instead construct common, non-tradable news factors for different AI-risk domains and test whether firms' heterogeneous exposures to innovations in these factors command cross-sectional risk premia. This approach fills a gap in the asset-pricing literature by shifting the focus from firm-level AI characteristics to the pricing of heterogeneous exposures to common AI-risk factors.

Third, this paper contributes to the literature that uses textual data to measure systematic risk. \citet{baker2016measuring} and \citet{manela2017news} use newspaper text to measure policy uncertainty and disaster concerns. \citet{engle2020hedging} extract innovations from climate news and construct hedge portfolios, while \citet{faccini2023dissecting} separate climate news into distinct risk factors and find that climate-policy risk is priced. \citet{bybee2023narrative} combine topic modeling with latent factor analysis to extract interpretable systematic factors from The Wall Street Journal. Using corporate risk disclosures, \citet{lopezlira2023risk} identify systematic risk factors and construct factor-mimicking portfolios. \citet{giglio2026biodiversity} construct an aggregate biodiversity news index and show that returns on exposure-sorted portfolios covary with its innovations. This paper extends these approaches to multiple AI-risk domains by combining data-driven topics with an external risk taxonomy.

The remainder of the paper proceeds as follows. Section~\ref{sec:conceptual_framework} develops the conceptual framework and hypothesis. Section~\ref{sec:data} describes the news and financial data. Section~\ref{sec:factor_construction} presents the topic model, taxonomy mapping, and factor construction. Section~\ref{sec:asset_pricing_test} reports the asset-pricing tests. Section~\ref{sec:economic_interpretation} develops the economic interpretation and additional analyses, and Section~\ref{sec:conclusion} concludes.

\section{Conceptual Framework and Hypothesis Development}
\label{sec:conceptual_framework}

\subsection{AI Risk and Its Domain Structure}
\label{subsec:ai_risk_domain_structure}

To study AI risk in an asset-pricing setting, I first define the scope of AI and AI risk. Following the definitions in the updated OECD Recommendation \citep{oecd2024explanatory} and the EU AI Act \citep[Article 3(1)]{europeanunion2024artificial}, I use AI to refer to machine-based systems that operate with some autonomy and generate predictions, content, recommendations, or decisions. Following the definition adopted by \citet{slattery2026airisk}, AI risk refers to the possibility of an adverse outcome associated with the development or deployment of these systems. 

Because AI risk encompasses many distinct forms of harm, I classify these harms using the Domain Taxonomy in the MIT AI Risk Repository \citep{slattery2026airisk}. Based on a systematic review, the Repository compiles 1,725 risks from 74 existing frameworks and organizes them into 24 subdomains within seven parent domains. The seven domains cover Discrimination and toxicity (D1), Privacy and security (D2), Misinformation (D3), Malicious actors and misuse (D4), Human--computer interaction (D5), Socioeconomic and environmental harm (D6), and AI system safety, failures, and limitations (D7). Appendix~\ref{app:mit_taxonomy} reports the complete domain and subdomain definitions. 

The empirical procedure in Section~\ref{sec:factor_construction} maps recurring news narratives to this external classification and determines which domains generate active news factors in the sample. The resulting factors measure how news coverage associated with the predefined risk domains varies over time. This approach converts a static classification into time-varying measures suitable for asset-pricing tests. Because the domain definitions come from an external taxonomy rather than being assigned solely through researcher interpretation of the topic-model output, this design disciplines the economic labeling of the topics and reduces researcher discretion in classifying AI-risk news.

\subsection{When Can AI Risk Be Priced?}
\label{subsec:when_ai_risk_priced}

For an AI-risk factor to be priced, it should capture a common, non-diversifiable state that affects a broad cross-section of firms. However, commonality alone is not sufficient. The common state must also be relevant to asset valuation, either by affecting firms' expected cash flows or by changing investors' required returns or demand for stocks with different exposures. Regulation, litigation exposure, reputation, product demand, production costs, and investment opportunities can affect expected cash flows, while investor preferences and hedging demand can affect required returns. 

Existing studies provide examples of such channels. In the context of sin stocks, \citet{hong2009price} show that norm-constrained institutions hold fewer sin stocks (alcohol, tobacco, and gaming) and these stocks earn higher expected returns. They interpret this return premium as consistent with both neglect by norm-constrained investors and greater litigation risk heightened by social norms. In the context of industrial pollution, \citet{hsu2023pollution} show that environmental policy uncertainty, measured by growth in litigation penalties, helps price portfolios sorted on pollution exposure. Similar channels may transmit common AI-risk states to asset prices. Firms can have different exposures to the common state. If high-beta stocks provide little protection when marginal utility is high or investment opportunities deteriorate, investors will require higher expected returns to hold them \citep{merton1973intertemporal,breeden1979intertemporal}. These conditions connect common AI-risk states to cross-sectional beta pricing.

The domain-level predictions in the next subsection follow this logic: they are based mainly on whether each domain is more likely to contain a common, non-diversifiable component rather than firm-specific risk. The predictions do not specify the exact channel through which a factor affects firm payoffs or investor demand, since it is difficult to determine ex ante whether a given domain operates through regulation, litigation, reputation, demand, or investor preferences. Nor do they determine ex ante whether high-beta stocks provide a hedge in bad states. I treat these mechanisms as ex post empirical questions and return to them when interpreting the results.

\subsection{Domain-Level Predictions and Main Hypothesis}
\label{subsec:domain_predictions}

The seven domains differ in how likely they are to contain a common, non-diversifiable component.
Discrimination and toxicity (D1) includes unfair discrimination and misrepresentation, exposure to toxic content, and unequal AI-system performance across groups. Such events can have substantial consequences for affected users and firms, especially through litigation and reputation. However, many observed incidents are associated with a particular model, application, employer, or customer group. Their effects may therefore remain largely firm-specific unless similar practices or regulatory responses affect many firms simultaneously.

Privacy and security (D2) can create broad losses through data breaches, unauthorized data use, AI system security vulnerabilities and attacks. At the same time, conventional cybersecurity exposure has already been shown to be priced in the cross-section of stock returns \citep{florackis2023cybersecurity}. AI-related privacy and security news may overlap with this broader cybersecurity state. Even if D2 is economically important, isolating a distinct AI-specific priced component may therefore be difficult.

Misinformation (D3) covers false or misleading information, pollution of information ecosystem, and loss of consensus reality. 
AI-generated or amplified misinformation can weaken the shared information environment, reduce trust, affect political processes, and prompt policy responses that extend across firms and industries. Policy uncertainty can influence both valuations and risk premia \citep{pastor2013political}. To the extent that misinformation changes the policy environment or the reliability of information used by firms and investors, D3 is relatively likely to contain a systematic component. 

Malicious actors and misuse (D4) contains both potentially systematic and highly idiosyncratic events. Large-scale disinformation, malicious surveillance, autonomous weapons, and geopolitical uses of AI may produce common shocks. By contrast, fraud, scams, targeted manipulation, or the misuse of a particular system may primarily affect individual firms or users. The mixture of broad and incident-specific narratives makes the pricing implication of D4 less clear.

Human--computer interaction (D5) includes overreliance, loss of human agency, and harms arising from user interaction with AI systems. These concerns can become widespread as adoption grows, but many realized incidents are tied to a particular interface, organization, or user. D5 may therefore contain a weaker common component during a period in which firms differ substantially in their reliance on human--AI interaction.

Socioeconomic and environmental harm (D6) covers labor-market displacement, inequality, resource use, energy demand, and environmental damage. These risks can affect production costs, consumer demand, corporate investment, and policy responses across multiple industries. D6 may therefore contain a common, non-diversifiable component. However, the domain combines several distinct economic narratives. Labor-market disruption, data-center investment, and environmental costs need not affect firms in the same direction. Aggregation may therefore either strengthen a common signal or dilute it through heterogeneous and offsetting effects.

AI system safety, failures, and limitations (D7) includes unreliable performance, lack of robustness and interpretability, multi-agent risks, and potentially severe system-level failures. These risks can ultimately be economy-wide, particularly if highly capable systems become deeply integrated into production and critical infrastructure. However, during an earlier stage of adoption many technical failures may remain concentrated among particular systems and users. At the current stage of AI adoption, a D7 news factor is therefore less likely to carry a nonzero price of risk.

Taken together, D3 appears most likely ex ante to contain a common, non-diversifiable component. D6 may also contain a common component, but the aggregation of economically heterogeneous narratives makes its pricing implication less clear. D2 and D4 have mixed implications because their associated risks may overlap with established sources of risk or combine systematic and incident-specific events. D1, D5, and D7 are more likely to be dominated by firm-, application-, or system-specific content during the sample period. Based on these domain differences, I test the non-directional hypothesis that:

\begin{quote}
\textbf{Main hypothesis.} At least one empirically active, domain-specific AI-risk factor carries a nonzero price of risk in the cross-section of U.S.\ stock returns.
\end{quote}

The factor-construction procedure determines which domains are empirically active in the daily news. Section~\ref{sec:asset_pricing_test} then translates this main hypothesis into formal joint null and alternative hypotheses for the resulting factors and uses the individual price-of-risk estimates to identify which domains account for any rejection of the joint null hypothesis.

\section{Data and Sample Construction}
\label{sec:data}
In this section, I describe the news data and the choice of sample period, explain how I construct the AI-risk news sample that provides the textual input for the domain-specific factors, and present the equity and other financial data used in the asset-pricing analyses.

\subsection{News Source and Sample Period}
\label{subsec:news_source}

Following \citet{boudoukh_information_2019}, who note that news articles are commonly used as a proxy for public information, I construct the news sample from the full-text archive of The Wall Street Journal (WSJ).\footnote{The text includes the article title and main body text.} The WSJ provides broad coverage of firms, financial markets, technology, and regulation, making it well suited to identifying news about AI-related risks and their economic implications.

\begin{figure}[H]
    \centering
    \includegraphics[width=0.82\textwidth]
    {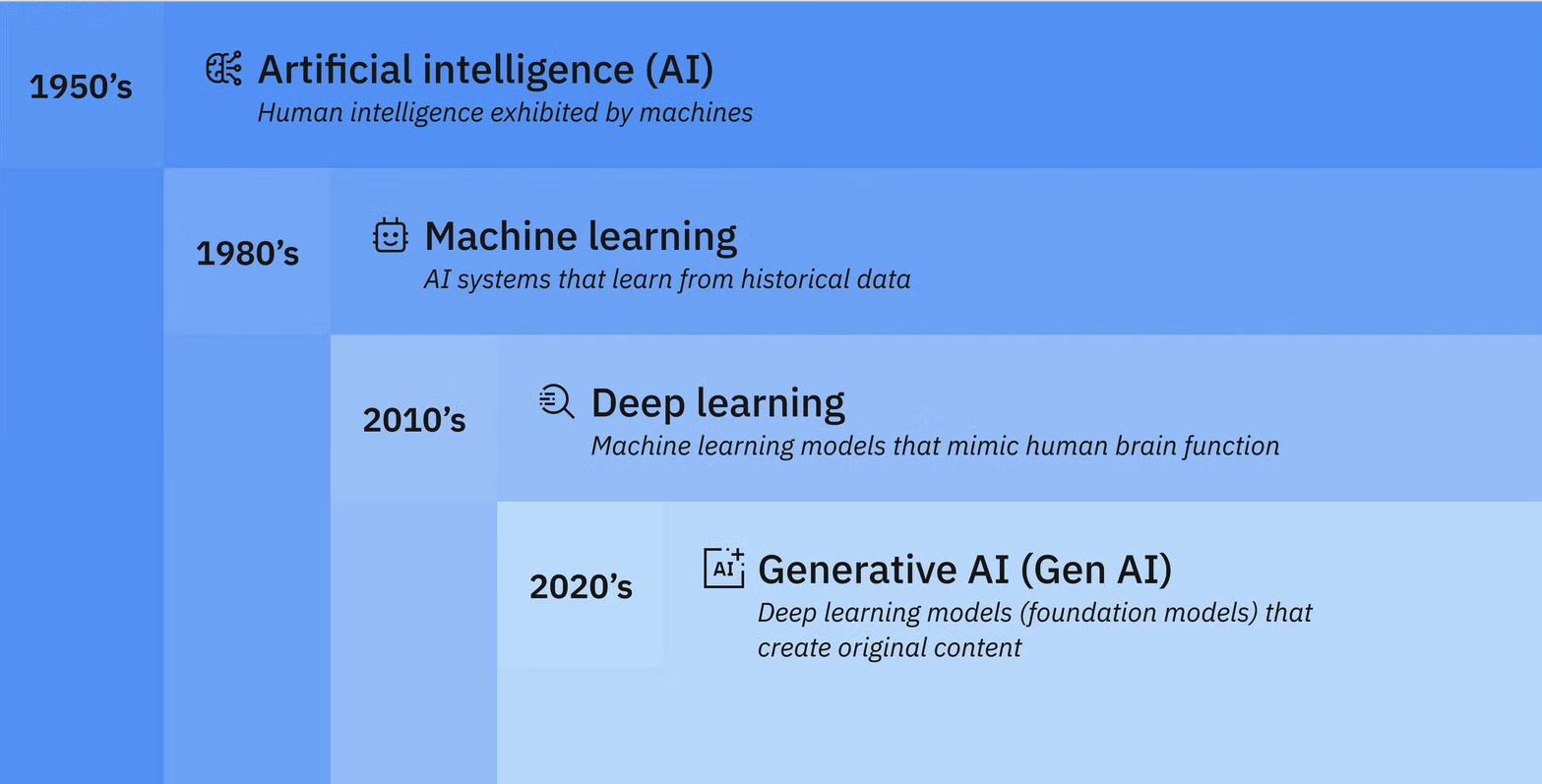}

    \captionsetup{
        justification=raggedright,
        singlelinecheck=false
    }
    \caption{\justifying \noindent \textbf{Evolution of Modern Artificial Intelligence.}
    This figure is sourced from IBM's website (\url{https://www.ibm.com/think/topics/artificial-intelligence})
    and presents a stylized timeline of the development of artificial
    intelligence, machine learning, deep learning, and generative AI.}
    \label{fig:ai_evolution}
\end{figure}

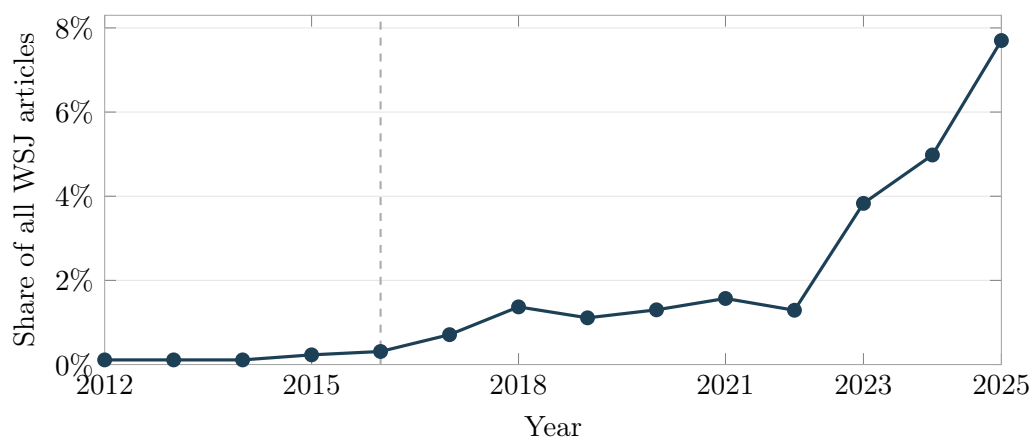
\begin{figure}[H]
\centering
\begin{tikzpicture}
\begin{axis}[
  width=0.84\textwidth,
  height=6.2cm,
  xmin=2012,
  xmax=2025,
  ymin=0,
  ymax=8.3,
  xtick={2012,2015,2018,2021,2023,2025},
  xticklabels={2012,2015,2018,2021,2023,2025},
  scaled x ticks=false,
  ytick={0,2,4,6,8},
  yticklabel={\pgfmathprintnumber{\tick}\%},
  tick label style={font=\small},
  axis line style={draw=gray!70},
  ymajorgrids=true,
  grid style={gray!20},
  xlabel={Year},
  ylabel={Share of all WSJ articles},
  label style={font=\small},
  clip=false
]

\addplot[
  darkblue,
  very thick,
  mark=*,
  mark size=2.2pt
] coordinates {
  (2012,0.11)
  (2013,0.11)
  (2014,0.11)
  (2015,0.23)
  (2016,0.31)
  (2017,0.71)
  (2018,1.37)
  (2019,1.11)
  (2020,1.30)
  (2021,1.57)
  (2022,1.29)
  (2023,3.83)
  (2024,4.98)
  (2025,7.70)
};

\addplot[
  gray!65,
  dashed,
  line width=0.8pt
] coordinates {
  (2016,0)
  (2016,8.3)
};

\end{axis}
\end{tikzpicture}

\captionsetup{
    justification=raggedright,
    singlelinecheck=false
}

\caption{\justifying \noindent \textbf{AI-Risk Coverage in The Wall Street Journal, 2012--2025.} The figure reports the annual number of WSJ articles meeting the AI-risk filtering as a percentage of all WSJ articles in that year. The years 2012--2015 are shown only to provide a pre-sample comparison. The main news sample begins in January 2016.}
\label{fig:wsj_ai_risk_coverage}

\end{figure}

I restrict the sample to articles dated between January 1, 2016 and December 31, 2025.\footnote{All article timestamps refer to the updated times displayed by the WSJ and are interpreted in Eastern Time (ET). The NYSE, AMEX, and Nasdaq also state their trading hours in ET. Using the same time convention keeps the news timestamps aligned with the U.S. equity-market trading calendar.} The sample begins in 2016, placing the analysis in the modern AI era shaped by the rise of deep learning in the 2010s. As illustrated in Figure~\ref{fig:ai_evolution}, deep learning developed as a subfield of machine learning and subsequently became a key technological foundation for the rapid expansion of generative AI in the 2020s. 
Figure~\ref{fig:wsj_ai_risk_coverage} provides additional evidence for this choice. The annual share of AI-risk-related WSJ articles remains low before 2016. It rises from 0.31\% in 2016 to 7.70\% in 2025, showing a clear upward trend and a particularly sharp increase after 2022. This pattern indicates that AI risk has become a prominent, persistent, and measurable topic in public attention since 2016.

\subsection{Construction of the AI-Risk News Sample}
\label{subsec:news_sample}

After excluding non-article records, including standalone video and image pages, the raw WSJ sample contains 387,051 articles published between January 1, 2016 and December 31, 2025. I then clean the retained content by removing advertising scripts, copyright notices and legal disclaimers, structured JSON metadata, author bylines and contact information, and other non-substantive website elements (e.g., navigation links, cookie notices). This text-cleaning step does not remove any additional articles from the sample.

Motivated by the dictionary-based approach to news measurement in \citet{baker2016measuring}, I identify AI-risk-related news articles by requiring the joint occurrence of terms from two predefined lexicons: an AI lexicon and a risk lexicon.\footnote{Following \citet{baker2016measuring}, I retain an article as AI-risk-related if it contains at least one term from the AI lexicon and at least one term from the risk lexicon.} Table~\ref{tab:keyword_lexicons} reports the complete list of keywords included in each lexicon. The AI lexicon contains 33 stable, low-ambiguity identifiers, first based on terminology used in ISO/IEC 22989, the OECD's updated definition of an AI system, the EU AI Act, and the NIST AI Risk Management Framework \citep{isoiec2022concepts,oecd2024explanatory,europeanunion2024artificial,tabassi2023airmf}, and then adapted to the language of general business news. The AI lexicon includes established terms such as  ``AI,'' ``A.I.,''\footnote{Stand-alone uses of ``AI'' and ``A.I.'' are matched only when both letters appear in uppercase. This restriction prevents the surname ``Ai'' and lowercase ``.ai'' domain names from being treated as AI keywords.} ``artificial intelligence,'' ``machine learning,'' ``deep learning,'' and ``neural network,'' as well as commonly recognized terms such as ``chatbot'' and ``deepfake.'' The term ``deepfake'' is retained because it is widely used in news coverage to describe AI-generated or manipulated audio, images, or video. In contrast, I exclude ambiguous standalone abbreviations, such as ``LLM,''\footnote{``LLM'' can refer to either a large language model or a Master of Laws degree.} and specific technical architecture terms, such as ``recurrent neural network'' and ``convolutional neural network,'' while retaining the broader phrase ``language model.''\footnote{The keyword ``language model'' also captures the phrase ``large language model.''} The risk lexicon contains 139 surface forms derived from two conceptual anchors, ``risk'' and ``uncertainty'', including synonyms and closely
related terms such as danger, threat, vulnerability, instability, fear,
and crisis, together with their common grammatical variants. After applying the two-lexicon filtering, I obtain a final sample of 7,787 AI-risk articles, approximately 2\% of the initial sample of 387,051 WSJ articles. This final sample provides the textual input for the topic model described in Section~\ref{sec:factor_construction}.

\subsection{Financial Data and Equity Sample}
\label{subsec:financial_data}

The data used in the asset-pricing analyses come from several sources. Daily and monthly equity data are from the Center for Research in Security Prices (CRSP). I use daily stock returns to estimate firm-level betas at the end of each formation month. In the reported portfolio-analysis specification, each beta is estimated using daily observations over the preceding three months. Monthly stock returns provide the one-month-ahead outcomes used in the second-pass portfolio analysis and Fama--MacBeth regressions. I adjust monthly stock returns for delistings to mitigate the potential upward bias in portfolio performance caused by omitted delisting returns \citep{shumway1997delisting}.\footnote{In CRSP's \texttt{msf\_v2} file, \texttt{mthdelflg} indicates whether the monthly aggregate return includes a delisting return. A value of \texttt{G} means that the delisting return is excluded because there is a gap between the delisting date and the available valuation date, while \texttt{M} means that the delisting return is missing. I set the monthly return to $-100\%$ in both cases.}
The equity sample consists of ordinary common stocks issued by U.S.-incorporated corporations and listed on the NYSE, AMEX, or Nasdaq.\footnote{In CRSP, I require \texttt{sharetype} to equal \texttt{NS}, \texttt{securitytype} to equal \texttt{EQTY}, \texttt{securitysubtype} to equal \texttt{COM}, and \texttt{usincflg} to equal \texttt{Y}. I also require \texttt{issuertype} to equal \texttt{ACOR} or \texttt{CORP} and \texttt{primaryexch} to equal \texttt{N}, \texttt{A}, or \texttt{Q}.} 

Daily and monthly returns on the market, size, value, profitability, investment, and momentum factors, together with the risk-free rate, are from the Kenneth R. French Data Library.\footnote{\url{https://mba.tuck.dartmouth.edu/pages/faculty/ken.french/data_library.html}} The daily factor returns enter the first-pass beta estimations, while the monthly factor returns are used to estimate risk-adjusted portfolio returns. To control for industry effects in the cross-sectional tests, I assign each stock to one of the Fama--French ten industries based on its CRSP SIC code.\footnote{The Fama--French ten-industry definitions are available from the Kenneth R. French Data Library at \url{https://mba.tuck.dartmouth.edu/pages/faculty/ken.french/Data_Library/det_10_ind_port.html}.} For the firm-characteristic analysis in Section~\ref{sec:economic_interpretation}, I supplement the CRSP data with accounting data from Compustat, analyst forecast data from the I/B/E/S Summary Statistics database, and VIX data from the Chicago Board Options Exchange (CBOE). I link the Compustat data to CRSP securities through the CRSP/Compustat Merged database. The I/B/E/S and VIX data are used to construct analyst forecast dispersion and firms' exposure to aggregate market volatility, respectively.

\section{Construction of Domain-Specific AI-Risk Factors}
\label{sec:factor_construction}

This section constructs domain-specific AI-risk factors from the 7,787 Wall Street Journal articles identified in Section~\ref{sec:data}. The construction proceeds through four main stages, which are developed across the subsections below. First, I transform the cleaned article text into a document-term representation and use latent Dirichlet allocation (LDA) to identify recurring news narratives. Second, I review the dominant articles of each topic and retain those with substantive AI-risk content. Third, I map the retained topics to the 24 subdomains in the MIT AI Risk Repository and aggregate the mapping weights into seven parent domains. This procedure produces four active domain-specific factors. Finally, I aggregate article-level topic weights into daily factor levels for each risk domain.

\subsection{Text Preprocessing and Document Representation}
\label{sec:text_preprocessing}

Before estimating the topic model, I transform each article into a standardized sequence of terms. Each article is lowercased, tokenized, and stripped of stopwords. The stopword list extends the standard NLTK English stopword set \citep{bird2009natural} with a supplement of generic, semantically weak terms prevalent in financial news including \textit{say, get, use, make}. Tokens are then lemmatized using the NLTK \texttt{WordNetLemmatizer} with part-of-speech tagging to obtain their appropriate base form (e.g., \textit{markets} $\to$ \textit{market}; \textit{rising} $\to$ \textit{rise}). The stopword filter is applied again after lemmatization to remove tokens whose lemmas appear in the stopword list (e.g., \textit{making} $\to$ \textit{make}, which is filtered post-lemmatization). Together, these steps reduce vocabulary dimensionality and decreases lexical sparsity, both of which improve topic coherence.

Next, frequently co-occurring token pairs and triples are identified and merged into compound tokens using Gensim's \texttt{Phrases} model (e.g., \textit{machine\_learn}, \textit{data\_center}, \textit{artificial\_intelligence}).\footnote{NLTK is used for token-level linguistic preprocessing, including stopword removal, part-of-speech tagging, and WordNet lemmatization. Gensim is used from phrase detection onward because it provides an integrated corpus-level pipeline for phrase formation, dictionary and bag-of-words construction, and subsequent LDA estimation.} Two adjacent words must occur together at least 10 times across the full corpus to be considered as a bigram, while three-word sequences must occur at least 5 times to be considered as trigrams. This step is critical for topic interpretability: without phrase detection, the model would treat the components of composite concepts as independent signals, attenuating the discriminative power of compound terminology that is particularly prevalent in AI news.

Finally, each processed document is represented as a sparse bag-of-words (BoW) vector via a Gensim \texttt{Dictionary}. I remove terms that appear in fewer than 10 documents because they provide limited information about recurring narratives. I also remove terms that appear in more than 50\% of the documents because they provide little discriminative power across topics. The resulting BoW dictionary contains 23,031 unique terms. The final corpus consists of 7,787 BoW vectors, one for each article.

\subsection{Topic Modeling and Selection of Optimal Topic Count}
\label{sec:lda_model}
Topic modeling summarizes a corpus through a set of latent themes. Among available methods, latent Dirichlet allocation (LDA) \citep{blei2003latent} is widely used and represents documents in a corpus as mixtures of latent topics. Each topic is a probability distribution over the vocabulary, while each document is a probability distribution over topics. This soft-assignment structure is useful for financial news because a single article can discuss several related narratives. For example, an article about an AI data center may discuss investment, electricity demand, employment, and environmental effects at the same time. 

LDA has been widely adopted in economics and finance. \cite{larsen2019value} and \cite{thorsrud2020words} apply LDA to Norwegian newspaper data to forecast macroeconomic variables. \cite{hansen2018transparency} use LDA to study Federal Reserve communication. \cite{dyer2017evolution} quantify disclosure trends in 10-K filings using LDA. \cite{adammer2020forecasting} extract topics from newspaper articles and use them to forecast the equity premium. 

In my setting, LDA provides an intermediate layer between individual news articles and the externally predefined MIT taxonomy. By reducing the article-level text to a small set of recurring topics, this step lowers the dimensionality of the subsequent semantic mapping and makes it less sensitive to article-specific wording. More importantly, the LDA layer improves the interpretability of the
classification. I can assess the validity of the resulting
topic-to-domain mappings by examining each topic's top words and
dominant articles. The topics also provide a data-driven description of the main narratives in the AI-risk news sample.

Let \(K\) denote the number of topics and \(V\) the number of unique terms in the BoW dictionary. Topic \(k\) is characterized by the topic-word distribution
\(\boldsymbol{\beta}_{k}=(\beta_{k,1},\ldots,\beta_{k,V})\), and article \(d\) is characterized by the topic mixture
\(\boldsymbol{\theta}_{d}=(\theta_{d,1},\ldots,\theta_{d,K})\). Let \(w_{d,n}\) denote the \(n\)-th token in article \(d\), and let \(v\in\{1,\ldots,V\}\) index a term in the BoW dictionary. Conditional on the article's topic mixture and the topic-word distributions, the probability that \(w_{d,n}\) corresponds to dictionary term \(v\) is
\begin{equation}
\Pr(w_{d,n}=v
\mid \boldsymbol{\theta}_{d},
\boldsymbol{\beta}_{1},\ldots,\boldsymbol{\beta}_{K})
=
\sum_{k=1}^{K}\theta_{d,k}\beta_{k,v}.
\end{equation}
Here, \(\theta_{d,k}\) is the weight of topic \(k\) in article \(d\), while \(\beta_{k,v}\) is the probability assigned to term \(v\) by topic \(k\). The topic weights satisfy
\(\theta_{d,k}\geq 0\) and \(\sum_{k=1}^{K}\theta_{d,k}=1\). Similarly, the term probabilities satisfy
\(\beta_{k,v}\geq 0\) and \(\sum_{v=1}^{V}\beta_{k,v}=1\). 

LDA places symmetric Dirichlet priors on the article-topic and topic-word distributions, such that
\(\boldsymbol{\theta}_{d}\sim\operatorname{Dirichlet}(\alpha)\)
and
\(\boldsymbol{\beta}_{k}\sim\operatorname{Dirichlet}(\eta)\),
respectively. Estimation identifies distributions that jointly assign high probability to the observed word occurrences while remaining regularized by these priors. I estimate the model using Gensim's online variational Bayes algorithm with 10 passes over the corpus, a chunk size of 2,000 articles, and 200 inference iterations per chunk. Following \cite{heinrich2005parameter}, I set the article-topic prior to \(\alpha=1/K\) and the topic-word prior to \(\eta=0.1\). The relatively low value of \(\alpha\) encourages each article’s topic mixture to concentrate on a smaller number of topics, while the relatively low value of \(\eta\) encourages each topic’s word distribution to concentrate on a smaller number of terms.

The number of topics, \(K\), is a hyperparameter that must be specified in advance. I select its optimal value based on three rules, including coherence, topic diversity, and cross-seed stability. Specifically, I estimate models for 16 candidate values, $K \in \{10, 12, \ldots, 40\}$
and use five random seeds for each candidate:
$\{7, 21, 42, 84, 2025\}$. This grid covers models ranging from broad narratives to more detailed topic structures. For each candidate \(K\), I first calculate the mean \(c_v\) coherence score across the five seeds \citep{roder2015exploring}. Coherence measures the degree to which the highest-probability terms within a topic tend to co-occur in similar contexts: a higher score indicates that the topics are more semantically coherent and interpretable. I next calculate mean topic diversity across seeds, defined as the percentage of unique words in the top 25 words of all topics \citep{dieng2020topic}. A higher value indicates greater diversity across topics. Finally, following \citet{greene2014how}, I assess cross-seed stability by optimally matching topics across each pair of seed-specific models and calculating the Average Jaccard similarity between their ranked top-25 term lists. A higher stability score indicates that models estimated with different random seeds produce topics with similar top terms, suggesting that the resulting topic structure is less sensitive to random initialization.

\begin{figure}[p]
    \centering
    \includegraphics[width=\textwidth]{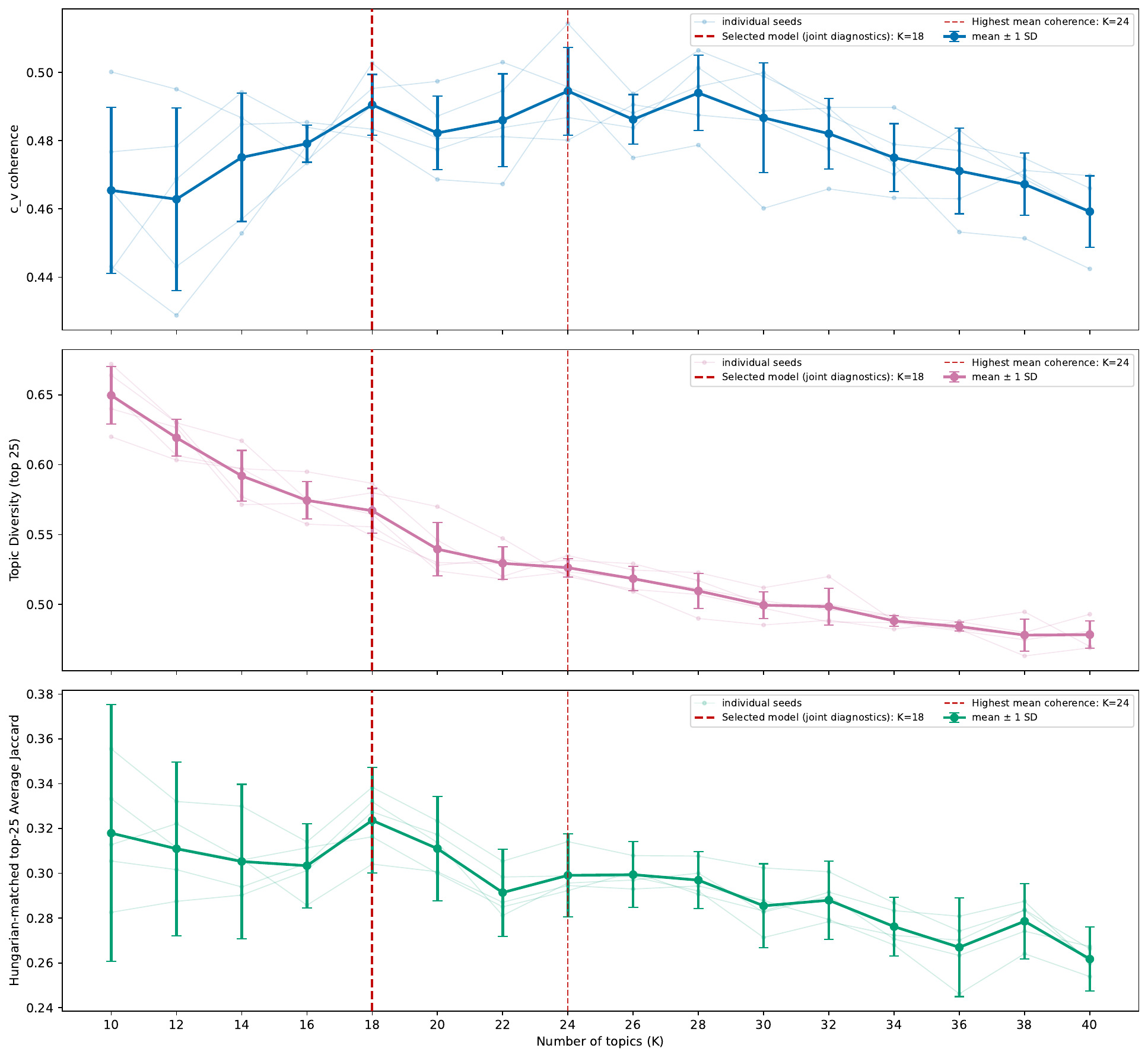}
   \captionsetup{
        justification=raggedright,
        singlelinecheck=false
    }
    \caption{\justifying \noindent \textbf{Selection of the number of LDA topics.}
    The top, middle, and bottom panels report mean \(c_v\) coherence,
    topic diversity based on the top 25 terms, and cross-seed topic
    stability, respectively, for \(K\in\{10,12,\ldots,40\}\).
    Light lines show estimates for the five individual seeds, while
    the darker lines and error bars report the mean plus or minus one
    standard deviation. The vertical lines mark the selected model
    \(K=18\) and the coherence-maximizing model \(K=24\).}
    \label{fig:k_selection_diagnostics}
\end{figure}

\begin{figure}[p]
    \centering
    \includegraphics[width=0.85\textwidth]
    {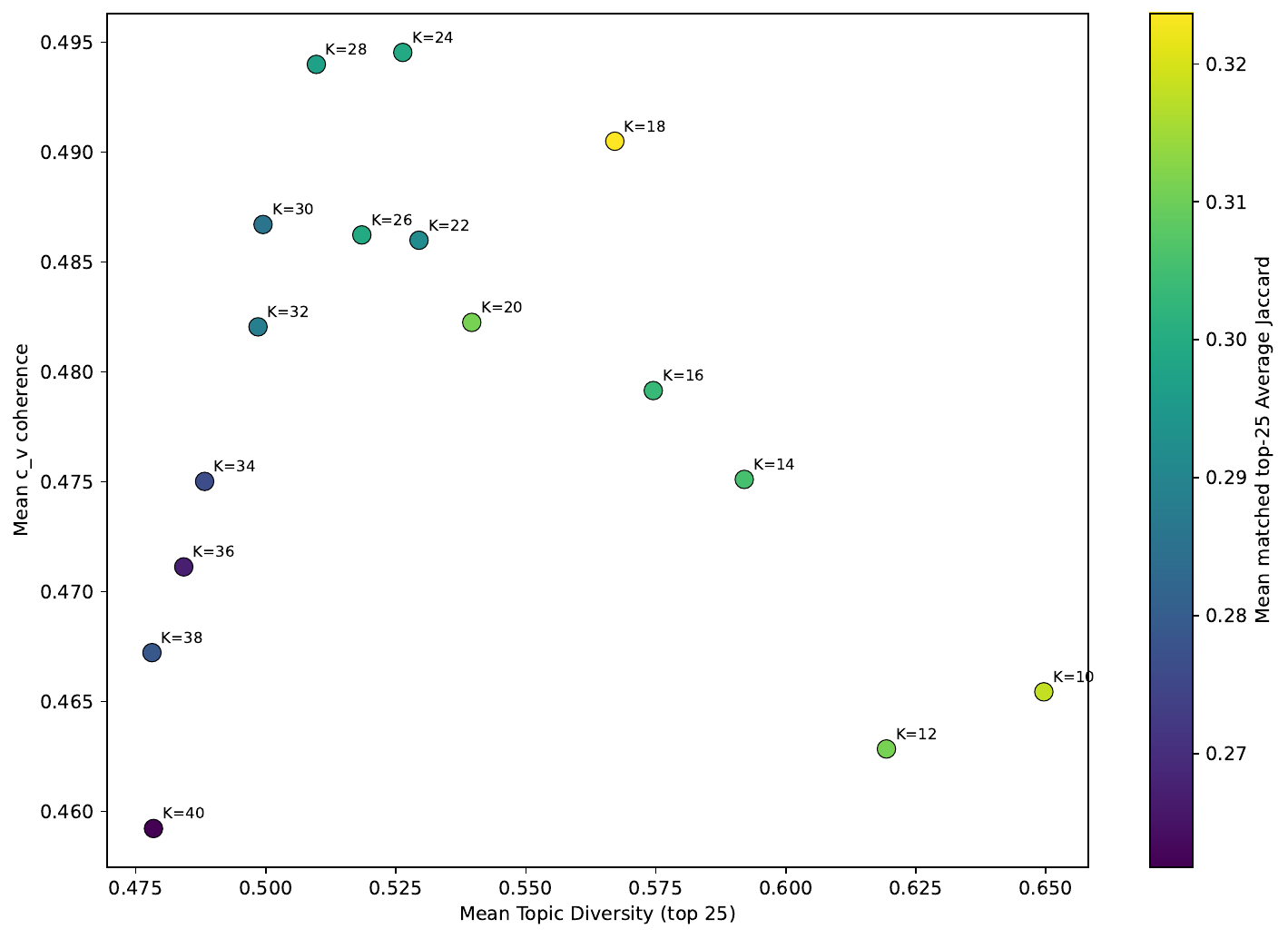}

    \captionsetup{
        justification=raggedright,
        singlelinecheck=false
    }
    \caption{\justifying \noindent \textbf{Trade-offs among LDA model-selection diagnostics.}
    Each point represents a candidate number of topics. The horizontal
    axis reports mean topic diversity, the vertical axis reports mean
    \(c_v\) coherence, and color indicates mean cross-seed stability.
    Relative to the coherence-maximizing model \(K=24\), \(K=18\)
    retains near-maximal coherence while providing greater topic
    diversity and the highest cross-seed stability.}
    \label{fig:k_selection_tradeoff}
\end{figure}

Figures~\ref{fig:k_selection_diagnostics} and
\ref{fig:k_selection_tradeoff} summarize the model-selection results across the full grid of candidate topic counts. Figure~\ref{fig:k_selection_diagnostics} shows that \(K=18\) attains the highest cross-seed stability in the grid and retains substantially greater topic diversity than \(K=24\). The coherence of \(K=18\) is only 0.004 lower than the maximum at \(K=24\). Figure~\ref{fig:k_selection_tradeoff} displays these three dimensions jointly: \(K=18\) lies close to the maximum-coherence region while combining relatively high topic diversity with the strongest cross-seed stability. I therefore select \(K=18\) as providing the best overall balance among semantic coherence, topic distinctiveness, and robustness to random initialization. Among the five \(K=18\) estimates, seed 84 has the highest average similarity to the other seed-specific models and is therefore used as the representative model in the remaining analysis.

\subsection{LDA Topics and AI-Risk Relevance Review}
\label{sec:topic_relevance}

\begin{sidewaysfigure}[p]
    \centering
    \includegraphics[
        width=0.95\textheight,
        keepaspectratio
    ]{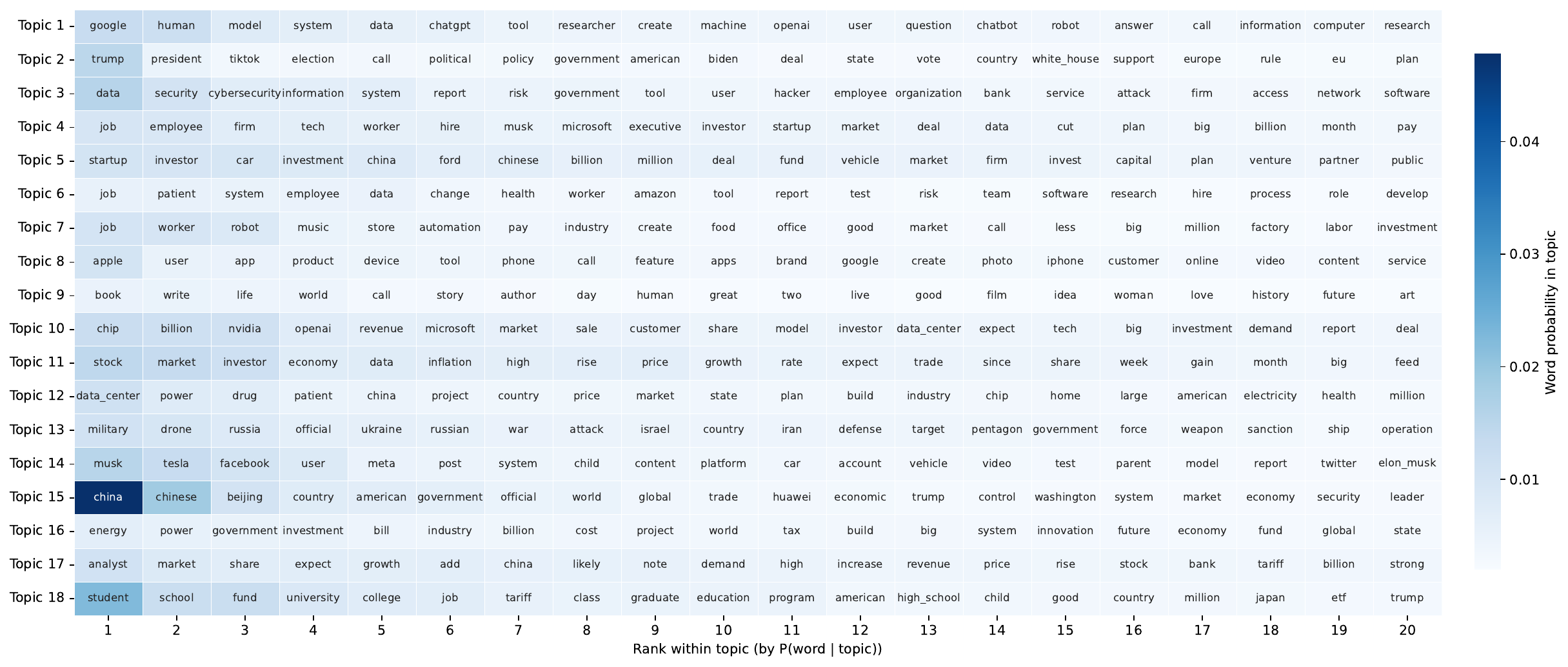}

    \captionsetup{
        width=0.95\textheight,
        justification=raggedright,
        singlelinecheck=false
    }
    \caption{\justifying \noindent \textbf{Top Terms of the 18 LDA Topics.}
    Each row corresponds to one LDA topic and displays its 20
    highest-probability terms. Terms are ordered from left to right by
    their estimated topic-word probabilities. Darker shading indicates
    a higher probability within the estimated topic-word distribution.}
    \label{fig:topic_heatmap}
\end{sidewaysfigure}

The selected LDA model identifies 18 recurring narratives in the article corpus. Figure~\ref{fig:topic_heatmap} reports the 20 terms with the highest estimated probabilities for each topic. Within each topic, terms are ordered by their topic-word probabilities, with darker shading indicating a higher probability. The heatmap obviously shows that some topics combine terms from several underlying narratives rather than correspond to a single, clearly defined AI-risk concept. For example, Topic 6 combines labor-market terms such as \textit{job}, \textit{employee}, and \textit{worker} with healthcare terms such as \textit{patient} and \textit{health}. Topic 12 similarly combines data-center and energy terms with healthcare terms such as \textit{drug} and \textit{patient}. This pattern arises because LDA groups terms according to their co-occurrence in the corpus; it does not determine whether the resulting topic represents a substantive form of AI risk. Although the keyword-based filtering in Section~\ref{sec:data} requires each article to contain both AI and risk-related language, it does not ensure that every latent topic recovered from the filtered corpus is itself centered on AI risk. I therefore conduct a separate relevance review after selecting \(K=18\). 

For each article \(d\), I identify the topic \(k\) with the highest posterior probability \(\theta_{d,k}\) as its dominant topic. Within each of the 18 topics, I rank the dominant articles by \(\theta_{d,k}\) and review the first 20 articles. This procedure produces 360 non-overlapping topic--article pairs. Each article receives a binary assessment indicating whether it substantively discusses an actual or potential harm associated with an AI capability, system, or use.

I adopt a title-first coding procedure. An article is coded as one when its title clearly identifies an AI-related harm or risk. Examples include ``A `Godfather of AI' Remains Concerned as Ever About Human Extinction,'' ``Microsoft Puts Caps on New Bing Usage After AI Chatbot Offered Unhinged Responses,'' and ``Hackers With AI Are Harder to Stop, Microsoft Says.'' I read the full text of all remaining articles, including those whose titles appear unrelated to AI risk and those whose titles do not provide enough information to determine their relevance. Full-text review is necessary because a title that does not signal AI risk may still be followed by a substantive discussion of AI-related harm. After reading the full text, I code an article as one only if at least one complete paragraph explains how the development or use of AI creates or increases a realized or potential harm. All other articles are coded as zero.


I apply this coding rule manually and use GPT-5.6 Sol as an additional reviewer.\footnote{GPT-5.6 Sol was the most advanced OpenAI model available at the time of the analysis.} The model independently applies the same rule and provides a provisional binary assessment together with potentially relevant passages. I compare its assessment with my manual coding, closely review all disagreements, and make the final classification. Table~\ref{tab:article_relevance_examples} presents four examples that require full-text review. The first two articles are coded as one because their full texts contain complete discussions of AI-related harm, even though their titles do not clearly signal such content. The latter two are coded as zero because their AI and risk-related language refers to separate subjects rather than to risks arising from AI.

After coding the 20 reviewed articles for each topic, I count the number classified as substantively related to AI risk. A topic is retained when at least five of its 20 reviewed articles contain substantive AI-risk content. Table~\ref{tab:topic_relevance_review} reports the number of qualifying articles and the final decision for each topic. Fourteen topics satisfy the threshold. Topics 9, 11, 15, and 17 are removed, with only 3, 1, 3, and 0 qualifying articles, respectively. The retained topics span AI capabilities, government and policy, cybersecurity, employment, healthcare, consumer technology, industrial investment, AI infrastructure, geopolitical conflict, platforms, and education. These 14 topics form the input to the semantic mapping described in the next subsection.

\subsection{Mapping Topics to the MIT AI Risk Repository}
\label{sec:taxonomy_mapping}

I use the Domain Taxonomy in the MIT AI Risk Repository to convert the retained LDA topics into economically interpretable risk domains \citep{slattery2026airisk}. This taxonomy captures the potential harms arising from the development and use of AI. It contains 24 subdomains organized under seven parent domains: Discrimination and toxicity (D1), Privacy and security (D2), Misinformation (D3), Malicious actors and misuse (D4), Human--computer interaction (D5), Socioeconomic and environmental harm (D6), and AI system safety, failures, and limitations (D7). Appendix~\ref{app:mit_taxonomy} provides the complete domain and subdomain definitions. 

Following the approach in \citet{hanley_dynamic_2019}, I map the retained topics to the MIT taxonomy by computing the cosine similarity between a vector representation constructed from each topic's 50 highest-probability terms and a vector representation of each taxonomy subdomain definition. The mapping is conducted at the topic level. This choice preserves the data-driven role of LDA while using the MIT taxonomy as an external source of risk labels. By reducing the semantic mapping from 7,787 individual articles
to 14 recurring narratives, the LDA layer substantially lowers the
computational cost of semantic mapping.

First, I construct a vector representation for each retained topic. For
topic \(k\), I use its 50 highest-probability terms.
Let \(\mathcal{T}_{k}\) denote this set and let \(\mathbf{e}_v\) denote the embedding of the \(v\)th
term in the BoW vocabulary. I construct the topic representation as the probability-weighted average
\begin{equation}
\mathbf{s}_k
=
\frac{
\sum_{v\in\mathcal{T}_k}\beta_{k,v}\mathbf{e}_v
}{
\sum_{v\in\mathcal{T}_k}\beta_{k,v}
}.
\label{eq:topic_embedding}
\end{equation}

The embeddings are produced by the pretrained
\texttt{sentence-transformers} model
\citep{reimers2019sentence}. The top-50 representation concentrates on the terms that most strongly define each topic and limits the influence of low-probability peripheral words.\footnote{As a robustness check, I also construct each topic embedding using its top 100 words rather than its top 50 words. The resulting mappings are very similar: across the 14 retained topics, the mean cosine similarity between the Top50 and Top100 domain-weight vectors is 0.9756. I use the top 50 words in the main analysis to reduce potential noise from lower-probability terms that are less representative of the topic.}

Second, I construct a vector representation of each of the 24 taxonomy subdomains by embedding its definition with the same pretrained model. Let \(\boldsymbol{r}_{j}\) denote the vector representation of the definition of subdomain \(j\). Because the topic representations and subdomain definitions are embedded using the same model, their vectors have the same dimensionality and lie in a common semantic space. This allows them to be compared directly. For each subdomain, I concatenate the title and definition reported in Appendix~\ref{app:mit_taxonomy} and pass the resulting text to the embedding model.

Third, I measure the raw similarity between topic \(k\) and subdomain \(j\) is
\begin{equation}
u_{kj}
=
\operatorname{cosine}
\left(\boldsymbol{s}_{k},\boldsymbol{r}_{j}\right).
\end{equation}
Raw cosine similarity may be high simply because both the topic and the
taxonomy definition contain words that are common in discussions of AI
risk. To separate topic-specific semantic similarity from this general
language overlap, I evaluate each observed similarity against a
matched-null benchmark constructed from the same WSJ BoW dictionary used
to estimate the LDA model. For each topic, I generate 5,000 randomized
representations by replacing each top-50 term with a dictionary term
matched on document-frequency decile and phrase status, while retaining
its original topic-word probability. I then compute the cosine similarity
between each randomized representation and every subdomain definition.
For each topic–subdomain pair, these randomized similarities form the null distribution: the level of similarity expected once a topic's specific semantic content is replaced by frequency- and phrase-matched noise. Let \(\mu^{\mathrm{null}}_{k,j}\) and
\(\sigma^{\mathrm{null}}_{k,j}\) denote the mean and standard deviation of
this null distribution. I standardize the observed similarity as
\begin{equation}
z_{k,j}
=
\frac{
u_{k,j}-\mu^{\mathrm{null}}_{k,j}
}{
\sigma^{\mathrm{null}}_{k,j}
}.
\label{eq:matched_null}
\end{equation}
A positive \(z_{k,j}\) indicates that topic \(k\) is more similar to
subdomain \(j\) than expected under the frequency-matched null, with larger values showing stronger topic-specific association net of general AI-risk language. Next, I apply Sparsemax to the 24 standardized scores for each topic
\citep{martins2016from}:
\begin{equation}
\boldsymbol{q}_{k}
=
\operatorname{Sparsemax}
\left(z_{k,1},\ldots,z_{k,24}\right).
\label{eq:sparsemax}
\end{equation}
The resulting vector \(\boldsymbol{q}_{k}\) is nonnegative, sums to one, and places exact zero weight on subdomains with relatively weak semantic support.\footnote{This sparse mapping improves interpretability by concentrating each topic's weight on the most strongly supported subdomains and prevents weak, diffuse semantic similarities from mechanically contributing to every domain-level factor.} I then aggregate the subdomain weights within each parent domain:
\begin{equation}
p_{k,g}
=
\sum_{j\in\mathcal{D}_{g}}q_{k,j},
\label{eq:domain_mapping}
\end{equation}
where \(\mathcal{D}_{g}\) is the set of subdomains belonging to parent domain \(g\). The mapping therefore allows a topic to contribute to several semantically related subdomains without requiring every subdomain to receive positive weight.

The final topic-domain weights are shown in Table~\ref{tab:topic_domain_mapping}. Four domains receive positive weight: D2 Privacy and security, D3 Misinformation, D4 Malicious actors and misuse, and D6 Socioeconomic and environmental harm. D1, D5, and D7 receive zero weight across all retained topics and therefore do not produce active daily factors.
The mapping does not assign every topic exclusively to a single domain. A topic can contribute to several domains, while a domain can combine contributions from several topics. For example, Topic 2 assigns 35.99\% of its weight to D3, 29.81\% to D4, and 34.20\% to D6. Because D3 receives positive weight only from Topic 2, the D3 daily factor is constructed solely from the component of Topic 2 associated with misinformation. In contrast, D2, D4, and D6 combine contributions from four, five, and ten retained topics, respectively. D6 therefore has the broadest underlying topic composition among the four active domains.

\subsection{Construction of Daily Domain-Specific Factors}
\label{sec:daily_factor_construction}

After developing the topic--domain mapping, I construct each daily domain-specific factor by aggregating the mapped topic weights of all articles published on that day. Let \(\mathcal{A}_{t}\) denote the set of articles in the AI-risk news corpus published on calendar date \(t\), let \(\theta_{d,k}\) denote the weight of topic \(k\) in article \(d\), and let \(p_{k,g}\) denote the mapping weight from topic \(k\) to parent domain \(g\) obtained from Equation~\eqref{eq:domain_mapping}. The daily factor level for domain \(g\) is
\begin{equation}
F_{g,t}
=
\sum_{d\in\mathcal{A}_{t}}
\sum_{k=1}^{14}
\theta_{d,k}p_{k,g}.
\label{eq:daily_domain_factor}
\end{equation}
An article can contribute to several domains when its topic mixture contains several narratives. Its contribution to each domain depends on both its article-topic probabilities and the fixed topic-domain mapping weights.
The factor \(F_{g,t}\) provide a dual signal of quantity and intensity of coverage: it increases when more AI-risk articles are published on date \(t\) or when those articles contain more probability mass associated with domain \(g\) on that day. 

Figure~\ref{fig:domain_factor_timeseries} plots the monthly averages of the four daily domain-specific factor levels from 2016 through 2025.\footnote{Monthly averaging makes the longer-run variation easier to read. For comparison, the asset pricing section includes a figure that plots the corresponding daily factor innovations used in the empirical tests.} All four series show larger or more frequent movements after late 2022, when generative AI became more widely used and reported. Before 2022, the four series are relatively stable, apart from a cluster of peaks in 2018 associated with data-privacy controversies, military and employment risks from AI, and early AI-governance initiatives. The differences in the timing and magnitude of the peaks indicate that the four factors capture distinct domain-specific news content despite the broader increase in AI-risk coverage. To understand which news events underlie the time-series variation in each domain factor, I examine the articles making the largest contributions to selected peaks in the corresponding series. For each peak month, I summarize the main event or group of related events and annotate the corresponding peak in the figure. I next discuss the peak events associated with each factor.

Figure~\ref{fig:d2_event} presents the Privacy and security factor (D2). The factor rises around events involving the unauthorized collection, breach, or exploitation of personal and confidential information. Early peaks include reporting on the CIA's exploitation of Samsung smart televisions, the Cambridge Analytica scandal involving Facebook, facial-recognition privacy disputes, and the Marriott data breach. Subsequent peaks reflect litigation over biometric-data collection and concerns about health data stored on cloud platforms. From 2023 onward, the series increasingly reflects AI-assisted phishing, corporate data leakage, threats to critical national infrastructure, deepfake impersonation, identity fraud, and credential theft. These events show how AI can increase the scale and credibility of existing privacy and cybersecurity threats. The domain--topic composition can also provide a complementary content check. Topics 1, 3, 8, and 14, which contribute to D2, collectively cover model and data risks, cybersecurity, consumer devices and applications, and risks involving users and online platforms. This topic composition is consistent with the privacy, data-security, device, and platform events observed at the peaks of D2.

Figure~\ref{fig:d3_event} presents the Misinformation factor (D3). Its early peaks are associated with European AI ethics initiatives, deepfake threats, synthetic-media regulation, and efforts by online platforms to restrict fake accounts and deceptive content. From 2021 onward, many of its largest movements are associated with institutional responses to AI-generated misinformation and synthetic content, as well as broader efforts to govern AI systems. The relevant events include UNESCO's framework on AI ethics, the development and passage of the EU AI Act, the White House Executive Order on AI, the Bletchley Declaration, voluntary safeguards adopted by technology firms, and enforcement actions involving AI washing. This event-based interpretation is consistent with the topic mapping. D3 receives positive mapped weight only from Topic 2, whose content centers on elections, politics, government policy, AI regulation, and online platforms. Together, the peak events and topic content indicate that D3 captures both AI-related information risks and the institutional responses to them, motivating the governance and policy-clarity interpretation examined later in the paper.

Figure~\ref{fig:d4_event} presents the Malicious actors and misuse factor (D4). Early peaks include reporting on the use of AI to detect terrorist propaganda and the international competition surrounding autonomous weapons. Later movements correspond to the exploitation of social-media algorithms to manipulate public opinion, AI-enabled surveillance, autonomous military systems, and AI-generated political content. The most recent peaks involve model jailbreaking and AI-assisted drone warfare in the conflict between Russia and Ukraine. These events capture the use or attempted use of AI systems by state and non-state actors for political, strategic, or harmful purposes. The topic mapping broadly supports this interpretation, especially for Topics 1, 2, 13, and 14. Their content centers on online platforms and content, political activity, military technologies, and geopolitical conflicts, respectively. 

Figure~\ref{fig:d6_event} presents the Socioeconomic and environmental harm factor (D6). Its early movements are associated with automation, employment, inequality, and changes in the organization of work. The factor rises during the COVID-19 period as firms increase their use of robotics and process automation. Later events include hiring freezes, concerns that AI systems may replace professional judgment in healthcare, and the growing resource demands of AI infrastructure. The largest peaks in 2024 and 2025 are associated with the electricity and water consumed by AI data centers, their carbon emissions, and concerns that large data-center projects create relatively few permanent jobs. D6 therefore captures both labor-market consequences and the environmental costs of expanding AI infrastructure. The topic mapping provides consistent evidence for the interpretation of D6. The domain receives contributions from ten topics, more than any other active domain. Topics 4, 6, and 7 primarily capture employment, workplace, and automation, while Topics 5 and 10 reflect technology investment, production, and computing markets. Topics 12 and 16 capture data-center expansion, electricity demand, energy investment, and infrastructure costs. Topics 2, 8, and 14 provide additional exposure to public policy, consumer technology, and online platforms. The large number of contributing topics also helps explain why D6 spans a wider range of values and reaches higher peaks than the other three factor series. At the same time, the breadth of D6 may have two competing implications for asset pricing. On the one hand, combining multiple related narratives may strengthen the underlying signal and make systematic firm exposure easier to detect. On the other hand, these narratives may affect firms through heterogeneous or offsetting channels, and thus weakening the common priced component of the factor. The asset pricing tests therefore determine whether the breadth of D6 strengthens or dilutes its economic signal.

\section{Asset Pricing Test}
\label{sec:asset_pricing_test}


This section tests the hypothesis stated in Section~\ref{subsec:domain_predictions}: at least one of the four empirically active, domain-specific AI-risk factors carries a nonzero price of risk. As discussed there, D3 appears ex ante most likely to contain a common, non-diversifiable component. D6 may also contain common variation, but its heterogeneous underlying narratives make its pricing implications less clear. The corresponding statistical hypotheses are
\begin{equation}
\begin{aligned}
H_{0}:&\quad
\lambda_{\mathrm{D2}}
=
\lambda_{\mathrm{D3}}
=
\lambda_{\mathrm{D4}}
=
\lambda_{\mathrm{D6}}
=
0,\\
H_{A}:&\quad
\lambda_{g}\neq 0
\quad
\text{for at least one }
g\in
\{\mathrm{D2},\mathrm{D3},\mathrm{D4},\mathrm{D6}\}.
\end{aligned}
\label{eq:asset_pricing_hypothesis}
\end{equation}
Rejecting \(H_{0}\) provides evidence that at least one active AI-risk domain is priced. The individual price-of-risk estimates identify which domain contributes to this rejection. 
In this section, I first construct innovations in the four domain-specific factors and examine their correlations with conventional asset-pricing factors and changes in market volatility. I then estimate firm-level AI-risk betas. Finally, I test the cross-sectional pricing of these betas using univariate portfolio analysis and Fama--MacBeth regressions.


\subsection{Construction of Factor Innovations}
\label{sec:factor_innovations}

The factor levels constructed in Section~\ref{sec:factor_construction}
contain both predictable persistence and unexpected changes in domain-specific
AI-risk news.
Because the asset pricing tests examine firms' exposures to unexpected
changes in these news measures, I extract an innovation from each domain-level factor.\footnote{For expositional convenience, unless otherwise stated, subsequent references to AI-risk factors refer to their innovations rather than their levels.}

I use AR(1) as the baseline innovation model. Because it is parsimonious
and produces a transparent one-step-ahead forecast, AR(1) is commonly used
in empirical asset pricing to separate unexpected changes from persistent
movements in nontradable factors. For example, AR(1) residuals have
been used to construct innovations in climate-news and biodiversity-risk
measures \citep{engle2020hedging,pastor2022dissecting,
giglio2026biodiversity}.

More generally, both autoregressive (AR) and autoregressive moving-average
(ARMA) models provide standard methods for extracting innovations from
persistent time series \citep{box2015time}. An AR(\(p\)) model predicts the
current value of a series using its preceding \(p\) values. An ARMA(\(p,q\)) model additionally includes the preceding \(q\)
forecast errors, thereby allowing past unexpected shocks to continue
affecting the current factor realization.

I compare four candidate innovation models: AR(1), AR(7), ARMA(1,1), and
ARMA(2,1). AR(1) provides the parsimonious baseline. AR(7) includes the
preceding seven calendar days and allows factor persistence to extend over
a complete calendar week. ARMA(1,1) adds the preceding forecast error to the AR(1) model, which provides a parsimonious way to capture the short-run continuation of an unexpected shock. ARMA(2,1) further adds a second autoregressive lag.
For AR(1) and AR(7), I reestimate the coefficients each calendar day by
ordinary least squares using the preceding 365 calendar-day observations.
For ARMA(1,1) and ARMA(2,1), I estimate the parameters by maximum
likelihood using the preceding 365 calendar days at the beginning of each
month and update the model state as new daily observations become
available.\footnote{Unlike AR models, which can be reestimated directly by ordinary least squares, maximum-likelihood estimation of ARMA models generally requires iterative numerical optimization. Daily ARMA refitting would require repeated
nonlinear optimization over highly overlapping samples, substantially
increasing computation and making convergence less stable.} 

I evaluate the relative performance of the four models using the Akaike information criterion (AIC) and Bayesian information criterion (BIC). Both criteria balance model fit
against the number of estimated parameters, with BIC imposing a stronger
penalty for additional complexity. 
Table~\ref{tab:innovation_model_selection} reports the results. The AIC results are mixed: ARMA(1,1) has the lowest mean AIC rank, whereas AR(1) has the lowest AIC in the largest number of domain-month pairs. BIC provides clearer support for AR(1): it has both the lowest mean BIC rank and the largest number of BIC wins. This BIC evidence, together with the parsimony of AR(1), supports its selection as the baseline innovation model. I report results based on AR(7),
ARMA(1,1), and ARMA(2,1) innovations in the robustness analysis.

\begin{figure}[p]
    \centering

    \captionsetup{
        justification=raggedright,
        singlelinecheck=false
    }

    \captionsetup[subfigure]{
        justification=centering,
        singlelinecheck=true,
        font=small,
        skip=2pt
    }

    \begin{subfigure}[t]{0.84\textwidth}
        \centering
        \includegraphics[
            width=\linewidth,
            keepaspectratio
        ]{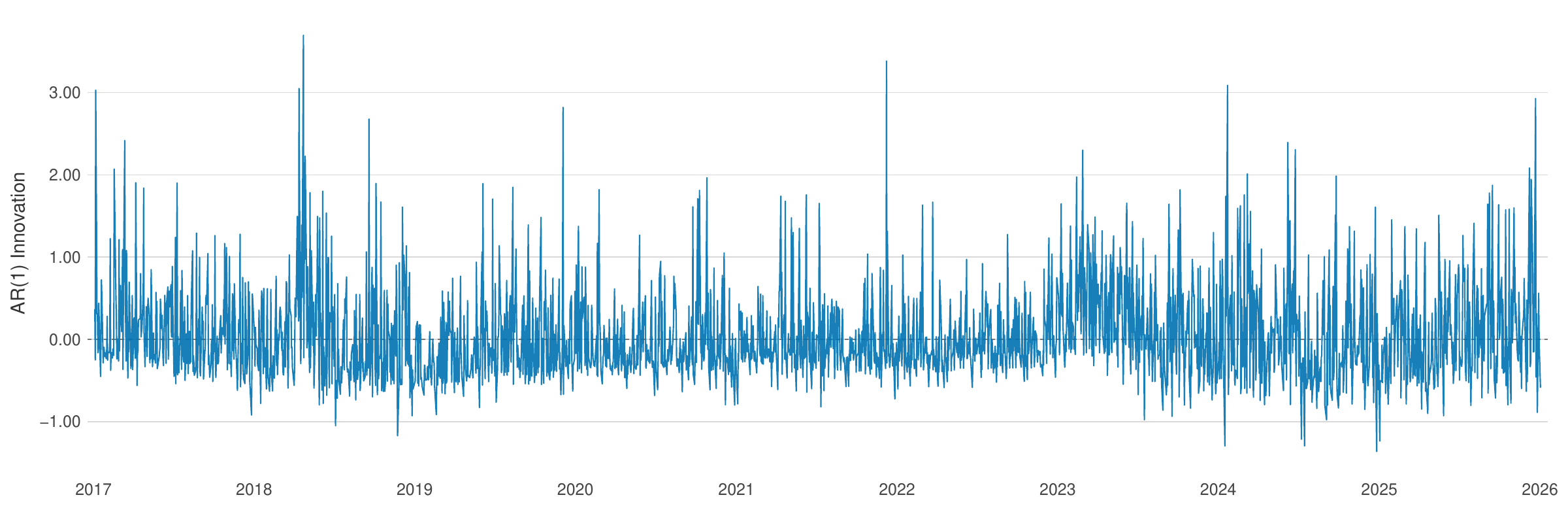}
        \caption{Privacy and security (D2)}
        \label{fig:ar1_innovation_d2}
    \end{subfigure}

    \vspace{0.4em}

    \begin{subfigure}[t]{0.84\textwidth}
        \centering
        \includegraphics[
            width=\linewidth,
            keepaspectratio
        ]{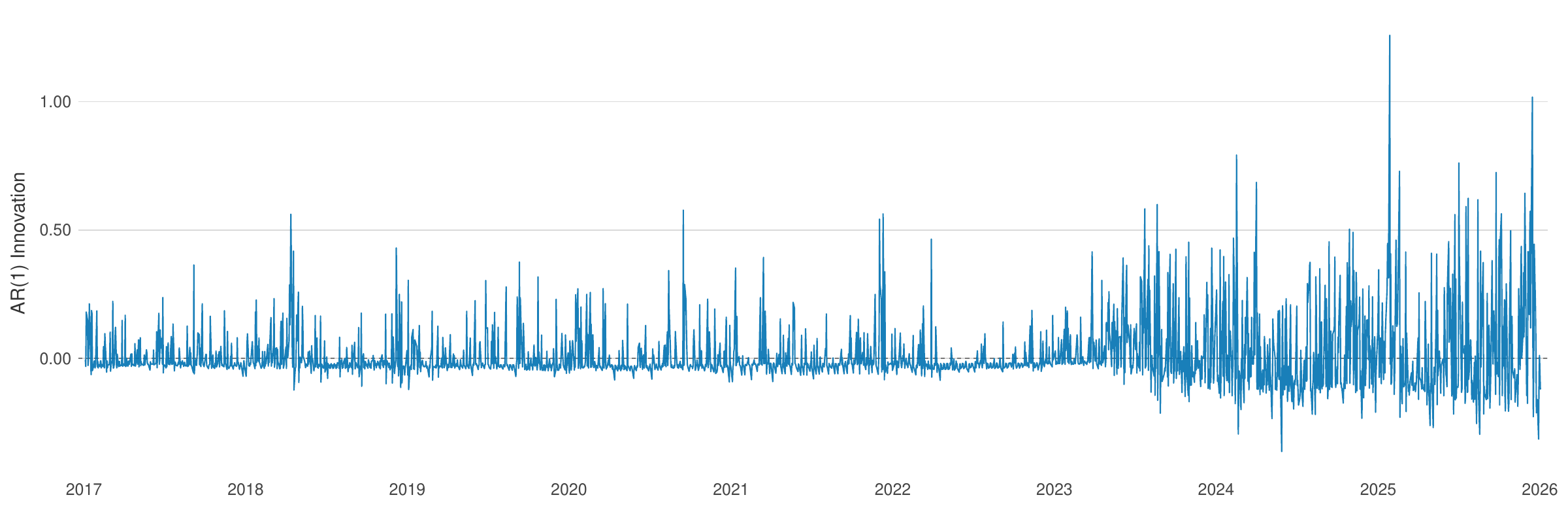}
        \caption{Misinformation (D3)}
        \label{fig:ar1_innovation_d3}
    \end{subfigure}

    \vspace{1.8em}

    \begin{subfigure}[t]{0.84\textwidth}
        \centering
        \includegraphics[
            width=\linewidth,
            keepaspectratio
        ]{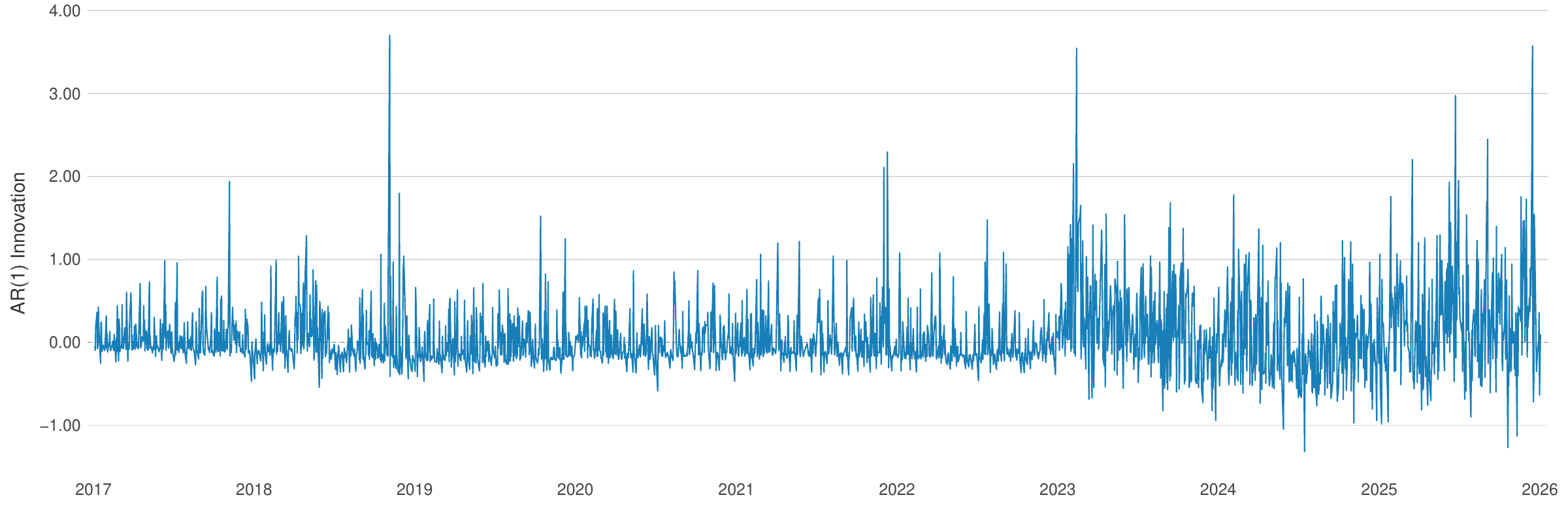}
        \caption{Malicious actors and misuse (D4)}
        \label{fig:ar1_innovation_d4}
    \end{subfigure}

    \vspace{1.8em}

    \begin{subfigure}[t]{0.84\textwidth}
        \centering
        \includegraphics[
            width=\linewidth,
            keepaspectratio
        ]{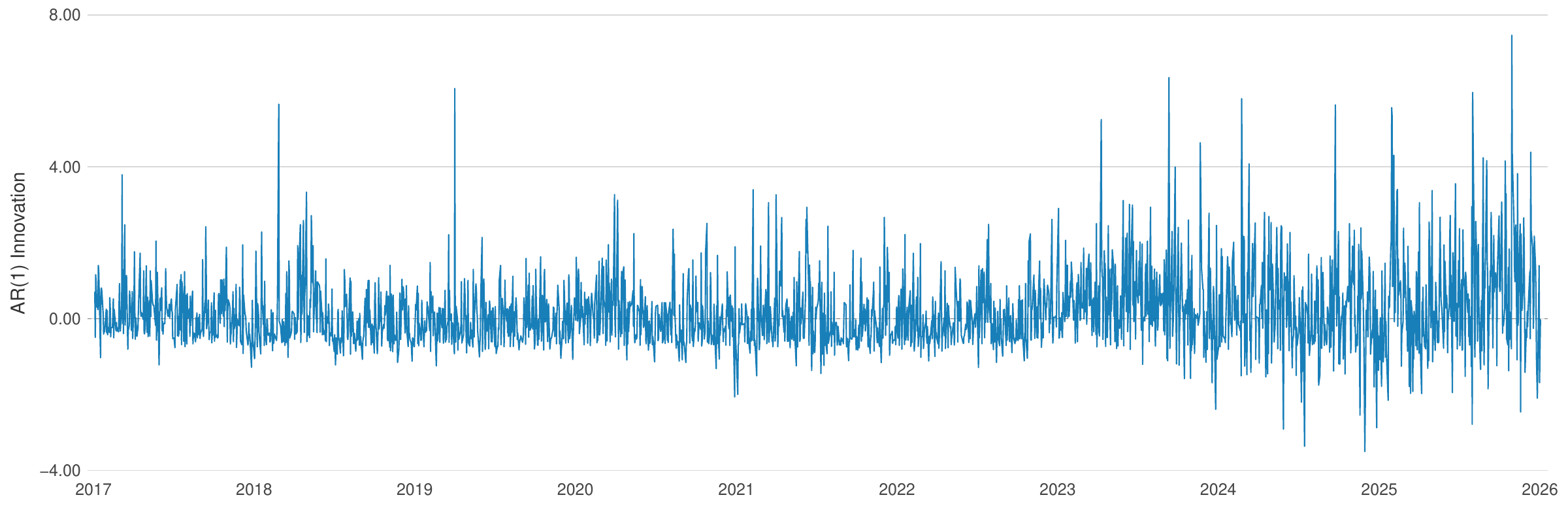}
        \caption{Socioeconomic and environmental harm (D6)}
        \label{fig:ar1_innovation_d6}
    \end{subfigure}

    \caption{\justifying \noindent \textbf{Daily Innovations in Domain-Specific AI-Risk Factors.}
    Panels (a)--(d) plot the AR(1) innovations after
trading-day alignment for Privacy and security
    (D2), Misinformation (D3), Malicious actors and misuse (D4), and
    Socioeconomic and environmental harm (D6), respectively. For each
    domain, the calendar-day innovation is the residual
    from a rolling AR(1) model estimated using the preceding 365 calendar
    days, an intercept, and weekday indicators. Because the AR(1) estimation window requires 365 prior calendar days, the plotted innovation series begins in January 2017.}
    \label{fig:ar1_innovations}
\end{figure}

The factor innovations are initially constructed at the calendar-day
frequency, whereas stock returns are observed on U.S. trading days. To
avoid look-ahead bias, I construct an aligned innovation for each trading
day using only calendar-day innovations dated before that trading day. For example, an ordinary Monday return is matched with the sum of
the innovations from the preceding Friday, Saturday, and Sunday. If Monday
is a market holiday, the Tuesday return is matched with the summed innovations
from Friday through Monday. An innovation dated on a trading day is
therefore matched with the stock return on the following trading day. Figure~\ref{fig:ar1_innovations} plots the four trading-day AR(1) innovation series.
The series fluctuate around zero and contain several large movements associated with changes in the underlying AI-risk news environment. The size and timing of these movements differ across domains, which are consistent with the domain-specific event patterns documented in Section~\ref{sec:daily_factor_construction}.

Finally, I calculate Pearson correlations among the four trading-day
innovations, the six conventional asset-pricing factors, and the first
differences of VIX and the Economic Policy Uncertainty (EPU) index.
I use the first difference of VIX following \citet{ang2006cross}.
EPU is the daily news-based index of \citet{baker2016measuring}, and
I use its first difference following the contemporaneous-return
specification in \citet{brogaard2015assetpricing}.
Table~\ref{tab:innovation_correlations} reports the results. The four
AI-risk innovations are positively correlated with one another, with
correlations ranging from 0.183 to 0.411. This common variation is
consistent with some news events affecting several AI-risk domains at
the same time. The correlations remain well below one, indicating that
the four innovations retain distinct domain-specific information.
The AI-risk innovations are nearly orthogonal to the six conventional
factors and to changes in VIX and EPU. Across these variables, the
largest absolute correlation with an AI-risk innovation is 0.045.
The innovations therefore contain information that is largely distinct
from established market, size, value, profitability, investment, and
momentum factors, as well as changes in market volatility and economic
policy uncertainty.

\subsection{Firm-Level Factor Betas}
\label{sec:firm_level_betas}
I estimate firm-level betas at the end of each month \(m\) using daily stock returns, daily AI-risk factor innovations, and daily conventional factor returns. The estimation window ends on the last trading day of month \(m\), so the estimated betas use only information available by the portfolio formation date. I then match these month-end beta estimates to each stock's delisting-adjusted excess return in month \(m+1\). The factor-innovation series begins in January 2017 because each AR(1) innovation is estimated using the preceding 365 calendar days. Beta estimation therefore also begins in January 2017; for the three-month window, the first holding month is April 2017.

I use two regression specifications to estimate firm-level AI-risk betas. 
First, at the end of each formation month \(m\), for each domain \(g\), I separately estimate
\begin{equation}
\begin{aligned}
R_{i,t}-R_{f,t}
&=
\alpha_i
+
\beta_{i,g}^{AI}\widetilde{F}_{g,t}
+
\boldsymbol{\gamma}_i^{(s)\prime}\boldsymbol{C}_t^{(s)}
+
\varepsilon_{i,t},
\qquad g\in\mathcal{G},
\\[4pt]
\text{where}\qquad
\mathcal{G}
&=
\{\mathrm{D2},\mathrm{D3},\mathrm{D4},\mathrm{D6}\}.
\end{aligned}
\label{eq:single_domain_beta}
\end{equation}
Here, \(R_{i,t}-R_{f,t}\)\footnote{From this subsection onward, \(t\) denotes a U.S. trading day.} is the daily excess return of stock \(i\), \(\widetilde{F}_{g,t}\) is the trading-day AR(1) innovation in domain \(g\), and \(\boldsymbol{C}_t^{(s)}\) contains the established factors in control specification \(s\). The coefficient \(\beta_{i,g}^{AI}\) measures stock \(i\)'s exposure to domain \(g\), conditional on the established factor controls. I estimate Equation~\eqref{eq:single_domain_beta} separately for D2, D3, D4, and D6. The resulting single-domain betas are used to form the univariate portfolios in the next subsection.

Second, I jointly estimate each stock's domain-specific AI-risk betas by including all four AI-risk factors in a single time-series regression. At the end of each formation month \(m\), I estimate
\begin{equation}
\begin{aligned}
R_{i,t}-R_{f,t}
&=
\alpha_i
+
\sum_{g\in\mathcal{G}}
\beta_{i,g}^{AI}\widetilde{F}_{g,t}
+
\boldsymbol{\gamma}_i^{(s)\prime}\boldsymbol{C}_t^{(s)}
+
\varepsilon_{i,t},
\\[4pt]
\text{where}\qquad
\mathcal{G}
&=
\{\mathrm{D2},\mathrm{D3},\mathrm{D4},\mathrm{D6}\}.
\end{aligned}
\label{eq:joint_domain_betas}
\end{equation}
This specification follows the approach in \citet{chen1986economic}. They include several macroeconomic factors in the single first-stage time-series regression and jointly estimate each test asset's factor betas. These factor betas then enter the Fama--MacBeth second-stage cross-sectional regression together. I apply the same procedure to D2, D3, D4, and D6. Equation~\eqref{eq:joint_domain_betas} therefore estimates each AI-risk beta conditional on the other three domain factors and the established factors in \(\boldsymbol{C}_t^{(s)}\).
Both equations are reestimated for each stock at the end of every formation month. 

I use three beta-estimation windows. The three-month window requires at least 50 valid daily observations, the six-month window requires at least 100, and the twelve-month window requires at least 200. These windows and minimum-observation requirements follow the daily-beta procedures in \citet[Section~8.1]{bali2016empirical}. The five first-pass control specifications are the capital asset pricing model (CAPM), the Fama--French three-factor model, the Carhart four-factor model, the Fama--French five-factor model, and the Fama--French five-factor model augmented with momentum (FF5+UMD) \citep{sharpe1964capital,lintner1965valuation,fama1993common,jegadeesh1993returns,carhart1997persistence,fama2015five}. Combining the three estimation windows with the five control specifications yields 15 window--control specifications. I apply these 15 specifications to each of the four single-domain regressions and to the joint four-domain regression, resulting in 60 single-domain beta specifications and 15 joint-domain beta specifications.

\subsection{Univariate Portfolio Analysis}
\label{sec:univariate_portfolios}

I first use univariate portfolio analysis to examine whether firms with different exposures to each AI-risk domain earn different subsequent returns. At the end of each month \(m\), I calculate the 20th, 40th, 60th, and 80th percentiles of the estimated beta distribution using NYSE stocks. I then apply these breakpoints to all eligible NYSE, AMEX, and Nasdaq stocks. Portfolio \(P1\) contains stocks with the lowest betas, and portfolio \(P5\) contains stocks with the highest betas. The high-minus-low portfolio (\(P5-P1\)) is a long--short portfolio that buys \(P5\) and sells \(P1\). Each portfolio is held during month \(m+1\) and value-weighted using market capitalization measured at the end of month \(m\).\footnote{I use value weights to prevent small firms from receiving the same weight as large firms and thereby mitigate the influence of their noisier beta estimates.} 

For each portfolio, I report the average beta used for sorting, the average
monthly excess return, and alphas relative to five benchmark factor models.
Because portfolios are formed at the end of month \(m\) and held during month
\(m+1\), I estimate the portfolio alphas from
\begin{equation}
R^e_{p,m+1}
=
\alpha_p
+
\boldsymbol{b}_p^{(s)\prime}
\boldsymbol{C}_{m+1}^{(s)}
+
e_{p,m+1},
\label{eq:portfolio_alpha}
\end{equation}
where \(R^e_{p,m+1}\) is the excess return of beta-sorted portfolio \(p\) during month
\(m+1\), and \(\boldsymbol{C}_{m+1}^{(s)}\) contains the benchmark factors
observed during the same month. The five benchmark models are the same as
the control specifications used to estimate the firm-level betas: the CAPM,
the Fama--French three-factor model, the Carhart four-factor model, the
Fama--French five-factor model, and the FF5+UMD model. Statistical inference
uses Newey--West standard errors with six lags \citep{newey1987simple}.

To examine whether the portfolios retain different realized exposures during the holding month, I construct next-month post-ranking portfolio betas in two steps. First, for each portfolio formed at the end of month \(m\), I reestimate every constituent stock's beta using daily observations during month \(m+1\), the corresponding domain innovation, and conventional factor controls. Each stock-level regression requires at least 15 valid daily observations, consistent with the one-month daily-beta implementation in \citet[Section~8.1]{bali2016empirical}.
Second, I calculate the portfolio's post-ranking beta as the value-weighted average of these stock-level estimates, with market-capitalization weights fixed at the end of month \(m\). The post-ranking sample contains 105 consecutive holding months from April 2017 through December 2025.

Table~\ref{tab:univariate_portfolio_results} reports all five quantile portfolios (\(P1\)--\(P5\)) and the \(P5-P1\) spread for each domain, based on the three-month beta-estimation window and FF5+UMD first-pass controls. The four domains display different post-ranking exposure and return patterns. Although all post-ranking beta estimates are small in magnitude, only D3 preserves a consistent monotonic ordering from \(P1\) to \(P5\); the other three domains do not. The factor-adjusted alpha measures the portion of excess return that cannot be explained by the portfolio's exposures to established factors. Because the portfolios are formed on AI-risk betas, a positive and statistically significant alpha indicates that high-beta stocks earn higher subsequent returns than low-beta stocks even after accounting for established factors. Such a result provides evidence that the corresponding AI-risk beta contains incremental cross-sectional pricing information. D3 is the only domain whose \(P5-P1\) alphas are positive and statistically significant across the five factor models in the reported specification. Neither the excess-return spread nor the factor-adjusted alpha spreads are statistically significant for D2, D4, or D6. D3 therefore provides the strongest univariate pricing evidence among the four domains.

For D3, the formation beta increases monotonically from \(-0.0450\) for \(P1\) to \(0.0465\) for \(P5\). The corresponding post-ranking beta also increases monotonically from \(-0.00253\) to \(0.00188\), preserving the same cross-portfolio pattern. The return results display a similar ordering. Under the Carhart four-factor, Fama--French five-factor, and FF5+UMD models, portfolio alphas increase monotonically from \(P1\) to \(P5\). The corresponding \(P5-P1\) alphas are \(0.538\%\), \(0.526\%\), and \(0.575\%\) per month, with \(t\)-statistics of \(2.16\), \(1.94\), and \(2.23\). The D3 alpha spread is positive under all five factor models and significant at the 10\% level in every case; it reaches the 5\% level under the Carhart four-factor and FF5+UMD models. By comparison, the raw excess-return spread is \(0.429\%\) per month with a \(t\)-statistic of \(1.52\). The D3 portfolio evidence therefore comes primarily from risk-adjusted returns rather than a statistically significant raw return spread.

The other three domains do not show clear pricing evidence in the portfolio analysis. For D2, the post-ranking beta rises from \(-0.00024\) for \(P1\) to \(0.00065\) for \(P5\), but the \(P5-P1\) excess return is \(-0.471\%\) per month and the five alpha estimates range from \(-0.374\%\) to \(-0.327\%\); none is statistically significant. For D4, the betas of the two extreme portfolios reverse signs during the holding month: the post-ranking beta is \(0.00037\) for \(P1\) and \(-0.00025\) for \(P5\). Its \(P5-P1\) excess return and alphas are positive, but none is statistically significant. For D6, the post-ranking betas do not exhibit a clear monotonic increase from \(P1\) to \(P5\). Its excess-return spread and five alpha estimates are also statistically insignificant.

The other unreported specifications provide additional evidence. For each domain, the three beta-estimation windows and five first-pass control specifications produce 15 high-minus-low portfolios. I evaluate each portfolio using five alpha benchmarks, and yield 75 high-minus-low alpha estimates for each domain. 
For D3, 24 of the 75 alpha estimates are statistically significant. Among these, 10 estimates are significant at the 5\% level, while the remaining 14 are significant at the 10\% level. All statistically significant D3 alphas are positive and come from the three- or six-month beta-estimation windows. In contrast, I find no statistically significant alpha estimates for D2 or D6, even at the 10\% level. D4 produces only 5 alpha estimates that are significant at the 10\% level, all of which arise from the three-month CAPM first-pass specification. The univariate portfolio analysis therefore identifies D3 as the only domain with a recurring positive relation between beta and subsequent risk-adjusted returns.

\subsection{Fama--MacBeth Regressions}
\label{sec:fama_macbeth}

I next use Fama--MacBeth regressions to test whether the four
domain-specific AI-risk betas are priced in the cross section of stock
returns and to estimate their associated prices of risk \citep{fama_risk_1973}. The first-pass regressions include the four AI-risk innovations jointly, as described in Equation~\eqref{eq:joint_domain_betas}. 

At the end of month \(m\), I match the estimated joint AI-risk betas and conventional factor betas to stock \(i\)'s delisting-adjusted excess return in month \(m+1\). For each return month, I estimate
\begin{equation}
\begin{aligned}
r^e_{i,m+1}
={}&
\lambda_{0,m+1}
+
\sum_{g\in\mathcal{G}}
\lambda_{g,m+1}\widehat{\beta}^{AI}_{i,g,m}
+
\boldsymbol{\psi}_{m+1}^{\prime}
\widehat{\boldsymbol{\gamma}}_{i,m}
+
\boldsymbol{\delta}_{m+1}^{\prime}
\boldsymbol{I}_{i,m}
+
\nu_{i,m+1},
\\[4pt]
\text{where}\qquad
\mathcal{G}
&=
\{\mathrm{D2},\mathrm{D3},\mathrm{D4},\mathrm{D6}\}.
\end{aligned}
\label{eq:fama_macbeth}
\end{equation}
Here, \(\widehat{\beta}^{AI}_{i,g,m}\) is the estimated beta for AI-risk domain \(g\) in Equation~\eqref{eq:joint_domain_betas}, and \(\widehat{\boldsymbol{\gamma}}_{i,m}\) contains the estimated conventional factor betas from the same first-pass regression. The vector \(\boldsymbol{I}_{i,m}\) contains nine Fama--French ten-industry\footnote{The Fama--French ten-industry
classification groups firms by four-digit SIC codes into the following
categories: NoDur (consumer nondurables) includes food, tobacco, textiles,
apparel, leather, and toys; Durbl (consumer durables) includes cars,
televisions, furniture, and household appliances; Manuf (manufacturing)
includes machinery, trucks, planes, chemicals, office furniture, paper,
and commercial printing; Enrgy includes oil, gas, and coal extraction and products;
HiTec (business equipment) includes computers, software, and electronic
equipment; Telcm includes telephone and television transmission; Shops includes wholesale,
retail, and some services (laundries and repair shops); Hlth includes healthcare, medical equipment, and pharmaceuticals; Utils represents utilities; Other includes mining, construction, building materials, transportation,
hotels, business services, entertainment, finance, and industries not
assigned to the preceding categories.} indicators when industry controls are included. The residual industry (i.e., Other) is the omitted category.
The estimated price of domain \(g\) is the time-series mean of its monthly cross-sectional coefficients:
\begin{equation}
\widehat{\lambda}_{g}
=
\frac{1}{T}
\sum_{m=1}^{T}
\widehat{\lambda}_{g,m}.
\label{eq:average_risk_price}
\end{equation}
The three-month, six-month, and twelve-month specifications contain 105, 102, and 96 return months, respectively. Their average monthly cross-sectional sample sizes are approximately 3,750, 3,707, and 3,621 stocks.

Table~\ref{tab:fama_macbeth_3m_full} reports the results based on the
three-month beta-estimation window. The estimated D3 price of risk is
positive in all ten specifications. Without industry controls, the D3
estimates range from \(0.0403\) to \(0.0496\), with Newey--West
\(t\)-statistics between \(2.73\) and \(3.49\). With industry controls,
the estimates range from \(0.0369\) to \(0.0429\), with \(t\)-statistics
between \(2.85\) and \(3.56\). All ten D3 estimates are statistically
significant at the 1\% level. In contrast, none of the estimated prices
associated with D2, D4, or D6 is statistically significant. The joint Wald test rejects the null that all four domain prices are zero
at the 5\% level in nine of the ten specifications and at the 10\% level
in the remaining CAPM specification with industry controls, for which the
\(p\)-value is \(0.0533\).
The three-month specifications also produce several positive coefficients
on the market and RMW betas. The market-beta
coefficient is positive and statistically significant at the 10\% level
under the FF5 and FF5+UMD specifications, both with and without industry
controls. It reaches the 5\% level only under FF5 with industry controls.
The RMW-beta coefficient is also positive and statistically significant
at the 10\% level in all four FF5 and FF5+UMD specifications, although
none of these estimates reaches the 5\% level. Among the industry
indicators, the Manuf coefficient is positive and significant at the
10\% level in all five specifications and at the 5\% level under the
CAPM and Carhart four-factor specifications. Because Other is the
omitted industry, these coefficients indicate higher conditional returns
for manufacturing firms relative to firms in the Other category. 

Tables~\ref{tab:fama_macbeth_6m_full} and
\ref{tab:fama_macbeth_12m_full} confirm that the main result is not
specific to the three-month beta-estimation window. The D3 price remains
positive and statistically significant in all 20 longer-window
specifications, while none of the D2, D4, or D6 estimates is significant
even at the 10\% level. All ten six-month D3 estimates are significant at
the 1\% level. The twelve-month estimates have lower \(t\)-statistics,
ranging from \(2.04\) to \(2.20\), but remain significant at the 5\%
level. The joint Wald test rejects the null at the 5\% level in all
six-month specifications. Under the twelve-month window, it rejects at
the 5\% level in seven of the ten specifications and at the 10\% level
in all ten. Among the established-factor betas, the longer-window specifications yield
a similar pattern: the coefficients on the market and RMW betas remain
positive and statistically significant, with significance holding across
a broader set of specifications.
Manuf remains the most consistently positive industry coefficient, and
HiTec becomes significant in several twelve-month specifications.

Across the three beta-estimation windows, all 30 D3 estimates are positive
and statistically significant at the 5\% level. All 20 estimates based on
the three- and six-month windows are significant at the 1\% level. In contrast, none of the 90 estimates associated with D2, D4, or D6 is
statistically significant even at the 10\% level.
The Fama--MacBeth results
therefore reinforce the univariate portfolio evidence. The
pricing evidence for D3 is more uniform across specifications and remains
significant when the four domain betas are estimated jointly and when the
cross-sectional regressions control for established-factor betas and broad
industry effects. These results therefore support the hypothesis that at least one empirically active AI-risk domain
carries a nonzero price of risk and identify D3 as the only domain with
robust pricing evidence in my sample. The next section develops the economic interpretation of this result,
examines the characteristics of high-D3-beta firms, and evaluates its
sensitivity to alternative innovation models.

The asset-pricing tests provide no reliable evidence that D6 is priced. One
possible explanation is that D6 combines heterogeneous narratives concerning
employment displacement, automation, healthcare consequences, data-center
energy demand, and environmental costs. These mixed narratives may have offsetting implications for expected returns.
Aggregating them into a single domain factor may therefore weaken their
common component and obscure pricing effects operating through different
economic channels. The absence of a significant D6 price does not imply that these narratives
are economically unimportant. It shows only that exposure to the combined
D6 factor is not reliably priced in the cross section of stock returns.

\section{Economic Interpretation and Additional Analyses}
\label{sec:economic_interpretation}

The asset-pricing results identify D3 Misinformation as the only active AI-risk domain with consistent pricing evidence. This section develops an economic interpretation of its positive price of risk. I then examine the characteristics of high-D3-beta firms, test its sensitivity to alternative innovation models, and assess whether it remains priced in the pre-ChatGPT subsample.

\subsection{Economic Interpretation of the D3 Pricing Result}
\label{sec:d3_economic_interpretation}

The interpretation of the positive D3 price depends on what an increase in the factor represents. Within the MIT AI Risk Repository, the D3 factor maps to subdomain 3.2, ``Pollution of information ecosystem and loss of consensus reality.'' This subdomain concerns AI-generated misinformation that weakens shared beliefs, social cohesion, and political processes. As shown in Table~\ref{tab:topic_domain_mapping}, D3 receives positive mapped weight only from Topic 2. The empirical content of this topic centers on elections, government policy, AI regulation, political activity, and online platforms. The D3 factor should therefore be interpreted as a news-based state variable that captures both risks to the information environment and institutional responses to those risks. 

The events associated with the prominent D3 peaks help identify the economic direction of this state variable. As shown in Figure~\ref{fig:d3_event}, these peaks are primarily driven by news coverage of major developments in AI governance and policy. These events include the development and passage of the EU AI Act, U.S.\ executive orders and state legislation on AI, the Bletchley Declaration, voluntary safety commitments by technology firms, and enforcement actions involving AI-generated content. These developments are generally intended to address AI-related risks and provide greater clarity about the rules governing the development and use of AI systems. Greater clarity can reduce uncertainty about firms' compliance costs, legal exposure, and product-market decisions. Changes in the information and policy environment may also affect investor demand for firms with different exposures to AI-related information risk. A high D3 factor therefore tends to represent a good state with greater policy attention, more regulatory clarity, and lower uncertainty about the AI-policy environment. Because the asset-pricing tests use factor innovations, a positive D3 innovation indicates an unexpected increase in this news state relative to its predicted level.

This interpretation helps explain the positive relation between D3 beta
and subsequent stock returns. Stocks with high D3 betas perform relatively well when the D3 innovation is positive. If positive D3 innovations tend to indicate greater AI-policy clarity, these stocks perform well when uncertainty about the AI-policy environment declines. However, they provide less protection when regulatory clarity is low or AI-policy uncertainty increases. The higher expected returns on high-D3-beta stocks are therefore consistent with investors requiring compensation for this exposure. This interpretation is consistent with the intertemporal asset-pricing framework of \citet{merton1973intertemporal}, in which expected returns compensate investors for assets that perform poorly when investment opportunities deteriorate. It is also consistent with evidence that government-policy uncertainty affects stock valuations and risk premia \citep{pastor2013political}. Taken together, the content of Topic 2 and the D3 peak-event evidence support an interpretation of D3 as a state variable related to AI-policy clarity and uncertainty.

\subsection{Characteristics of High-D3-Beta Firms}
\label{sec:d3_characteristics}

I next examine which types of firms have greater exposure to D3 innovations. This analysis serves two purposes. First, it provides an economic description of high-D3-beta firms. Second, it assesses whether D3 beta is closely associated with established risk exposures and firm characteristics or concentrated in specific industries.

Following \citet{bali2017economic}, I estimate monthly cross-sectional regressions in which the dependent variable is the firm-level D3 beta. I use the D3 beta estimated from the three-month window with FF5+UMD controls. For each formation month \(m\) from March 2017 through November 2025, the general specification is
\begin{equation}
\widehat{\beta}^{AI}_{i,\mathrm{D3},m}
=
a_m
+
\boldsymbol{\rho}_{m}^{\prime}\boldsymbol{X}_{i,m}
+
\boldsymbol{\phi}_{m}^{\prime}\boldsymbol{I}_{i,m}
+
\zeta_{i,m},
\label{eq:d3_characteristics}
\end{equation}
where \(\boldsymbol{X}_{i,m}\) contains the stock-level characteristics and \(\boldsymbol{I}_{i,m}\) contains the Fama--French ten-industry indicators. I first estimate separate regressions that include one characteristic or one industry indicator at a time. I then estimate a joint regression that includes all characteristics and the complete set of industry indicators. The reported coefficients are the time-series averages of the monthly cross-sectional estimates. 

The explanatory variables cover several dimensions of firm exposure. They include market and volatility exposures, size and value, past returns, liquidity, higher-moment and idiosyncratic risk, analyst forecast dispersion, investment, profitability, lottery-like returns, and industry indicators. I construct these variables based on the definitions in \citet{bali2017economic}, with adaptations for data availability and sample coverage. Appendix~\ref{app:stock_characteristics} provides the detailed definitions and estimation windows.

Table~\ref{tab:d3_characteristics} reports the results. Because of space constraints, I report only Energy, HiTec, and Utility, the three industries most closely related to AI technology and infrastructure. The coefficients on the remaining Fama--French industry indicators are also statistically insignificant. Columns~(1)--(16) present univariate specifications that include one explanatory variable at a time. Among all continuous characteristics and industry indicators, \(\mathrm{I/A}\) is the only variable that is statistically significant at the 1\% level. Its coefficient is \(0.0021\), with a \(t\)-statistic of \(3.72\). High-D3-beta firms therefore tend to have higher asset growth, consistent with greater investment. At the 10\% level, the coefficient on ROE is negative and the coefficient on MAX is positive. These estimates provide weaker evidence that high-D3-beta firms are less profitable and exhibit more lottery-like returns.

Column~(17) presents the joint specification, which includes all continuous characteristics and the full set of Fama--French ten-industry indicators. The coefficient on \(\mathrm{I/A}\) remains positive and statistically significant. Its estimate is \(0.0027\), with a \(t\)-statistic of \(3.67\). In contrast, the coefficients on ROE and MAX become statistically insignificant. The coefficient on ILLIQ becomes positive and marginally significant at the 10\% level, with an estimate of \(0.0098\) and a \(t\)-statistic of \(1.92\). This result provides some evidence that high-D3-beta firms are less liquid after controlling for the other characteristics. This pattern is consistent with \citet{amihud2002illiquidity}, who documents a positive relation between illiquidity and expected stock returns. Overall, higher asset growth is the most stable characteristic of high-D3-beta firms.

The investment result is consistent with the policy-clarity interpretation of D3. \citet{gulen2016policy} document a negative relation between policy uncertainty and corporate investment. In my setting, a higher D3 factor tends to indicate greater AI-policy clarity and lower uncertainty. Firms with stronger investment activity may benefit more when uncertainty about AI regulation and product-market decisions declines, which may help explain why their returns covary positively with D3 innovations.

\subsection{Robustness Checks}
\label{sec:d3_robustness}

I first test whether the D3 pricing result depends on the AR(1) innovation model. I reconstruct the four domain-specific innovations using AR(7), ARMA(1,1), and ARMA(2,1) models. For each alternative model, I repeat the joint beta estimation and Fama--MacBeth regressions using the three-month beta window, the five established factor-control specifications, and Fama--French ten-industry controls.

Panel A of Table~\ref{tab:d3_robustness} reports the results. The estimated D3 price remains positive and statistically significant under all three alternative innovation models. Across the 15 specifications, the D3 estimates range from \(0.0380\) to \(0.0479\), with \(t\)-statistics between \(2.98\) and \(4.00\). All 15 estimates are statistically significant at the 1\% level. In contrast, none of the prices associated with D2, D4, or D6 is statistically significant even at the 10\% level. The D3 result is therefore not driven by the lag length or error structure used to extract the factor innovations.

I next examine whether the pricing evidence is concentrated in the period following the public release of ChatGPT on November 30, 2022. AI-risk news coverage rose sharply after late 2022, raising the concern that the full-sample results may simply reflect the post-ChatGPT surge in public attention to AI. I address this concern by restricting the sample to the pre-ChatGPT period ending in November 2022. Specifically, I restrict the baseline AR(1) innovations and three-month beta estimates to the pre-ChatGPT period. I then recompute the Fama--MacBeth results with Fama--French ten-industry controls using only pre-ChatGPT return months.

Panel B of Table~\ref{tab:d3_robustness} presents the subsample results. The estimated D3 price ranges from \(0.0346\) to \(0.0412\), with \(t\)-statistics between \(2.36\) and \(3.42\). All five estimates are positive and statistically significant. The CAPM estimate is significant at the 5\% level, while the remaining four estimates are significant at the 1\% level. The estimated prices of D2, D4, and D6 remain statistically indistinguishable from zero. The D3 pricing result is therefore present before the release of ChatGPT and is not driven solely by the subsequent increase in AI-related news.

Taken together, the additional tests show that the positive D3 price is stable across alternative innovation models and the pre-ChatGPT subsample. The characteristic analysis also identifies a consistent positive relation between D3 beta and firm investment. These findings support the interpretation of D3 as a systematic factor related to the AI information and policy environment.

\section{Conclusion}
\label{sec:conclusion}

This paper examines whether firms' exposures to common AI-risk states are priced in the cross-section of U.S.\ stock returns. I identify 7,787 AI-risk articles published by The Wall Street Journal between 2016 and 2025. I use LDA to identify recurring news narratives and map the retained topics to the MIT AI Risk Repository. This procedure produces four active domain-specific factors: Privacy and security, Misinformation, Malicious actors and misuse, and Socioeconomic and environmental harm. I then estimate firm-level betas to innovations in these factors and test whether these beta exposures are priced using univariate portfolio analysis and Fama--MacBeth regressions.

The results show that the pricing of AI risk differs across domains. D3 Misinformation is the only domain with consistent pricing evidence. In the portfolio analysis, the value-weighted high-minus-low D3-beta portfolio earns monthly factor-adjusted alphas of \(0.49\%\) to \(0.57\%\) under the reported three-month specification. The Fama--MacBeth regressions provide consistent cross-sectional evidence: the coefficient on D3 beta is positive and statistically significant across all three beta-estimation windows, all five established factor-control specifications, and specifications both with and without industry controls. The D3 coefficient also remains positive and statistically significant when the factor innovations are reconstructed using alternative time-series models and when the sample is restricted to the pre-ChatGPT period. By contrast, D2, D4, and D6 do not show reliable individual pricing evidence. These findings support the hypothesis that at least one empirically active AI-risk domain carries a nonzero price of risk, and indicate that AI-risk pricing is domain-specific rather than common across all domains.

The empirical content of D3 provides an economic interpretation of its positive price of risk. D3's underlying news topic and the events associated with its peaks are closely related to the AI information environment, governance, and regulation. A high D3 factor tends to signal greater policy attention, more regulatory clarity, and lower AI-policy uncertainty. High-D3-beta stocks perform relatively well when this state improves but provide less protection when policy uncertainty is high. Their positive risk premium is therefore consistent with investors requiring compensation for assets that perform poorly when investment opportunities deteriorate. The characteristic analysis provides additional evidence for this interpretation. High-D3-beta firms have higher asset growth, and this relation remains statistically significant after controlling for other characteristics and industry effects. 

The domain-level taxonomy separates broad categories of AI risk, but individual domains may still contain narratives with different economic implications. This issue is particularly relevant for D6. Although the narratives within D6 have economy-wide implications, the combined D6 factor does not show reliable pricing evidence. One possibility is that D6 aggregates distinct narratives related to labor-market disruption, inequality, resource use, and environmental damage that affect firms through different or offsetting channels. Future research could further separate these narratives into narrower factors and examine whether their exposures have distinct cross-sectional pricing implications. Such an analysis would help determine whether the aggregate D6 result arises because none of its individual narratives is priced or because their effects offset one another.

\newpage

\bibliographystyle{apalike}  
\bibliography{references}

\begin{table}[p]
\centering

\caption{Keyword Lexicons for AI-Risk News Identification}
\label{tab:keyword_lexicons}

\TableDescription{\justifying \noindent
Panel A lists the 33 AI surface forms used in the keyword-based filtering.
Panel B reports one representative root term for each of the 22 word
families used to construct the risk lexicon. Each root represents its
predefined grammatical and derived forms. For example, the
\textit{danger} family includes forms such as \textit{dangerous} and
\textit{endanger}. The 22 word families produce 139 surface forms in
the implemented risk lexicon.
}

\begin{tabularx}{\textwidth}{
    >{\raggedright\arraybackslash}X
}
\toprule

Panel A: AI Lexicon\\[-0.3em]
\midrule

AI; A.I.; artificial intelligence; machine intelligence; machine learning;
deep learning; neural network; neural networks; language model; language
models; foundation model; foundation models; artificial general intelligence;
artificial narrow intelligence; computer vision; genAI; natural language
processing; speech recognition; voice recognition; image recognition; facial
recognition; face recognition; machine translation; machine translations;
chatbot; chatbots; deepfake; deepfakes; deep fake; deep fakes; expert system;
expert systems; cognitive computing
\\[0.5em]

\midrule

Panel B: Risk Lexicon\\[-0.3em]
\midrule

risk; uncertainty; danger; hazard; peril; jeopardy; threat; menace; vulnerability;
susceptibility; insecurity; precariousness; instability; unsafe;
unreliability; unpredictability; doubt; crisis; worry; fear; alarm; disturbance
\\

\bottomrule
\end{tabularx}
\end{table}



\begin{table}[p]
\centering
\caption{Illustrative Full-Text Assessments of AI-Risk Relevance}
\label{tab:article_relevance_examples}
\small
\begin{tabularx}{\textwidth}{
    @{}
    >{\raggedright\arraybackslash}p{0.3\textwidth}
    @{\hspace{0.8cm}}
    >{\raggedright\arraybackslash}p{0.15\textwidth}
    @{\hspace{0.4cm}}
    >{\raggedright\arraybackslash}X
    @{}
}
\toprule
Article title & Binary assessment & Reason for classification \\
\midrule

``The Best Ways to Ask ChatGPT Questions''
&
1
&
Although the title suggests a practical guide to prompting, the article discusses the unpredictability of large language models and explains that certain prompting strategies, including impolite prompts, can increase bias and produce incorrect answers. These passages describe potential harms arising from the use of AI.
\\
\addlinespace

``Hollywood Actors Join Writers on Strike''
&
1
&
The article devotes a complete paragraph to concerns that generative AI could replace writers and allow performers' likenesses to be used without their consent or compensation. It therefore describes a substantive employment and appropriation risk arising from AI.
\\
\addlinespace

``Microsoft to Buy Activision Blizzard in All-Cash Deal Valued at \$75 Billion''
&
0
&
The article centers on Microsoft's acquisition of Activision Blizzard. AI appears only as background information about Microsoft's earlier acquisition of Nuance, while the reported threats and misconduct are unrelated to harm caused by AI.
\\
\addlinespace

``Harvard's Endowment Jumps to \$56.9 Billion''
&
0
&
The article focuses on university endowment returns and threats to university funding and independence. The risk-related term “threat” refers to government actions against universities, while AI is mentioned only in MIT’s reference to the ``beneficial use of artificial intelligence.'' The threats discussed are therefore unrelated to harm caused by AI.
\\

\bottomrule
\end{tabularx}

\end{table}

\begin{table}[p]
\centering
\scriptsize
\caption{AI-Risk Relevance Review of the 18 LDA Topics}
\label{tab:topic_relevance_review}
\TableDescription{
\justifying \noindent
The second column reports the 10 terms with the highest estimated topic-word probabilities for each topic, ordered from highest to lowest. The relevance review covers the 20 articles with the highest dominant-topic probabilities. An article is coded as AI-risk-related when it substantively connects an AI capability, system, or use to a realized or potential harm. A topic is retained when at least 5 of its 20 reviewed articles meet this criterion.}

\renewcommand{\arraystretch}{1.08}
\setlength{\tabcolsep}{4pt}
\begin{tabularx}{\textwidth}{L{0.9cm} X L{1.75cm} L{1.3cm}}
\toprule
\textbf{Topic} &
\textbf{Top 10 words} &
\textbf{AI-risk articles} &
\textbf{Decision} \\
\midrule

1 &
google, human, model, system, data, chatgpt, tool, researcher, create, machine
& 18/20 & Retained \\

2 &
trump, president, tiktok, election, call, political, policy, government, american, biden
& 11/20 & Retained \\

3 &
data, security, cybersecurity, information, system, report, risk, government, tool, user
& 11/20 & Retained \\

4 &
job, employee, firm, tech, worker, hire, musk, microsoft, executive, investor
& 11/20 & Retained \\

5 &
startup, investor, car, investment, china, ford, chinese, billion, million, deal
& 6/20 & Retained \\

6 &
job, patient, system, employee, data, change, health, worker, amazon, tool
& 15/20 & Retained \\

7 &
job, worker, robot, music, store, automation, pay, industry, create, food
& 15/20 & Retained \\

8 &
apple, user, app, product, device, tool, phone, call, feature, apps
& 8/20 & Retained \\

9 &
book, write, life, world, call, story, author, day, human, great
& 3/20 & \textbf{Removed} \\

10 &
chip, billion, nvidia, openai, revenue, microsoft, market, sale, customer, share
& 10/20 & Retained \\

11 &
stock, market, investor, economy, data, inflation, high, rise, price, growth
& 1/20 & \textbf{Removed} \\

12 &
data\_center, power, drug, patient, china, project, country, price, market, state
& 14/20 & Retained \\

13 &
military, drone, russia, official, ukraine, russian, war, attack, israel, country
& 8/20 & Retained \\

14 &
musk, tesla, facebook, user, meta, post, system, child, content, platform
& 15/20 & Retained \\

15 &
china, chinese, beijing, country, american, government, official, world, global, trade
& 3/20 & \textbf{Removed} \\

16 &
energy, power, government, investment, bill, industry, billion, cost, project, world
& 11/20 & Retained \\

17 &
analyst, market, share, expect, growth, add, china, likely, note, demand
& 0/20 & \textbf{Removed} \\

18 &
student, school, fund, university, college, job, tariff, class, graduate, education
& 5/20 & Retained \\

\bottomrule
\end{tabularx}
\end{table}

\clearpage
\begin{table}[p]
\centering
\small
\caption{\textbf{Mapping of Retained LDA Topics to Active AI-Risk Domains}}
\label{tab:topic_domain_mapping}
\TableDescription{\justifying \noindent 
The table reports the mapping of the 14 retained LDA topics to the parent domains of the MIT AI Risk Repository. Each topic is represented by a probability-weighted combination of the embeddings of its 50 highest-probability words, while each subdomain is represented by an embedding of its title and definition. The cosine similarities between each topic representation and the 24 subdomain representations are standardized, converted into subdomain weights, and summed within each parent domain. D1, D5, and D7 receive zero weight across all retained topics and are omitted. Each row sums to 100\% across the complete seven-domain taxonomy.}
\begin{tabular}{L{1cm} L{3.2cm} L{3.2cm} L{3.2cm} L{3.2cm}}
\toprule
& \multicolumn{4}{c}{\textbf{Topic-domain weight (\%)}} \\
\cmidrule(lr){2-5}
\textbf{Topic} &
\textbf{D2 Privacy and security} &
\textbf{D3 Misinformation} &
\textbf{D4 Malicious actors and misuse} &
\textbf{D6 Socioeconomic and environmental harm} \\
\midrule
1  & 29.00 & 0.00  & 71.00 & 0.00 \\
2  & 0.00  & 35.99 & 29.81 & 34.20 \\
3  & 100.00& 0.00  & 0.00  & 0.00 \\
4  & 0.00  & 0.00  & 0.00  & 100.00 \\
5  & 0.00  & 0.00  & 0.00  & 100.00 \\
6  & 0.00  & 0.00  & 0.00  & 100.00 \\
7  & 0.00  & 0.00  & 0.00  & 100.00 \\
8  & 70.29 & 0.00  & 0.00  & 29.71 \\
10 & 0.00  & 0.00  & 0.00  & 100.00 \\
12 & 0.00  & 0.00  & 0.00  & 100.00 \\
13 & 0.00  & 0.00  & 100.00& 0.00 \\
14 & 59.57 & 0.00  & 9.45  & 30.98 \\
16 & 0.00  & 0.00  & 0.00  & 100.00 \\
18 & 0.00  & 0.00  & 100.00& 0.00 \\
\bottomrule
\end{tabular}
\end{table}

\clearpage
\begin{table}[p]
\centering
\caption{Selection of the Baseline Innovation Model}
\label{tab:innovation_model_selection}

\TableDescription{\justifying \noindent
The table compares four innovation models across 432 common domain-month
pairs, consisting of four AI-risk domains and 108 monthly estimation dates.
Within each pair, all four models are estimated by maximum likelihood using
the same preceding 365 calendar days, an intercept, and weekday indicators for Tuesday through Sunday. Models are ranked separately by AIC and
BIC from 1, indicating the lowest information criterion, to 4, indicating
the highest. The mean rank is the average of these rankings across the 432
pairs. A win indicates that a model has the lowest AIC or BIC within a pair.
The AIC results are mixed: ARMA(1,1) has the lowest mean AIC rank, whereas AR(1) has the lowest AIC in the largest number of domain-month pairs. BIC provides clearer support for AR(1): it has both the lowest mean BIC rank and the largest number of BIC wins. This BIC evidence, together with the parsimony of AR(1), supports its selection as the baseline innovation model.
}

\begin{tabular}{
    @{}
    L{2.5cm}
    L{3cm}
    L{3.2cm}
    L{3.2cm}
    L{2.4cm}
    @{}
}
\toprule
Model &
Mean AIC rank &
AIC wins &
Mean BIC rank &
BIC wins \\
\midrule
AR(1)     & 2.2824 & 176 (40.7\%) & 1.5394 & 291 (67.4\%) \\
AR(7)     & 3.2824 &  63 (14.6\%) & 3.8796 &   9 (2.1\%)  \\
ARMA(1,1) & 1.7917 & 160 (37.0\%) & 1.7269 & 128 (29.6\%) \\
ARMA(2,1) & 2.6435 &  33 (7.6\%)  & 2.8542 &   4 (0.9\%)  \\
\bottomrule
\end{tabular}
\end{table}

\begin{table}[p]
\centering
\captionsetup{
    justification=raggedright,
    singlelinecheck=false
}
\caption{Pairwise Correlations of AI-Risk Innovations and Other Variables}
\label{tab:innovation_correlations}

\TableDescription{\justifying \noindent
The table reports Pearson correlations among the four trading-day
AR(1) innovations, the six conventional asset-pricing factors, and the
first differences of VIX and EPU.
\(\Delta D2\), \(\Delta D3\), \(\Delta D4\), and \(\Delta D6\) denote
the AR(1) innovations in the corresponding domain-specific AI-risk
factors. VIX is obtained from the CBOE Indexes, and
\(\Delta\mathrm{VIX}\) is its first difference
\citep{ang2006cross}. EPU is the daily news-based Economic Policy
Uncertainty index of \citet{baker2016measuring}. Following the
contemporaneous-return specification in
\citet{brogaard2015assetpricing}, \(\Delta\mathrm{EPU}\) denotes its
first difference. All correlations involving at least one AI-risk
innovation are based on 2,253 overlapping trading days. Among the
remaining pairs, correlations involving \(\Delta\mathrm{VIX}\) are
based on 2,513 overlapping trading days, while the other correlations
are based on 2,514 overlapping trading days. Each correlation is
calculated using dates for which both variables are available.
}

\begingroup
\scriptsize
\setlength{\tabcolsep}{5pt}
\renewcommand{\arraystretch}{1.10}

\resizebox{\textwidth}{!}{%
\begin{tabular}{l*{12}{r}}
\toprule
&
\(\Delta D2\) &
\(\Delta D3\) &
\(\Delta D4\) &
\(\Delta D6\) &
\(\mathrm{MktRF}\) &
\(\mathrm{SMB}\) &
\(\mathrm{HML}\) &
\(\mathrm{RMW}\) &
\(\mathrm{CMA}\) &
\(\mathrm{UMD}\) &
\(\Delta\mathrm{VIX}\) &
\(\Delta\mathrm{EPU}\) \\
\midrule

\(\Delta D2\)
  & 1.000 \\

\(\Delta D3\)
  & 0.183 & 1.000 \\

\(\Delta D4\)
  & 0.380 & 0.411 & 1.000 \\

\(\Delta D6\)
  & 0.361 & 0.251 & 0.324 & 1.000 \\

\(\mathrm{MktRF}\)
  & -0.021 & -0.009 & -0.003 & -0.012
  & 1.000 \\

\(\mathrm{SMB}\)
  & -0.027 & -0.015 & -0.043 & -0.038
  & 0.204 & 1.000 \\

\(\mathrm{HML}\)
  & -0.020 & 0.019 & 0.006 & -0.034
  & -0.088 & 0.291 & 1.000 \\

\(\mathrm{RMW}\)
  & -0.018 & 0.005 & 0.005 & 0.007
  & -0.216 & -0.268 & 0.327 & 1.000 \\

\(\mathrm{CMA}\)
  & 0.008 & -0.015 & 0.000 & -0.028
  & -0.260 & 0.041 & 0.552 & 0.270
  & 1.000 \\

\(\mathrm{UMD}\)
  & 0.010 & -0.024 & -0.003 & 0.011
  & -0.130 & -0.330 & -0.308 & -0.034
  & 0.022 & 1.000 \\

\(\Delta\mathrm{VIX}\)
  & 0.007 & 0.005 & -0.004 & 0.024
  & -0.786 & -0.119 & 0.064 & 0.178
  & 0.148 & 0.018 & 1.000 \\

\(\Delta\mathrm{EPU}\)
  & -0.034 & -0.045 & -0.040 & -0.022
  & 0.004 & -0.018 & -0.061 & -0.021
  & -0.029 & 0.035 & -0.008 & 1.000 \\

\bottomrule
\end{tabular}%
}

\endgroup
\end{table}

\begin{table}[p]
\vspace*{-0.7cm}
\centering
\caption{Univariate Portfolios Sorted on Domain-Specific AI-Risk Betas}
\label{tab:univariate_portfolio_results}

\TableDescription{\justifying \noindent
The table reports value-weighted quintile portfolios formed from the
three-month, FF5+UMD single-domain beta specification. NYSE stocks determine
the quintile breakpoints, which are applied to all eligible NYSE, AMEX, and
Nasdaq stocks. Portfolio \(P1\) contains the lowest-beta stocks and \(P5\)
contains the highest-beta stocks. Portfolio \(P5-P1\) is a long--short portfolio that buys \(P5\) and sells \(P1\). The pre-ranking beta is the time-series average of each portfolio's beta measured at the end of formation month \(m\). The post-ranking beta is estimated from daily observations during holding month \(m+1\) for the formation-month constituents of each portfolio and aggregated using their month-\(m\) market-capitalization weights. Excess returns are monthly time-series averages, and factor-adjusted alphas are monthly regression intercepts; both are reported in percentages. \citet{newey1987simple} \(t\)-statistics based on six lags are reported in parentheses.
}

\footnotesize
\setlength{\tabcolsep}{3pt}
\renewcommand{\arraystretch}{0.85}

\makebox[\textwidth][c]{%
\begin{tabular*}{1\textwidth}
{@{\extracolsep{\fill}}lcccccccc@{}}
\toprule
Portfolio &
\(\beta_{\mathrm{pre}}\) &
\(\beta_{\mathrm{post}}\) &
\(\mathrm{RET}-\mathrm{RF}\) &
\(\alpha_{\mathrm{CAPM}}\) &
\(\alpha_{\mathrm{FF3}}\) &
\(\alpha_{\mathrm{Carhart4}}\) &
\(\alpha_{\mathrm{FF5}}\) &
\(\alpha_{\mathrm{FF5+UMD}}\) \\
\midrule

\multicolumn{9}{l}{Panel A: D2 Privacy and security}\\[-0.3em]
\midrule

\(P1\)
& -0.0067 & -0.00024
& 1.284 & 0.119 & 0.211 & 0.191 & 0.236 & 0.233\\[-0.2em]
& & &
(2.79) & (0.59) & (1.05) & (0.95) & (1.16) & (1.18)\\[0.1em]

\(P2\)
& -0.0022 & -0.00030
& 1.289 & 0.236 & 0.211 & 0.190 & 0.201 & 0.180\\[-0.2em]
& & &
(3.15) & (1.99) & (1.71) & (1.44) & (1.69) & (1.41)\\[0.1em]

\(P3\)
& 0.0001 & 0.00006
& 0.991 & -0.037 & -0.083 & -0.075 & -0.085 & -0.079\\[-0.2em]
& & &
(2.72) & (-0.30) & (-0.66) & (-0.59) & (-0.65) & (-0.61)\\[0.1em]

\(P4\)
& 0.0024 & 0.00013
& 0.838 & -0.183 & -0.175 & -0.142 & -0.195 & -0.172\\[-0.2em]
& & &
(2.34) & (-1.02) & (-0.99) & (-0.80) & (-1.14) & (-0.99)\\[0.1em]

\(P5\)
& 0.0070 & 0.00065
& 0.812 & -0.255 & -0.139 & -0.136 & -0.106 & -0.097\\[-0.2em]
& & &
(1.74) & (-1.21) & (-0.61) & (-0.61) & (-0.48) & (-0.44)\\[0.1em]

\textbf{\(P5-P1\)}
& 0.0137 & 0.00089
& -0.471 & -0.374 & -0.350 & -0.327 & -0.341 & -0.330\\[-0.2em]
& & &
(-1.31) & (-1.04) & (-0.91) & (-0.87) & (-0.88) & (-0.87)\\[0.1em]

\midrule
\multicolumn{9}{l}{Panel B: D3 Misinformation}\\[-0.3em]
\midrule

\(P1\)
& -0.0450 & -0.00253
& 0.683 & -0.470 & -0.369 & -0.374 & -0.348 & -0.345\\[-0.2em]
& & &
(1.36) & (-2.51) & (-2.24) & (-2.34) & (-2.05) & (-2.14)\\[0.1em]

\(P2\)
& -0.0147 & -0.00187
& 0.985 & -0.080 & -0.084 & -0.078 & -0.096 & -0.095\\[-0.2em]
& & &
(2.60) & (-0.65) & (-0.66) & (-0.60) & (-0.77) & (-0.76)\\[0.1em]

\(P3\)
& 0.0006 & 0.00026
& 1.071 & 0.055 & 0.039 & 0.029 & 0.031 & 0.017\\[-0.2em]
& & &
(3.36) & (0.41) & (0.27) & (0.20) & (0.22) & (0.12)\\[0.1em]

\(P4\)
& 0.0158 & 0.00077
& 1.198 & 0.186 & 0.151 & 0.136 & 0.130 & 0.112\\[-0.2em]
& & &
(3.47) & (1.62) & (1.24) & (1.06) & (0.99) & (0.82)\\[0.1em]

\(P5\)
& 0.0465 & 0.00188
& 1.112 & 0.017 & 0.129 & 0.164 & 0.178 & 0.230\\[-0.2em]
& & &
(2.47) & (0.09) & (0.67) & (0.89) & (0.96) & (1.32)\\[0.1em]

\textbf{\(P5-P1\)}
& 0.0916 & 0.00440
& 0.429 & 0.487 & 0.498 & 0.538 & 0.526 & 0.575\\[-0.2em]
& & &
(1.52) & (1.81) & (1.89) & (2.16) & (1.94) & (2.23)\\[0.1em]

\midrule
\multicolumn{9}{l}{Panel C: D4 Malicious actors and misuse}\\[-0.3em]
\midrule

\(P1\)
& -0.0105 & 0.00037
& 0.895 & -0.318 & -0.192 & -0.161 & -0.173 & -0.122\\[-0.2em]
& & &
(1.86) & (-1.60) & (-1.00) & (-0.90) & (-0.93) & (-0.71)\\[0.1em]

\(P2\)
& -0.0035 & 0.00001
& 1.132 & 0.109 & 0.054 & 0.047 & 0.058 & 0.054\\[-0.2em]
& & &
(2.84) & (0.85) & (0.39) & (0.33) & (0.44) & (0.38)\\[0.1em]

\(P3\)
& -0.0000 & 0.00002
& 0.788 & -0.217 & -0.240 & -0.248 & -0.246 & -0.258\\[-0.2em]
& & &
(2.44) & (-1.55) & (-1.74) & (-1.81) & (-1.83) & (-1.92)\\[0.1em]

\(P4\)
& 0.0034 & 0.00001
& 1.040 & 0.026 & 0.007 & -0.008 & -0.014 & -0.040\\[-0.2em]
& & &
(2.82) & (0.26) & (0.07) & (-0.08) & (-0.13) & (-0.39)\\[0.1em]

\(P5\)
& 0.0103 & -0.00025
& 1.264 & 0.119 & 0.239 & 0.258 & 0.293 & 0.315\\[-0.2em]
& & &
(2.76) & (0.54) & (1.13) & (1.29) & (1.50) & (1.73)\\[0.1em]

\textbf{\(P5-P1\)}
& 0.0208 & -0.00062
& 0.370 & 0.437 & 0.431 & 0.420 & 0.465 & 0.438\\[-0.2em]
& & &
(1.10) & (1.38) & (1.34) & (1.40) & (1.61) & (1.65)\\[0.1em]

\midrule
\multicolumn{9}{l}{Panel D: D6 Socioeconomic and environmental harm}\\[-0.3em]
\midrule

\(P1\)
& -0.0040 & -0.00015
& 1.109 & -0.054 & 0.060 & 0.040 & 0.086 & 0.078\\[-0.2em]
& & &
(2.23) & (-0.24) & (0.28) & (0.20) & (0.38) & (0.36)\\[0.1em]

\(P2\)
& -0.0013 & 0.00002
& 0.986 & -0.022 & -0.037 & -0.039 & -0.050 & -0.051\\[-0.2em]
& & &
(2.82) & (-0.23) & (-0.39) & (-0.40) & (-0.58) & (-0.59)\\[0.1em]

\(P3\)
& 0.0001 & 0.00001
& 0.981 & -0.021 & -0.082 & -0.105 & -0.101 & -0.127\\[-0.2em]
& & &
(2.82) & (-0.23) & (-0.83) & (-1.08) & (-1.13) & (-1.42)\\[0.1em]

\(P4\)
& 0.0014 & 0.00005
& 1.025 & -0.002 & -0.021 & 0.001 & -0.032 & -0.017\\[-0.2em]
& & &
(3.27) & (-0.01) & (-0.17) & (0.01) & (-0.25) & (-0.14)\\[0.1em]

\(P5\)
& 0.0042 & 0.00002
& 0.910 & -0.298 & -0.141 & -0.099 & -0.075 & -0.020\\[-0.2em]
& & &
(1.73) & (-1.18) & (-0.62) & (-0.44) & (-0.37) & (-0.10)\\[0.1em]

\textbf{\(P5-P1\)}
& 0.0082 & 0.00017
& -0.199 & -0.245 & -0.201 & -0.140 & -0.160 & -0.097\\[-0.2em]
& & &
(-0.52) & (-0.68) & (-0.60) & (-0.44) & (-0.47) & (-0.31)\\

\bottomrule
\end{tabular*}%
}

\end{table}

\clearpage
\begin{table}[p]
\centering
\caption{Three-Month Fama--MacBeth Regression Results}
\label{tab:fama_macbeth_3m_full}

\TableDescription{\justifying \noindent
The table reports the Fama--MacBeth regression results based on
three-month first-pass beta-estimation windows, with the same set of conventional factor controls used in both passes. The first five columns report regressions without industry controls. The
remaining five columns include nine Fama--French ten-industry indicators,
with Other as the omitted industry. The continuous independent variable is winsorized at the 0.5\% level on a monthly basis. \citet{newey1987simple} \(t\)-statistics based on six lags are reported in parentheses. The final row reports \(p\)-values from Wald tests of the joint null
hypothesis
\(H_0:
\lambda_{\mathrm{D2}}
=
\lambda_{\mathrm{D3}} 
=
\lambda_{\mathrm{D4}} 
=
\lambda_{\mathrm{D6}} 
=
0\). The sample period is April 2017--December 2025.
}

\footnotesize
\setlength{\tabcolsep}{2.2pt}
\renewcommand{\arraystretch}{1.02}

\resizebox{\textwidth}{!}{%
\begin{tabular}{lcccccccccc}
\toprule
&
\multicolumn{5}{c}{Without Industry Controls}
&
\multicolumn{5}{c}{With Fama--French Ten-Industry Controls}
\\
\cmidrule(lr){2-6}
\cmidrule(lr){7-11}
&
CAPM &
FF3 &
Carhart 4 &
FF5 &
FF5+UMD &
CAPM &
FF3 &
Carhart 4 &
FF5 &
FF5+UMD \\
\midrule

Intercept
& \FMcell{0.0008\\(0.15)}
& \FMcell{0.0006\\(0.14)}
& \FMcell{0.0005\\(0.10)}
& \FMcell{0.0006\\(0.13)}
& \FMcell{0.0007\\(0.17)}
& \FMcell{0.0007\\(0.15)}
& \FMcell{0.0005\\(0.11)}
& \FMcell{0.0004\\(0.09)}
& \FMcell{0.0003\\(0.06)}
& \FMcell{0.0005\\(0.11)} \\

AI-risk beta \(\Delta D2\)
& \FMcell{0.0644\\(0.86)}
& \FMcell{0.0384\\(0.61)}
& \FMcell{0.0312\\(0.52)}
& \FMcell{0.0268\\(0.43)}
& \FMcell{0.0196\\(0.33)}
& \FMcell{0.0456\\(0.65)}
& \FMcell{0.0239\\(0.39)}
& \FMcell{0.0179\\(0.30)}
& \FMcell{0.0224\\(0.37)}
& \FMcell{0.0154\\(0.26)} \\

AI-risk beta \(\Delta D3\)
& \FMcell{0.0496\\(2.73)}
& \FMcell{0.0447\\(3.34)}
& \FMcell{0.0414\\(3.12)}
& \FMcell{0.0415\\(3.49)}
& \FMcell{0.0403\\(3.41)}
& \FMcell{0.0429\\(2.85)}
& \FMcell{0.0403\\(3.42)}
& \FMcell{0.0380\\(3.23)}
& \FMcell{0.0373\\(3.56)}
& \FMcell{0.0369\\(3.52)} \\

AI-risk beta \(\Delta D4\)
& \FMcell{0.0834\\(1.27)}
& \FMcell{0.0487\\(0.88)}
& \FMcell{0.0397\\(0.74)}
& \FMcell{0.0406\\(0.78)}
& \FMcell{0.0362\\(0.71)}
& \FMcell{0.0715\\(1.22)}
& \FMcell{0.0478\\(0.96)}
& \FMcell{0.0404\\(0.84)}
& \FMcell{0.0428\\(0.91)}
& \FMcell{0.0402\\(0.88)} \\

AI-risk beta \(\Delta D6\)
& \FMcell{0.0369\\(0.24)}
& \FMcell{-0.0320\\(-0.24)}
& \FMcell{-0.0531\\(-0.40)}
& \FMcell{-0.0383\\(-0.32)}
& \FMcell{-0.0468\\(-0.40)}
& \FMcell{-0.0044\\(-0.03)}
& \FMcell{-0.0643\\(-0.52)}
& \FMcell{-0.0843\\(-0.68)}
& \FMcell{-0.0740\\(-0.67)}
& \FMcell{-0.0806\\(-0.74)} \\

\midrule

Market beta
& \FMcell{0.0018\\(0.72)}
& \FMcell{0.0030\\(1.47)}
& \FMcell{0.0031\\(1.61)}
& \FMcell{0.0036\\(1.93)}
& \FMcell{0.0035\\(1.88)}
& \FMcell{0.0019\\(0.80)}
& \FMcell{0.0030\\(1.53)}
& \FMcell{0.0030\\(1.66)}
& \FMcell{0.0036\\(2.03)}
& \FMcell{0.0034\\(1.98)} \\

SMB beta
& {}
& \FMcell{-0.0017\\(-0.99)}
& \FMcell{-0.0012\\(-0.76)}
& \FMcell{-0.0005\\(-0.40)}
& \FMcell{-0.0003\\(-0.24)}
& {}
& \FMcell{-0.0016\\(-1.00)}
& \FMcell{-0.0012\\(-0.79)}
& \FMcell{-0.0005\\(-0.40)}
& \FMcell{-0.0004\\(-0.28)} \\

HML beta
& {}
& \FMcell{0.0036\\(1.24)}
& \FMcell{0.0035\\(1.28)}
& \FMcell{0.0025\\(0.98)}
& \FMcell{0.0025\\(0.98)}
& {}
& \FMcell{0.0032\\(1.32)}
& \FMcell{0.0032\\(1.35)}
& \FMcell{0.0021\\(1.03)}
& \FMcell{0.0022\\(1.03)} \\

UMD beta
& {}
& {}
& \FMcell{0.0007\\(0.29)}
& {}
& \FMcell{0.0010\\(0.45)}
& {}
& {}
& \FMcell{0.0004\\(0.17)}
& {}
& \FMcell{0.0007\\(0.31)} \\

RMW beta
& {}
& {}
& {}
& \FMcell{0.0036\\(1.94)}
& \FMcell{0.0035\\(1.98)}
& {}
& {}
& {}
& \FMcell{0.0034\\(1.88)}
& \FMcell{0.0034\\(1.91)} \\

CMA beta
& {}
& {}
& {}
& \FMcell{0.0007\\(0.48)}
& \FMcell{0.0006\\(0.43)}
& {}
& {}
& {}
& \FMcell{0.0005\\(0.38)}
& \FMcell{0.0004\\(0.33)} \\

\midrule
\multicolumn{11}{l}{\textit{Fama--French industry indicators (Other omitted)}} \\

NoDur
& & & & &
& \FMcell{-0.0023\\(-1.38)}
& \FMcell{-0.0025\\(-1.47)}
& \FMcell{-0.0024\\(-1.44)}
& \FMcell{-0.0025\\(-1.48)}
& \FMcell{-0.0024\\(-1.40)} \\

Durbl
& & & & &
& \FMcell{-0.0021\\(-0.52)}
& \FMcell{-0.0017\\(-0.42)}
& \FMcell{-0.0015\\(-0.37)}
& \FMcell{-0.0021\\(-0.52)}
& \FMcell{-0.0020\\(-0.50)} \\

Manuf
& & & & &
& \FMcell{0.0039\\(2.15)}
& \FMcell{0.0032\\(1.93)}
& \FMcell{0.0032\\(1.99)}
& \FMcell{0.0031\\(1.84)}
& \FMcell{0.0030\\(1.85)} \\

Enrgy
& & & & &
& \FMcell{0.0006\\(0.07)}
& \FMcell{-0.0014\\(-0.18)}
& \FMcell{-0.0018\\(-0.24)}
& \FMcell{0.0010\\(0.12)}
& \FMcell{0.0003\\(0.04)} \\

HiTec
& & & & &
& \FMcell{0.0026\\(0.77)}
& \FMcell{0.0037\\(1.30)}
& \FMcell{0.0036\\(1.28)}
& \FMcell{0.0038\\(1.37)}
& \FMcell{0.0037\\(1.37)} \\

Telcm
& & & & &
& \FMcell{-0.0026\\(-0.89)}
& \FMcell{-0.0024\\(-0.82)}
& \FMcell{-0.0020\\(-0.68)}
& \FMcell{-0.0020\\(-0.67)}
& \FMcell{-0.0017\\(-0.58)} \\

Shops
& & & & &
& \FMcell{0.0003\\(0.11)}
& \FMcell{0.0006\\(0.20)}
& \FMcell{0.0007\\(0.26)}
& \FMcell{0.0001\\(0.05)}
& \FMcell{0.0003\\(0.10)} \\

Hlth
& & & & &
& \FMcell{-0.0043\\(-0.91)}
& \FMcell{-0.0026\\(-0.63)}
& \FMcell{-0.0029\\(-0.70)}
& \FMcell{-0.0022\\(-0.54)}
& \FMcell{-0.0022\\(-0.56)} \\

Utils
& & & & &
& \FMcell{0.0028\\(0.68)}
& \FMcell{0.0016\\(0.42)}
& \FMcell{0.0013\\(0.36)}
& \FMcell{0.0020\\(0.54)}
& \FMcell{0.0020\\(0.54)} \\

\midrule

Valid months
& \multicolumn{5}{c}{105}
& \multicolumn{5}{c}{105} \\

Average \# stocks
& \multicolumn{5}{c}{3,750}
& \multicolumn{5}{c}{3,750} \\

Adjusted \(R^{2}\)
& 2.31\%
& 3.82\%
& 4.08\%
& 4.37\%
& 4.57\%
& 4.27\%
& 5.33\%
& 5.53\%
& 5.74\%
& 5.89\% \\

Wald test \(p\)-value
& 0.0436
& 0.0108
& 0.0231
& 0.0103
& 0.0139
& 0.0533
& 0.0146
& 0.0254
& 0.0127
& 0.0143 \\
\bottomrule
\end{tabular}%
}
\end{table}

\clearpage
\begin{table}[p]
\centering
\caption{Six-Month Fama--MacBeth Regression Results}
\label{tab:fama_macbeth_6m_full}

\TableDescription{\justifying \noindent
The table reports the Fama--MacBeth regression results based on
six-month first-pass beta-estimation windows, with the same set of conventional factor controls used in both passes. The first five columns report regressions without industry controls. The
remaining five columns include nine Fama--French ten-industry indicators,
with Other as the omitted industry. The continuous independent variable is winsorized at the 0.5\% level on a monthly basis. \citet{newey1987simple} \(t\)-statistics based on six lags are reported in parentheses. The final row reports \(p\)-values from Wald tests of the joint null
hypothesis
\(H_0:
\lambda_{\mathrm{D2}}
=
\lambda_{\mathrm{D3}}
=
\lambda_{\mathrm{D4}}
=
\lambda_{\mathrm{D6}}
=
0\). The sample period is July 2017--December 2025.
}

\footnotesize
\setlength{\tabcolsep}{2.2pt}
\renewcommand{\arraystretch}{1.02}

\resizebox{\textwidth}{!}{%
\begin{tabular}{lcccccccccc}
\toprule
&
\multicolumn{5}{c}{Without Industry Controls} &
\multicolumn{5}{c}{With Fama--French Ten-Industry Controls} \\
\cmidrule(lr){2-6}\cmidrule(lr){7-11}
&
CAPM & FF3 & Carhart 4 & FF5 & FF5+UMD &
CAPM & FF3 & Carhart 4 & FF5 & FF5+UMD \\
\midrule

Intercept
& \FMcell{0.0002\\(0.05)}
& \FMcell{-0.0004\\(-0.10)}
& \FMcell{-0.0006\\(-0.15)}
& \FMcell{-0.0005\\(-0.11)}
& \FMcell{-0.0005\\(-0.13)}
& \FMcell{-0.0001\\(-0.02)}
& \FMcell{-0.0008\\(-0.21)}
& \FMcell{-0.0010\\(-0.25)}
& \FMcell{-0.0011\\(-0.28)}
& \FMcell{-0.0012\\(-0.29)} \\

AI-risk beta \(\Delta D2\)
& \FMcell{0.0218\\(0.18)}
& \FMcell{0.0753\\(0.76)}
& \FMcell{0.0780\\(0.78)}
& \FMcell{0.0685\\(0.69)}
& \FMcell{0.0784\\(0.79)}
& \FMcell{0.0221\\(0.19)}
& \FMcell{0.0574\\(0.57)}
& \FMcell{0.0619\\(0.60)}
& \FMcell{0.0654\\(0.67)}
& \FMcell{0.0746\\(0.75)} \\

AI-risk beta \(\Delta D3\)
& \FMcell{0.0823\\(2.85)}
& \FMcell{0.0600\\(2.93)}
& \FMcell{0.0569\\(2.87)}
& \FMcell{0.0535\\(3.00)}
& \FMcell{0.0539\\(3.08)}
& \FMcell{0.0704\\(2.73)}
& \FMcell{0.0569\\(2.88)}
& \FMcell{0.0544\\(2.83)}
& \FMcell{0.0503\\(2.85)}
& \FMcell{0.0509\\(2.94)} \\

AI-risk beta \(\Delta D4\)
& \FMcell{0.0745\\(0.85)}
& \FMcell{0.0045\\(0.06)}
& \FMcell{-0.0063\\(-0.08)}
& \FMcell{-0.0076\\(-0.11)}
& \FMcell{-0.0075\\(-0.11)}
& \FMcell{0.0772\\(0.96)}
& \FMcell{0.0286\\(0.40)}
& \FMcell{0.0178\\(0.25)}
& \FMcell{0.0166\\(0.25)}
& \FMcell{0.0151\\(0.22)} \\

AI-risk beta \(\Delta D6\)
& \FMcell{-0.1630\\(-0.63)}
& \FMcell{-0.0407\\(-0.19)}
& \FMcell{-0.0186\\(-0.09)}
& \FMcell{-0.0389\\(-0.20)}
& \FMcell{-0.0141\\(-0.07)}
& \FMcell{-0.1185\\(-0.50)}
& \FMcell{-0.0397\\(-0.19)}
& \FMcell{-0.0231\\(-0.11)}
& \FMcell{-0.0481\\(-0.26)}
& \FMcell{-0.0282\\(-0.15)} \\

\midrule

Market beta
& \FMcell{0.0027\\(0.83)}
& \FMcell{0.0046\\(1.96)}
& \FMcell{0.0049\\(2.23)}
& \FMcell{0.0055\\(2.48)}
& \FMcell{0.0055\\(2.59)}
& \FMcell{0.0029\\(0.92)}
& \FMcell{0.0045\\(1.97)}
& \FMcell{0.0047\\(2.21)}
& \FMcell{0.0053\\(2.51)}
& \FMcell{0.0054\\(2.60)} \\

SMB beta
&
& \FMcell{-0.0020\\(-0.98)}
& \FMcell{-0.0014\\(-0.71)}
& \FMcell{-0.0007\\(-0.42)}
& \FMcell{-0.0003\\(-0.17)}
&
& \FMcell{-0.0022\\(-1.09)}
& \FMcell{-0.0016\\(-0.83)}
& \FMcell{-0.0008\\(-0.47)}
& \FMcell{-0.0004\\(-0.25)} \\

HML beta
&
& \FMcell{0.0040\\(1.19)}
& \FMcell{0.0038\\(1.16)}
& \FMcell{0.0028\\(0.91)}
& \FMcell{0.0028\\(0.95)}
&
& \FMcell{0.0037\\(1.30)}
& \FMcell{0.0036\\(1.28)}
& \FMcell{0.0025\\(1.01)}
& \FMcell{0.0026\\(1.07)} \\

UMD beta
&
&
& \FMcell{0.0022\\(0.68)}
&
& \FMcell{0.0020\\(0.65)}
&
&
& \FMcell{0.0017\\(0.57)}
&
& \FMcell{0.0015\\(0.51)} \\

RMW beta
&
&
&
& \FMcell{0.0043\\(1.92)}
& \FMcell{0.0042\\(1.89)}
&
&
&
& \FMcell{0.0043\\(1.93)}
& \FMcell{0.0043\\(1.91)} \\

CMA beta
&
&
&
& \FMcell{0.0004\\(0.22)}
& \FMcell{0.0003\\(0.16)}
&
&
&
& \FMcell{0.0001\\(0.06)}
& \FMcell{0.0001\\(0.03)} \\

\midrule
\multicolumn{11}{l}{\textit{Fama--French industry indicators (Other omitted)}} \\

NoDur
& & & & &
& \FMcell{-0.0018\\(-1.09)}
& \FMcell{-0.0019\\(-1.11)}
& \FMcell{-0.0018\\(-1.04)}
& \FMcell{-0.0021\\(-1.16)}
& \FMcell{-0.0019\\(-1.09)} \\

Durbl
& & & & &
& \FMcell{-0.0024\\(-0.56)}
& \FMcell{-0.0021\\(-0.50)}
& \FMcell{-0.0016\\(-0.39)}
& \FMcell{-0.0027\\(-0.67)}
& \FMcell{-0.0026\\(-0.63)} \\

Manuf
& & & & &
& \FMcell{0.0041\\(2.08)}
& \FMcell{0.0033\\(1.87)}
& \FMcell{0.0032\\(1.92)}
& \FMcell{0.0031\\(1.80)}
& \FMcell{0.0029\\(1.78)} \\

Enrgy
& & & & &
& \FMcell{0.0017\\(0.19)}
& \FMcell{-0.0002\\(-0.02)}
& \FMcell{-0.0003\\(-0.04)}
& \FMcell{0.0022\\(0.27)}
& \FMcell{0.0018\\(0.23)} \\

HiTec
& & & & &
& \FMcell{0.0029\\(0.86)}
& \FMcell{0.0041\\(1.45)}
& \FMcell{0.0040\\(1.45)}
& \FMcell{0.0043\\(1.60)}
& \FMcell{0.0043\\(1.66)} \\

Telcm
& & & & &
& \FMcell{-0.0017\\(-0.61)}
& \FMcell{-0.0014\\(-0.52)}
& \FMcell{-0.0009\\(-0.31)}
& \FMcell{-0.0009\\(-0.30)}
& \FMcell{-0.0003\\(-0.09)} \\

Shops
& & & & &
& \FMcell{0.0008\\(0.28)}
& \FMcell{0.0012\\(0.43)}
& \FMcell{0.0012\\(0.44)}
& \FMcell{0.0006\\(0.21)}
& \FMcell{0.0005\\(0.18)} \\

Hlth
& & & & &
& \FMcell{-0.0037\\(-0.76)}
& \FMcell{-0.0013\\(-0.32)}
& \FMcell{-0.0015\\(-0.37)}
& \FMcell{-0.0005\\(-0.12)}
& \FMcell{-0.0005\\(-0.14)} \\

Utils
& & & & &
& \FMcell{0.0029\\(0.73)}
& \FMcell{0.0020\\(0.54)}
& \FMcell{0.0021\\(0.56)}
& \FMcell{0.0025\\(0.70)}
& \FMcell{0.0027\\(0.73)} \\

\midrule

Valid months
& \multicolumn{5}{c}{102}
& \multicolumn{5}{c}{102} \\

Average \# stocks
& \multicolumn{5}{c}{3,707}
& \multicolumn{5}{c}{3,707} \\

Adjusted \(R^2\)
& 2.52\% & 4.38\% & 4.75\% & 5.12\% & 5.40\%
& 4.50\% & 5.77\% & 6.02\% & 6.30\% & 6.50\% \\

Wald test \(p\)-value
& 0.0064
& 0.0097
& 0.0051
& 0.0123
& 0.0071
& 0.0123
& 0.0185
& 0.0140
& 0.0285
& 0.0197 \\
\bottomrule
\end{tabular}%
}
\end{table}

\clearpage
\begin{table}[p]
\centering
\caption{Twelve-Month Fama--MacBeth Regression Results}
\label{tab:fama_macbeth_12m_full}

\TableDescription{
\justifying \noindent
The table reports the Fama--MacBeth regression results based on
twelve-month first-pass beta-estimation windows, with the same set of conventional factor controls used in both passes. The first five columns report regressions without industry controls. The
remaining five columns include nine Fama--French ten-industry indicators,
with Other as the omitted industry. The continuous independent variable is winsorized at the 0.5\% level on a monthly basis. \citet{newey1987simple} \(t\)-statistics based on six lags are reported in parentheses. The final row reports \(p\)-values from Wald tests of the joint null
hypothesis
\(H_0:
\lambda_{\mathrm{D2}}
=
\lambda_{\mathrm{D3}}
=
\lambda_{\mathrm{D4}}
=
\lambda_{\mathrm{D6}}
=
0\). The sample period is January 2018--December 2025.
}

\footnotesize
\setlength{\tabcolsep}{2.2pt}
\renewcommand{\arraystretch}{1.02}

\resizebox{\textwidth}{!}{%
\begin{tabular}{lcccccccccc}
\toprule
&
\multicolumn{5}{c}{Without Industry Controls} &
\multicolumn{5}{c}{With Fama--French Ten-Industry Controls} \\
\cmidrule(lr){2-6}
\cmidrule(lr){7-11}
&
CAPM &
FF3 &
Carhart 4 &
FF5 &
FF5+UMD &
CAPM &
FF3 &
Carhart 4 &
FF5 &
FF5+UMD \\
\midrule

Intercept
& \FMcell{-0.0011\\(-0.26)}
& \FMcell{-0.0025\\(-0.53)}
& \FMcell{-0.0028\\(-0.62)}
& \FMcell{-0.0023\\(-0.46)}
& \FMcell{-0.0027\\(-0.56)}
& \FMcell{-0.0015\\(-0.37)}
& \FMcell{-0.0034\\(-0.76)}
& \FMcell{-0.0037\\(-0.85)}
& \FMcell{-0.0034\\(-0.74)}
& \FMcell{-0.0037\\(-0.83)} \\

AI-risk beta \(\Delta D2\)
& \FMcell{0.0344\\(0.17)}
& \FMcell{0.0765\\(0.43)}
& \FMcell{0.0756\\(0.42)}
& \FMcell{0.0857\\(0.53)}
& \FMcell{0.0972\\(0.58)}
& \FMcell{0.0283\\(0.14)}
& \FMcell{0.0617\\(0.34)}
& \FMcell{0.0575\\(0.32)}
& \FMcell{0.0813\\(0.50)}
& \FMcell{0.0864\\(0.52)} \\

AI-risk beta \(\Delta D3\)
& \FMcell{0.1042\\(2.15)}
& \FMcell{0.0842\\(2.16)}
& \FMcell{0.0783\\(2.04)}
& \FMcell{0.0799\\(2.20)}
& \FMcell{0.0731\\(2.06)}
& \FMcell{0.0928\\(2.12)}
& \FMcell{0.0822\\(2.20)}
& \FMcell{0.0774\\(2.11)}
& \FMcell{0.0747\\(2.15)}
& \FMcell{0.0692\\(2.04)} \\

AI-risk beta \(\Delta D4\)
& \FMcell{-0.0091\\(-0.06)}
& \FMcell{-0.0728\\(-0.60)}
& \FMcell{-0.0706\\(-0.58)}
& \FMcell{-0.0687\\(-0.61)}
& \FMcell{-0.0642\\(-0.58)}
& \FMcell{0.0310\\(0.23)}
& \FMcell{-0.0254\\(-0.22)}
& \FMcell{-0.0262\\(-0.22)}
& \FMcell{-0.0349\\(-0.33)}
& \FMcell{-0.0314\\(-0.29)} \\

AI-risk beta \(\Delta D6\)
& \FMcell{-0.1652\\(-0.51)}
& \FMcell{0.1096\\(0.45)}
& \FMcell{0.1627\\(0.68)}
& \FMcell{0.1150\\(0.53)}
& \FMcell{0.1619\\(0.76)}
& \FMcell{-0.1002\\(-0.35)}
& \FMcell{0.0807\\(0.33)}
& \FMcell{0.1284\\(0.55)}
& \FMcell{0.1007\\(0.47)}
& \FMcell{0.1431\\(0.68)} \\

\midrule

Market beta
& \FMcell{0.0028\\(0.71)}
& \FMcell{0.0059\\(2.24)}
& \FMcell{0.0064\\(2.40)}
& \FMcell{0.0068\\(2.79)}
& \FMcell{0.0072\\(2.96)}
& \FMcell{0.0030\\(0.82)}
& \FMcell{0.0057\\(2.29)}
& \FMcell{0.0062\\(2.48)}
& \FMcell{0.0065\\(2.76)}
& \FMcell{0.0069\\(2.94)} \\

SMB beta
&
& \FMcell{-0.0024\\(-0.91)}
& \FMcell{-0.0015\\(-0.57)}
& \FMcell{-0.0006\\(-0.25)}
& \FMcell{0.0000\\(-0.01)}
&
& \FMcell{-0.0026\\(-1.05)}
& \FMcell{-0.0018\\(-0.72)}
& \FMcell{-0.0005\\(-0.23)}
& \FMcell{0.0000\\(0.01)} \\

HML beta
&
& \FMcell{0.0051\\(1.29)}
& \FMcell{0.0049\\(1.24)}
& \FMcell{0.0031\\(0.82)}
& \FMcell{0.0030\\(0.80)}
&
& \FMcell{0.0054\\(1.53)}
& \FMcell{0.0053\\(1.51)}
& \FMcell{0.0031\\(1.00)}
& \FMcell{0.0031\\(1.02)} \\

UMD beta
&
&
& \FMcell{0.0028\\(0.66)}
&
& \FMcell{0.0025\\(0.63)}
&
&
& \FMcell{0.0024\\(0.59)}
&
& \FMcell{0.0022\\(0.59)} \\

RMW beta
&
&
&
& \FMcell{0.0063\\(2.52)}
& \FMcell{0.0061\\(2.43)}
&
&
&
& \FMcell{0.0064\\(2.49)}
& \FMcell{0.0062\\(2.41)} \\

CMA beta
&
&
&
& \FMcell{-0.0008\\(-0.33)}
& \FMcell{-0.0011\\(-0.45)}
&
&
&
& \FMcell{-0.0011\\(-0.52)}
& \FMcell{-0.0013\\(-0.62)} \\

\midrule
\multicolumn{11}{l}{
\textit{Fama--French industry indicators (Other omitted)}
} \\

NoDur
&
&
&
&
&
& \FMcell{-0.0011\\(-0.58)}
& \FMcell{-0.0010\\(-0.58)}
& \FMcell{-0.0004\\(-0.23)}
& \FMcell{-0.0014\\(-0.73)}
& \FMcell{-0.0009\\(-0.49)} \\

Durbl
&
&
&
&
&
& \FMcell{-0.0019\\(-0.41)}
& \FMcell{-0.0015\\(-0.34)}
& \FMcell{-0.0011\\(-0.26)}
& \FMcell{-0.0027\\(-0.64)}
& \FMcell{-0.0025\\(-0.59)} \\

Manuf
&
&
&
&
&
& \FMcell{0.0045\\(2.11)}
& \FMcell{0.0036\\(1.96)}
& \FMcell{0.0031\\(1.85)}
& \FMcell{0.0031\\(1.72)}
& \FMcell{0.0027\\(1.57)} \\

Enrgy
&
&
&
&
&
& \FMcell{0.0037\\(0.38)}
& \FMcell{0.0005\\(0.05)}
& \FMcell{-0.0014\\(-0.18)}
& \FMcell{0.0055\\(0.66)}
& \FMcell{0.0034\\(0.42)} \\

HiTec
&
&
&
&
&
& \FMcell{0.0036\\(1.00)}
& \FMcell{0.0057\\(1.95)}
& \FMcell{0.0053\\(1.88)}
& \FMcell{0.0055\\(2.07)}
& \FMcell{0.0052\\(2.02)} \\

Telcm
&
&
&
&
&
& \FMcell{-0.0006\\(-0.22)}
& \FMcell{0.0000\\(0.01)}
& \FMcell{0.0013\\(0.45)}
& \FMcell{0.0006\\(0.19)}
& \FMcell{0.0014\\(0.48)} \\

Shops
&
&
&
&
&
& \FMcell{0.0007\\(0.22)}
& \FMcell{0.0013\\(0.46)}
& \FMcell{0.0015\\(0.52)}
& \FMcell{0.0002\\(0.08)}
& \FMcell{0.0003\\(0.09)} \\

Hlth
&
&
&
&
&
& \FMcell{-0.0037\\(-0.70)}
& \FMcell{0.0001\\(0.02)}
& \FMcell{-0.0001\\(-0.04)}
& \FMcell{0.0014\\(0.37)}
& \FMcell{0.0011\\(0.30)} \\

Utils
&
&
&
&
&
& \FMcell{0.0036\\(0.91)}
& \FMcell{0.0034\\(0.96)}
& \FMcell{0.0045\\(1.22)}
& \FMcell{0.0046\\(1.35)}
& \FMcell{0.0053\\(1.54)} \\

\midrule

Valid months
& \multicolumn{5}{c}{96}
& \multicolumn{5}{c}{96} \\

Average \# stocks
& \multicolumn{5}{c}{3,621}
& \multicolumn{5}{c}{3,621} \\

Adjusted \(R^2\)
& 2.53\%
& 4.85\%
& 5.18\%
& 5.90\%
& 6.16\%
& 4.64\%
& 6.22\%
& 6.43\%
& 6.99\%
& 7.15\% \\

Wald test \(p\)-value
& 0.0314
& 0.0127
& 0.0065
& 0.0209
& 0.0181
& 0.0772
& 0.0493
& 0.0359
& 0.0656
& 0.0650 \\
\bottomrule
\end{tabular}%
}
\end{table}

\begin{sidewaystable}[p]
\centering

\captionsetup{
    width=0.96\textheight,
    justification=raggedright,
    singlelinecheck=false
}
\caption{Stock-Level Characteristics for D3 Betas}
\label{tab:d3_characteristics}

\begin{minipage}{0.96\textheight}
\footnotesize
\raggedright
\justifying \noindent
The table reports monthly cross-sectional regressions of firm-level D3
betas on industry indicators, established risk exposures, and stock-level
characteristics. The dependent variable is the D3 beta estimated using the
three-month window and FF5+UMD controls. Columns (1)--(16) report separate
regressions that include one explanatory variable at a time.
Column (17) reports a joint regression that includes all
characteristics and the complete set of Fama--French ten-industry
indicators. For compactness, only the Energy, HiTec, and Utility industry
coefficients are displayed. The continuous independent variable is winsorized at the 0.5\% level on a monthly basis. Coefficients are time-series averages of the monthly cross-sectional
estimates. \citet{newey1987simple} \(t\)-statistics based on six lags are
reported in parentheses. The sample covers 105 portfolio formation months from March 2017 through November 2025.
\end{minipage}

\vspace{0.6em}

\scriptsize
\setlength{\tabcolsep}{2.2pt}
\renewcommand{\arraystretch}{0.96}

\resizebox{0.96\textheight}{!}{%
\begin{tabular}{@{}l*{17}{c}@{}}
\toprule
Variable
& (1) & (2) & (3) & (4) & (5) & (6)
& (7) & (8) & (9) & (10) & (11) & (12)
& (13) & (14) & (15) & (16) & (17) \\
\midrule

Intercept
& \FMcell{0.0010\\(1.52)}
& \FMcell{0.0012\\(2.08)}
& \FMcell{0.0010\\(1.55)}
& \FMcell{-0.0004\\(-0.33)}
& \FMcell{0.0002\\(0.20)}
& \FMcell{0.0028\\(1.51)}
& \FMcell{0.0003\\(0.56)}
& \FMcell{0.0009\\(1.63)}
& \FMcell{0.0006\\(0.98)}
& \FMcell{0.0004\\(0.54)}
& \FMcell{0.0009\\(1.39)}
& \FMcell{-0.0006\\(-0.82)}
& \FMcell{-0.0000\\(-0.04)}
& \FMcell{0.0001\\(0.18)}
& \FMcell{0.0005\\(0.65)}
& \FMcell{-0.0012\\(-1.21)}
& \FMcell{-0.0061\\(-2.03)}
\\[1pt]

Energy
& \FMcell{0.0038\\(1.01)}
& \multicolumn{15}{c}{}
& \FMcell{0.0036\\(1.00)}
\\[1pt]

HiTec
& & \FMcell{-0.0009\\(-0.92)}
& \multicolumn{14}{c}{}
& \FMcell{-0.0007\\(-0.77)}
\\[1pt]

Utility
& \multicolumn{2}{c}{}
& \FMcell{0.0017\\(0.68)}
& \multicolumn{13}{c}{}
& \FMcell{0.0042\\(1.19)}
\\[1pt]

\(\beta^{MKT}\)
& \multicolumn{3}{c}{}
& \FMcell{0.0011\\(0.84)}
& \multicolumn{12}{c}{}
& \FMcell{0.0002\\(0.30)}
\\[1pt]

\(\beta^{VIX}\)
& \multicolumn{4}{c}{}
& \FMcell{-0.0005\\(-0.74)}
& \multicolumn{11}{c}{}
& \FMcell{-0.0007\\(-0.92)}
\\[1pt]

SIZE
& \multicolumn{5}{c}{}
& \FMcell{-0.0003\\(-1.14)}
& \multicolumn{10}{c}{}
& \FMcell{0.0003\\(1.44)}
\\[1pt]

BM
& \multicolumn{6}{c}{}
& \FMcell{-0.0003\\(-0.53)}
& \multicolumn{9}{c}{}
& \FMcell{0.0000\\(0.01)}
\\[1pt]

MOM
& \multicolumn{7}{c}{}
& \FMcell{-0.0018\\(-1.55)}
& \multicolumn{8}{c}{}
& \FMcell{0.0004\\(0.43)}
\\[1pt]

REV
& \multicolumn{8}{c}{}
& \FMcell{-0.0017\\(-0.38)}
& \multicolumn{7}{c}{}
& \FMcell{0.0018\\(0.40)}
\\[1pt]

ILLIQ
& \multicolumn{9}{c}{}
& \FMcell{0.0000\\(0.04)}
& \multicolumn{6}{c}{}
& \FMcell{0.0098\\(1.92)}
\\[1pt]

COSKEW
& \multicolumn{10}{c}{}
& \FMcell{0.0028\\(1.17)}
& \multicolumn{5}{c}{}
& \FMcell{0.0008\\(0.54)}
\\[1pt]

IVOL
& \multicolumn{11}{c}{}
& \FMcell{0.0759\\(1.62)}
& \multicolumn{4}{c}{}
& \FMcell{-0.0140\\(-0.21)}
\\[1pt]

DISP
& \multicolumn{12}{c}{}
& \FMcell{0.0021\\(1.24)}
& \multicolumn{3}{c}{}
& \FMcell{0.0016\\(0.88)}
\\[1pt]

\(\mathrm{I/A}\)
& \multicolumn{13}{c}{}
& \FMcell{0.0021\\(3.72)}
& \multicolumn{2}{c}{}
& \FMcell{0.0027\\(3.67)}
\\[1pt]

ROE
& \multicolumn{14}{c}{}
& \FMcell{-0.0083\\(-1.86)}
&
& \FMcell{0.0010\\(0.18)}
\\[1pt]

MAX
& \multicolumn{15}{c}{}
& \FMcell{0.0635\\(1.86)}
& \FMcell{0.0531\\(1.04)}
\\

\midrule

Avg.\ stocks
& 3753 & 3753 & 3753 & 4054 & 8423 & 4759
& 4378 & 4418 & 4759 & 8261 & 4054 & 4750
& 3608 & 4574 & 4461 & 4750 & 2316
\\

Avg.\ monthly \(R^2\)
& 0.24\% & 0.08\% & 0.05\% & 0.19\% & 0.91\% & 0.36\%
& 0.10\% & 0.35\% & 1.07\% & 0.05\% & 0.13\% & 0.63\%
& 0.14\% & 0.08\% & 0.40\% & 0.73\% & 7.18\%
\\

\bottomrule
\end{tabular}%
}

\end{sidewaystable}

\begin{table}[p]
\centering

\caption{Robustness Tests of Domain-Specific Prices of Risk}
\label{tab:d3_robustness}

\TableDescription{\justifying \noindent
The table reports Fama--MacBeth estimates of the prices of risk associated with the four domain-specific AI-risk factors. All specifications use three-month first-pass beta-estimation windows and control for industry effects in the monthly cross-sectional regressions. The five columns correspond to the conventional factor controls used in the first-pass beta estimations.
Panel A reconstructs the factor innovations using AR(7), ARMA(1,1), and
ARMA(2,1) models over the full sample. Panel B uses the baseline AR(1)
innovations and restricts the sample to the pre-ChatGPT period ending in
November 2022. Coefficients are time-series averages of monthly
cross-sectional prices of risk. \citet{newey1987simple} \(t\)-statistics
based on six lags are reported in parentheses.
}

\scriptsize
\setlength{\tabcolsep}{3.5pt}
\renewcommand{\arraystretch}{1.00}

\begin{tabular*}{\textwidth}{@{\extracolsep{\fill}}llccccc@{}}
\toprule
Innovation model &
Domain &
CAPM &
FF3 &
Carhart 4 &
FF5 &
FF5+UMD \\
\midrule

\multicolumn{7}{l}{Panel A: Alternative Innovation Models}\\[-0.3em]
\midrule

AR(7)
& D2
& \FMcell{0.0617\\(0.82)}
& \FMcell{0.0401\\(0.59)}
& \FMcell{0.0337\\(0.51)}
& \FMcell{0.0454\\(0.65)}
& \FMcell{0.0387\\(0.57)} \\

&
D3
& \FMcell{0.0442\\(2.98)}
& \FMcell{0.0420\\(3.41)}
& \FMcell{0.0394\\(3.13)}
& \FMcell{0.0402\\(3.40)}
& \FMcell{0.0393\\(3.26)} \\

&
D4
& \FMcell{0.0667\\(1.13)}
& \FMcell{0.0472\\(0.93)}
& \FMcell{0.0416\\(0.85)}
& \FMcell{0.0474\\(0.98)}
& \FMcell{0.0451\\(0.95)} \\

&
D6
& \FMcell{0.0120\\(0.08)}
& \FMcell{-0.0449\\(-0.36)}
& \FMcell{-0.0708\\(-0.57)}
& \FMcell{-0.0476\\(-0.43)}
& \FMcell{-0.0582\\(-0.52)} \\

\midrule

ARMA(1,1)
& D2
& \FMcell{0.0538\\(0.74)}
& \FMcell{0.0271\\(0.43)}
& \FMcell{0.0258\\(0.41)}
& \FMcell{0.0238\\(0.37)}
& \FMcell{0.0214\\(0.34)} \\

&
D3
& \FMcell{0.0479\\(3.15)}
& \FMcell{0.0456\\(3.89)}
& \FMcell{0.0433\\(3.68)}
& \FMcell{0.0427\\(4.00)}
& \FMcell{0.0422\\(3.93)} \\

&
D4
& \FMcell{0.0702\\(1.15)}
& \FMcell{0.0449\\(0.86)}
& \FMcell{0.0418\\(0.83)}
& \FMcell{0.0370\\(0.76)}
& \FMcell{0.0394\\(0.84)} \\

&
D6
& \FMcell{0.0525\\(0.37)}
& \FMcell{-0.0131\\(-0.11)}
& \FMcell{-0.0393\\(-0.32)}
& \FMcell{-0.0380\\(-0.34)}
& \FMcell{-0.0473\\(-0.43)} \\

\midrule

ARMA(2,1)
& D2
& \FMcell{0.0555\\(0.79)}
& \FMcell{0.0289\\(0.48)}
& \FMcell{0.0242\\(0.40)}
& \FMcell{0.0271\\(0.43)}
& \FMcell{0.0209\\(0.34)} \\

&
D3
& \FMcell{0.0452\\(3.04)}
& \FMcell{0.0416\\(3.54)}
& \FMcell{0.0388\\(3.27)}
& \FMcell{0.0386\\(3.55)}
& \FMcell{0.0380\\(3.46)} \\

&
D4
& \FMcell{0.0738\\(1.19)}
& \FMcell{0.0496\\(0.93)}
& \FMcell{0.0439\\(0.85)}
& \FMcell{0.0430\\(0.85)}
& \FMcell{0.0428\\(0.88)} \\

&
D6
& \FMcell{0.0454\\(0.31)}
& \FMcell{-0.0194\\(-0.16)}
& \FMcell{-0.0465\\(-0.37)}
& \FMcell{-0.0372\\(-0.34)}
& \FMcell{-0.0488\\(-0.45)} \\

\midrule

\multicolumn{7}{l}{Panel B: Pre-ChatGPT Subsample}\\[-0.3em]
\midrule

AR(1)
& D2
& \FMcell{0.0156\\(0.23)}
& \FMcell{0.0096\\(0.15)}
& \FMcell{0.0061\\(0.10)}
& \FMcell{0.0127\\(0.20)}
& \FMcell{0.0101\\(0.15)} \\

&
D3
& \FMcell{0.0412\\(2.36)}
& \FMcell{0.0387\\(3.39)}
& \FMcell{0.0383\\(3.37)}
& \FMcell{0.0346\\(3.35)}
& \FMcell{0.0354\\(3.42)} \\

&
D4
& \FMcell{0.0482\\(0.86)}
& \FMcell{0.0322\\(0.75)}
& \FMcell{0.0272\\(0.67)}
& \FMcell{0.0291\\(0.70)}
& \FMcell{0.0268\\(0.70)} \\

&
D6
& \FMcell{-0.0874\\(-0.56)}
& \FMcell{-0.1249\\(-0.85)}
& \FMcell{-0.1399\\(-0.94)}
& \FMcell{-0.0965\\(-0.70)}
& \FMcell{-0.1029\\(-0.75)} \\

\bottomrule
\end{tabular*}%

\end{table}

\clearpage

\vspace{3cm}

\begin{figure}[p]
    \centering

    \begin{subfigure}[t]{\textwidth}
        \centering
        \includegraphics[width=\textwidth]{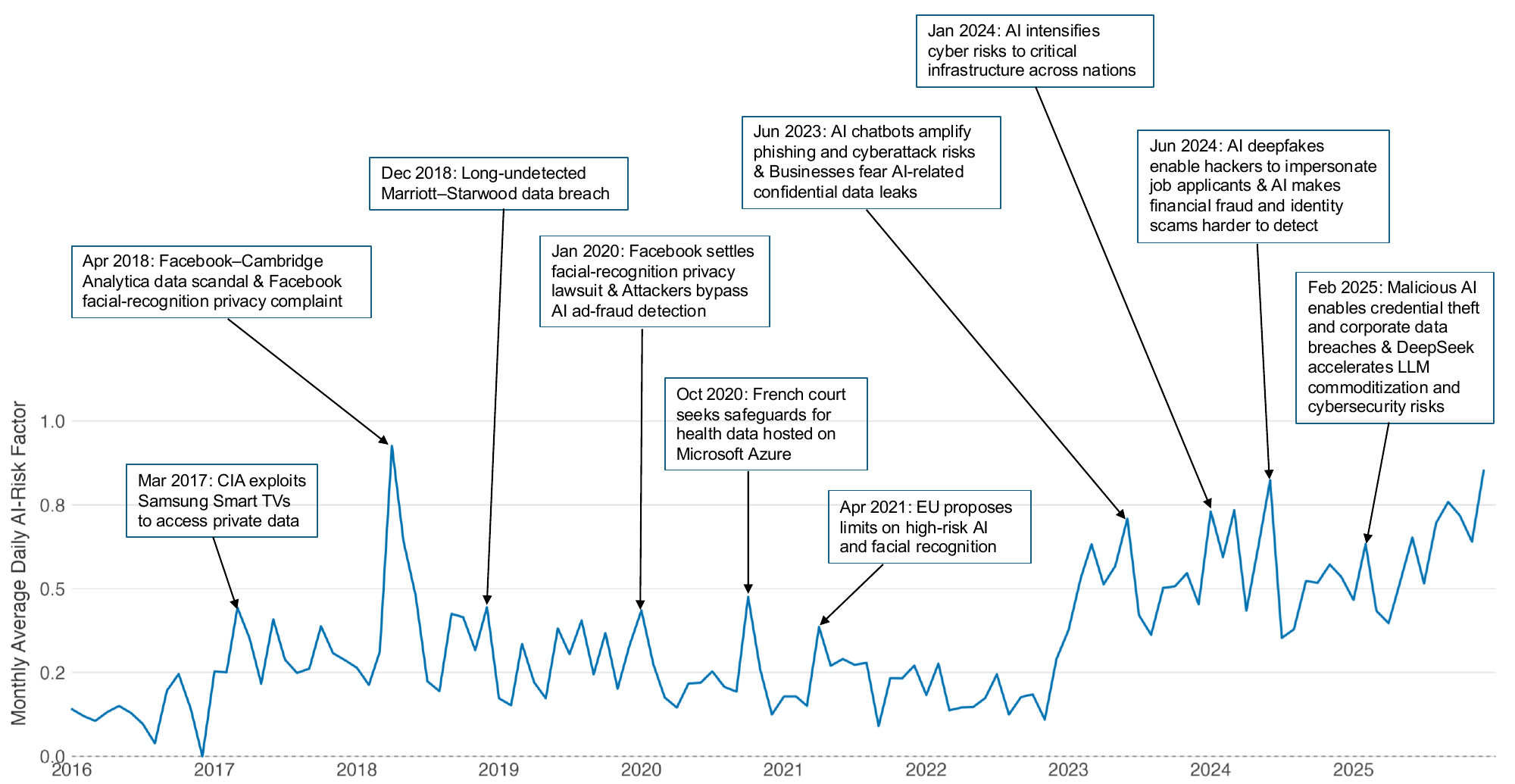}
        \caption{Privacy and security (D2)}
        \label{fig:d2_event}
    \end{subfigure}

    \vspace{3cm}

    \begin{subfigure}[t]{\textwidth}
        \centering
        \includegraphics[width=\textwidth]{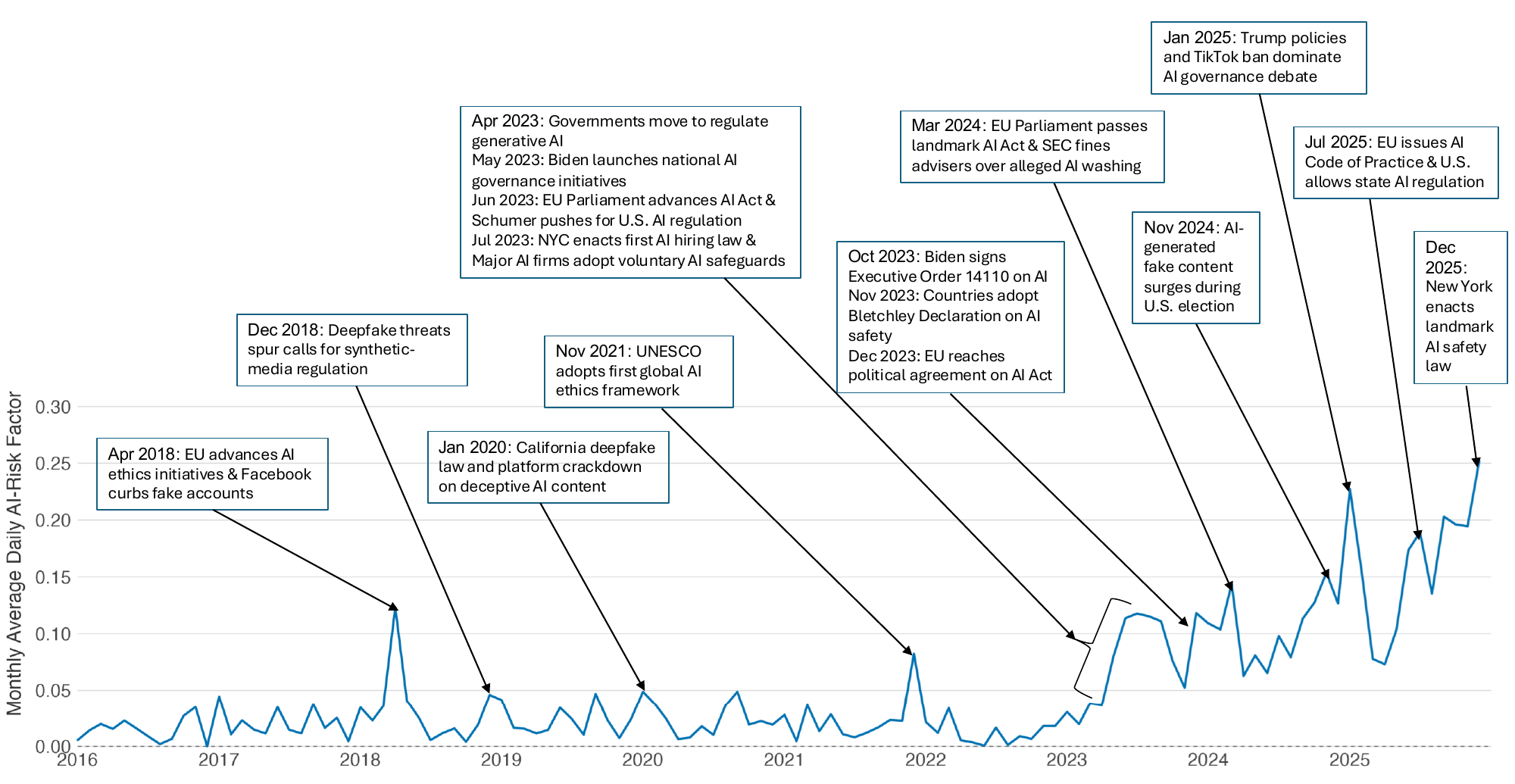}
        \caption{Misinformation (D3)}
        \label{fig:d3_event}
    \end{subfigure}
\captionsetup{
    justification=raggedright,
    singlelinecheck=false,
    skip=40pt
}
    \caption{\justifying \noindent \textbf{Time Series of Domain-Specific AI-Risk Factors.}
    This figure plots the monthly averages of the four daily domain-specific
    AI-risk factors. Panels (a)--(d) correspond to Privacy and security (D2),
    Misinformation (D3), Malicious actors and misuse (D4), and Socioeconomic
    and environmental harm (D6), respectively. The annotations identify
    major news events associated with selected peaks in each series.}
    \label{fig:domain_factor_timeseries}
\end{figure}

\clearpage

\begin{figure}[p]
    \ContinuedFloat
    \centering
    \setcounter{subfigure}{2}

    \begin{subfigure}[t]{\textwidth}
        \centering
        \includegraphics[width=\textwidth]{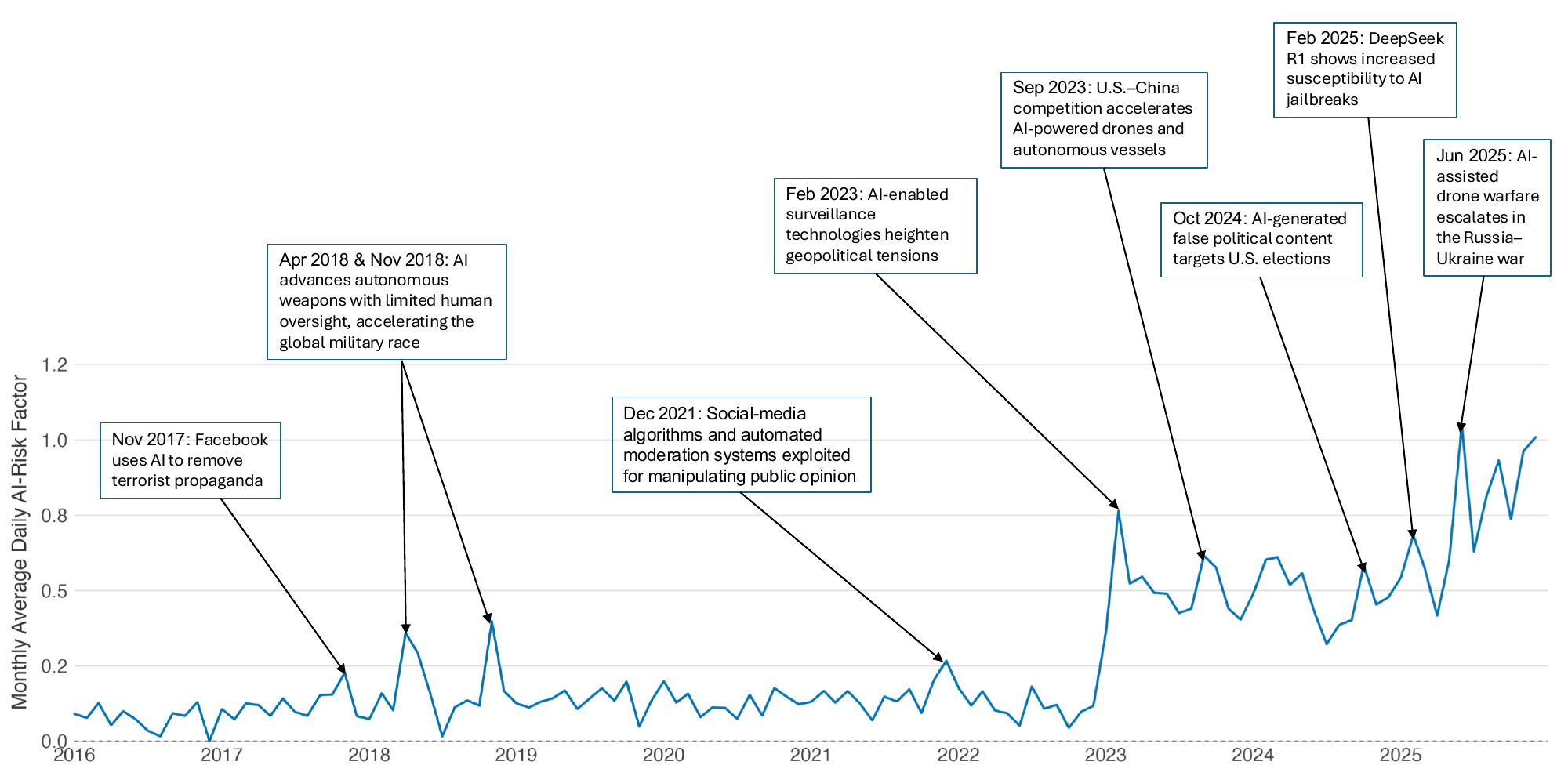}
        \caption{Malicious actors and misuse (D4)}
        \label{fig:d4_event}
    \end{subfigure}

    \vspace{3cm}

    \begin{subfigure}[t]{\textwidth}
        \centering
        \includegraphics[width=\textwidth]{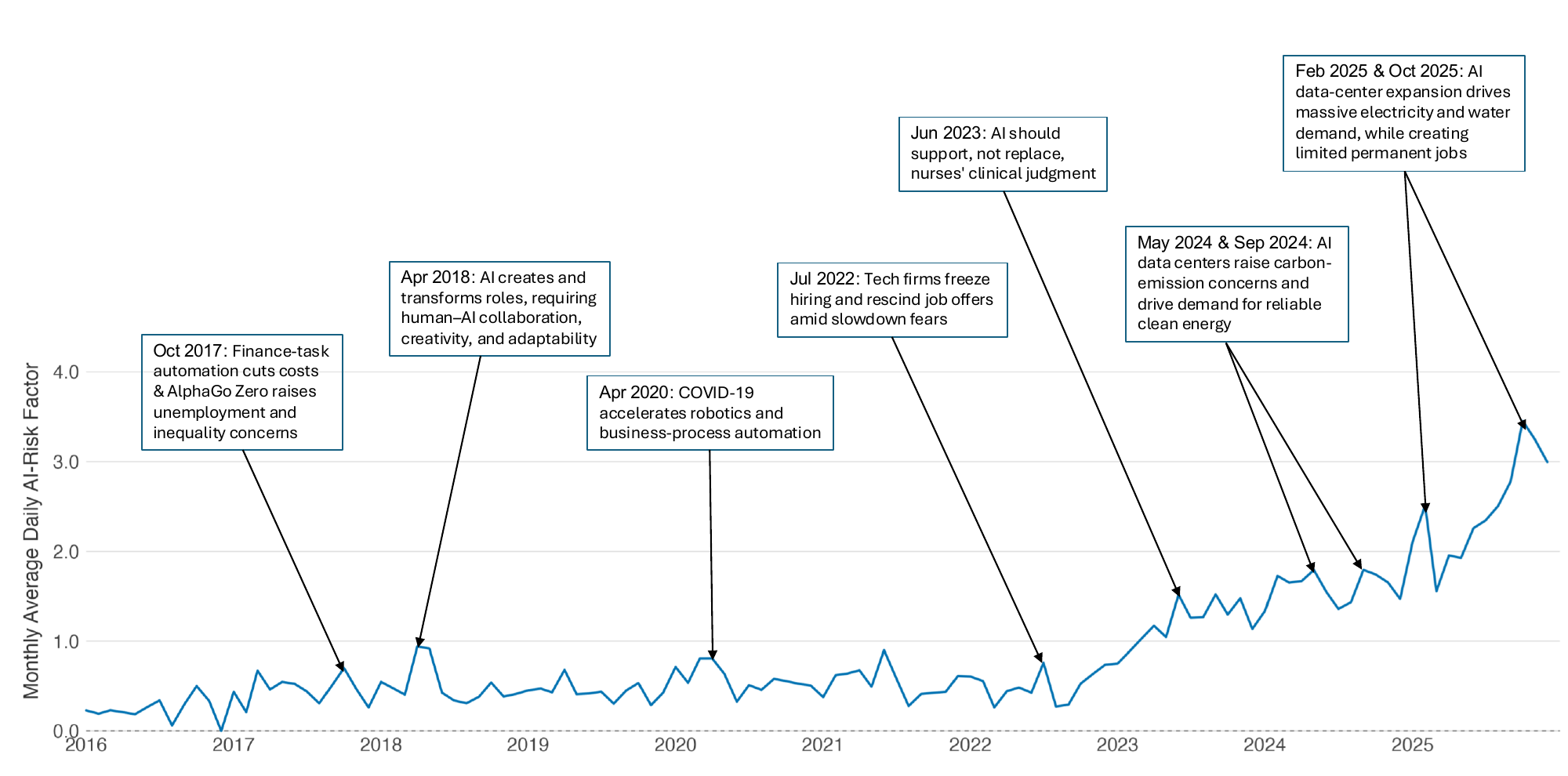}
        \caption{Socioeconomic and environmental harm (D6)}
        \label{fig:d6_event}
    \end{subfigure}
\captionsetup{
    justification=raggedright,
    singlelinecheck=false,
    skip=40pt
}
    \caption[]{\justifying \noindent \textbf{Time Series of Domain-Specific AI-Risk Factors
    (continued).} Panels (c) and (d) present the Malicious actors and misuse
    (D4) and Socioeconomic and environmental harm (D6) factors, respectively.}
\end{figure}

\clearpage
\appendix

\clearpage
\appendix

\section{MIT AI Risk Repository Domain and Subdomain Definitions}
\label{app:mit_taxonomy}

This appendix presents the 7 domains and 24 subdomains in the Domain
Taxonomy of the MIT AI Risk Repository. The definitions are based on \citet[Table~2]{slattery2026airisk}.

\begingroup
\small
\begin{spacing}{1.05}

\newcommand{\AIRiskSubdomain}[3]{%
  \noindent\textbf{#1 #2.} #3\par\vspace{0.55em}
}

\subsection*{D1 Discrimination and toxicity}

\AIRiskSubdomain
{1.1}
{Unfair discrimination and misrepresentation}
{Unequal treatment of individuals or groups by AI, often based on race,
gender, or other sensitive characteristics, resulting in unfair outcomes
and unfair representation of those groups.}

\AIRiskSubdomain
{1.2}
{Exposure to toxic content}
{AI that exposes users to harmful, abusive, unsafe. or inappropriate
content. May involve providing advice or encouraging action. Examples of toxic content include hate speech, violence, extremism, illegal acts, or child sexual abuse material, as well as content that violates community norms such as profanity, inflammatory political speech, or pornography.}

\AIRiskSubdomain
{1.3}
{Unequal performance across groups}
{Accuracy and effectiveness of AI decisions and actions is dependent on group membership, where decisions in AI system design and biased training data lead to unequal outcomes, reduced benefits, increased effort, and alienation of users.}

\subsection*{D2 Privacy and security}

\AIRiskSubdomain
{2.1}
{Compromise of privacy by obtaining, leaking, or correctly inferring
sensitive information}
{AI systems that memorize and leak sensitive personal data or infer private information about individuals without their consent. Unexpected or unauthorized sharing of data and information can compromise user expectation of privacy, assist identity theft, or cause loss of confidential intellectual property.}

\AIRiskSubdomain
{2.2}
{AI system security vulnerabilities and attacks}
{Vulnerabilities that can be exploited in AI systems, software
development toolchains, and hardware, resulting in unauthorized
access, data and privacy breaches, or system manipulation
causing unsafe outputs or behavior.}

\subsection*{D3 Misinformation}

\AIRiskSubdomain
{3.1}
{False or misleading information}
{AI systems that inadvertently generate or spread incorrect or deceptive information, which can lead to inaccurate beliefs in users and undermine their autonomy. Humans that make decisions based on false beliefs can experience physical, emotional, or material harms.}

\AIRiskSubdomain
{3.2}
{Pollution of information ecosystem and loss of consensus reality}
{Highly personalized AI-generated misinformation that creates ``filter bubbles'' where individuals only see what matches their existing beliefs, undermining shared reality and weakening social cohesion and political processes.}

\subsection*{D4 Malicious actors and misuse}

\AIRiskSubdomain
{4.1}
{Disinformation, surveillance, and influence at scale}
{Using AI systems to conduct large-scale disinformation campaigns, malicious surveillance, or targeted and sophisticated automated censorship and propaganda, with the aim of manipulating political processes, public opinion, and behavior.}

\AIRiskSubdomain
{4.2}
{Cyberattacks, weapon development or use, and mass harm}
{Using AI systems to develop cyber weapons (e.g., by coding cheaper, more effective malware), develop new or enhance existing weapons (e.g., lethal autonomous weapons or chemical, biological, radiological, nuclear, and high-yield explosives), or use weapons to cause mass harm.}

\AIRiskSubdomain
{4.3}
{Fraud, scams, and targeted manipulation}
{Using AI systems to gain a personal advantage over others such as through cheating, fraud, scams, blackmail, or targeted manipulation of beliefs or behavior. Examples include AI-facilitated plagiarism for research or education, impersonating a trusted or fake individual for illegitimate financial benefit, or creating humiliating or sexual imagery.}

\subsection*{D5 Human--computer interaction}

\AIRiskSubdomain
{5.1}
{Overreliance and unsafe use}
{Anthropomorphizing, trusting, or relying on AI systems by users, leading to emotional or material dependence and to inappropriate relationships with, or expectations of, AI systems. Trust can be exploited by malicious actors (e.g., to harvest information or enable manipulation) or result in harm from inappropriate use of AI in critical situations (e.g., medical emergency). Over-reliance on AI systems can compromise autonomy and weaken social ties.}

\AIRiskSubdomain
{5.2}
{Loss of human agency and autonomy}
{Delegating by humans of key decisions to AI systems, or AI systems that make decisions that diminish human control and autonomy, potentially leading to humans feeling disempowered, losing the ability to shape a fulfilling life trajectory or becoming cognitively enfeebled.}

\subsection*{D6 Socioeconomic and environmental harm}

\AIRiskSubdomain
{6.1}
{Power centralization and unfair distribution of benefits}
{AI-driven concentration of power and resources within certain entities or groups, especially those with access to or ownership of powerful AI systems, leading to inequitable distribution of benefits and increased societal inequality.}

\AIRiskSubdomain
{6.2}
{Increased inequality and decline in employment quality}
{Social and economic inequalities caused by widespread use of AI, such as by automating jobs, reducing the quality of employment, or producing exploitative dependencies between workers and their employers.}

\AIRiskSubdomain
{6.3}
{Economic and cultural devaluation of human effort}
{AI systems capable of creating economic or cultural value, including through reproduction of human innovation or creativity (e.g., art, music, writing, coding, and invention), destabilizing economic and social systems that rely on human effort. The ubiquity of AI-generated content may lead to reduced appreciation for human skills, disruption of creative and knowledge-based industries, and homogenization of cultural experiences.}

\AIRiskSubdomain
{6.4}
{Competitive dynamics}
{Competition by AI developers or state-like actors in an AI ``race'' by rapidly developing, deploying, and applying AI systems to maximize strategic or economic advantage, increasing the risk they release unsafe and error-prone systems.}

\AIRiskSubdomain
{6.5}
{Governance failure}
{Inadequate regulatory frameworks and oversight mechanisms that fail to keep pace with AI development, leading to ineffective governance and the inability to manage AI risks appropriately.}

\AIRiskSubdomain
{6.6}
{Environmental harm}
{The development and operation of AI systems that cause environmental harm, such as through energy consumption of data centers or the materials and carbon footprints associated with AI hardware.}

\subsection*{D7 AI system safety, failures, and limitations}

\AIRiskSubdomain
{7.1}
{AI pursuing its own goals in conflict with human goals or values}
{AI systems that act in conflict with ethical standards or human goals or values, especially the goals of designers or users. These misaligned behaviors may be introduced by humans during design and development, such as through reward hacking and goal misgeneralization, and may result in AI using dangerous capabilities such as manipulation, deception, or situational awareness to seek power, self-proliferate, or achieve other goals.}

\AIRiskSubdomain
{7.2}
{AI possessing dangerous capabilities}
{AI systems that develop, access, or are provided with capabilities that increase their potential to cause mass harm through deception, weapons development and acquisition, persuasion and manipulation, political strategy, cyber-offense, AI development, situational awareness, and self-proliferation. These capabilities may cause mass harm due to malicious human actors, misaligned AI systems, or failure in the AI system.}

\AIRiskSubdomain
{7.3}
{Lack of capability or robustness}
{AI systems that fail to perform reliably or effectively under varying conditions, exposing them to errors and failures that can have significant consequences, especially in critical applications or areas that require moral reasoning.}

\AIRiskSubdomain
{7.4}
{Lack of transparency or interpretability}
{Challenges in understanding or explaining the decision-making processes of AI systems, which can lead to mistrust, difficulty in enforcing compliance standards or holding relevant actors accountable for harms, and the inability to identify and correct errors.}

\AIRiskSubdomain
{7.5}
{AI welfare and rights}
{Ethical considerations regarding the treatment of potentially sentient AI entities, including discussions around their potential rights and welfare, particularly as AI systems become more advanced and autonomous.}

\AIRiskSubdomain
{7.6}
{Multi-agent risks}
{Risks from multi-agent interactions due to incentives (which can lead to conflict or collusion) and/or the structure of multi-agent systems, which can create cascading failures, selection pressures, new security vulnerabilities, and a lack of shared information and trust.}

\end{spacing}
\endgroup

\clearpage
\section{Construction of Stock-Level Characteristics}
\label{app:stock_characteristics}

This appendix describes the construction of the variables used to examine
the characteristics of high-D3-beta firms. The definitions follow
\citet{bali2017economic} where possible. I replace VXO with VIX because VXO
does not cover the full 2016--2025 sample. All characteristics are measured at the stock-month level, although their underlying data frequencies and estimation windows differ.
\begingroup
\small
\setlength{\tabcolsep}{4pt}
\renewcommand{\arraystretch}{1.15}

\begin{longtable}{
    L{0.13\textwidth}
    L{0.5\textwidth}
    L{0.31\textwidth}
}
\caption{Construction of Stock-Level Characteristics}
\label{tab:stock_characteristic_definitions}\\

\toprule
Variable & Construction & Estimation window \\
\midrule
\endfirsthead

\multicolumn{3}{l}{\textit{Table \thetable\ continued}}\\
\toprule
Variable & Construction & Estimation window \\
\midrule
\endhead

\midrule
\multicolumn{3}{r}{\textit{Continued on next page}}\\
\endfoot

\bottomrule
\endlastfoot

\(\beta^{MKT}\)
&
The slope from a regression of the stock's monthly excess return on the
market excess return.
&
Estimated over the preceding 60 months, with at least 24 valid monthly observations.
\\

\(\beta^{VIX}\)
&
The coefficient on the daily change in the VIX-based variance measure,
estimated each month by regressing the stock's daily excess return on
the market excess return and the change in this variance measure. The
VIX-based variance measure is obtained by converting the daily closing
CBOE VIX from annualized volatility to annualized variance.
&
Estimated within month \(m\), with at least 15 valid daily observations.
\\

SIZE
&
The natural logarithm of the stock's market capitalization measured in
millions of dollars.
&
Measured in month \(m\).
\\

BM
&
The natural logarithm of book equity divided by the corresponding
December market equity. Book equity equals stockholders' equity plus
deferred taxes and investment tax credits, when available, minus preferred
stock.
&
Fiscal-year information ending in calendar year \(y\) is used from July
of year \(y+1\) through June of year \(y+2\).
\\

MOM
&
The cumulative stock return from months \(m-12\) through \(m-2\).
&
An 11-month return window that excludes month \(m-1\).
\\

REV
&
The stock return in month \(m-1\).
&
One month before month \(m\).
\\

ILLIQ
&
The monthly average of the absolute daily stock return divided by daily
dollar trading volume:
\[
\mathrm{ILLIQ}_{i,m}
=
10^6
\mathop{\mathrm{Avg}}\limits_{t\in m}
\left(
\frac{|R_{i,t}|}{\mathrm{VOLD}_{i,t}}
\right).
\]
Daily dollar trading volume is price multiplied by share volume.
&
Calculated within month \(m\), with at least 15 valid daily observations.
\\

COSKEW
&
The stock's monthly co-skewness is calculated as
\[
\mathrm{COSKEW}_{i,m}
=
\frac{
E\!\left[\epsilon_{i,\tau}(R^{e}_{M,\tau})^2\right]
}{
\sqrt{E[\epsilon_{i,\tau}^{2}]}\,
E[(R^{e}_{M,\tau})^2]
},
\]
where \(\tau\) indexes months in the estimation window,
\(R^{e}_{M,\tau}\) is the market excess return,
\(\epsilon_{i,\tau}\) is the residual from regressing the stock's
monthly excess return on the market excess return, and \(E[\cdot]\)
denotes the sample average over the estimation window.
&
Estimated over the preceding 60 months, with at least 24 valid monthly observations.
\\

IVOL
&
The stock's monthly idiosyncratic volatility is calculated as the standard deviation of the residuals from a daily Fama--French
three-factor regression of the stock's excess return.
&
Estimated within month \(m\), with at least 15 valid daily observations.
\\

MAX
&
The average of the five highest daily stock returns during the month,
used as a proxy for lottery-like payoff characteristics.
&
Calculated within month \(m\), with at least 15 valid daily observations.
\\

DISP
&
The standard deviation of analysts' EPS forecasts for the nearest
unreported fiscal year, divided by the absolute value of their
mean forecast.
&
Calculated using the most recent I/B/E/S forecast summary available at
the start of month \(m\), provided that the summary is no more than
three months old.
\\

\(\mathrm{I/A}\)
&
The annual growth rate of total assets is measured by the change in book assets divided by lagged book assets.
&
Fiscal-year information ending in calendar year \(y\) is used from July
of year \(y+1\) through June of year \(y+2\).
\\

ROE
&
The quarterly operating proﬁtability is measured by quarterly income before extraordinary items divided by one-quarter-lagged
book equity.
&
Calculated using the most recent quarterly financial information available
by the end of month \(m\), provided that it was announced no more than
180 days earlier.
\\

\end{longtable}
\endgroup

\end{document}